\documentclass[12pt, a4paper, oneside, openright, abstract=on]{article}
\pdfoutput=1
\usepackage{amsmath}
\usepackage{amssymb}
\usepackage[english]{babel}
\usepackage{bbm}
\usepackage{bm}
\usepackage{booktabs}
\usepackage[font=small,labelfont=bf,format=plain]{caption}
\usepackage{cite}
\usepackage{dsfont}
\usepackage{emptypage}
\usepackage[T1]{fontenc}        
\usepackage{textcomp,gensymb}
\usepackage[paper=a4paper, inner=20mm, outer=20mm, top=30mm, bottom=30mm]{geometry}
\usepackage{graphicx}
\usepackage{icomma}
\usepackage{ifthen}
\usepackage[utf8x]{inputenc}    
\usepackage{listings}
\usepackage{longtable}
\usepackage{lmodern}
\usepackage{mathrsfs}
\usepackage{mathtools}
\usepackage{mdframed}
\usepackage{multirow}
\usepackage{physics}
\usepackage[onehalfspacing]{setspace}
\usepackage{slashed}
\usepackage{soul}
\usepackage{stmaryrd}
\usepackage[list=true]{subcaption}
\usepackage{subdepth}
\usepackage{tabularx}
\usepackage[many]{tcolorbox}
\usepackage[absolute]{textpos}
\usepackage{verbatim}
\usepackage{xparse}
\usepackage{xspace}
\usepackage{wrapfig}
\usepackage{hyperref}
\usepackage{tocbasic}
\DeclareTOCStyleEntry[
  beforeskip=.7em plus 1pt,
  pagenumberformat=\textbf
]{tocline}{section}

\allowdisplaybreaks

\newcommand{\eqn}[1]{Eq.\,(\ref{#1})}
\newcommand{\eqs}[1]{Eqs.\,(\ref{#1})}
\newcommand{\fig}[1]{Fig.\,\ref{#1}}
\newcommand{\figs}[1]{Figs.\,\ref{#1}}

\newcommand{\sct}[1]{Sect.\,\ref{#1}}
\newcommand{\scts}[1]{Sects.\,\ref{#1}}
\newcommand{\cha}[1]{Sec.\,\ref{#1}}
\newcommand{\appx}[1]{App.\,\ref{#1}}
\newcommand{\citere}[1]{Ref.~\cite{#1}}
\newcommand{\citeres}[1]{Refs.~\cite{#1}}

\newcommand{\OS}{\text{OS}}
\newcommand{\MOM}{\text{MOM}}

\newcommand{\MSbar}{{\overline{\text{MS}}}}
\newcommand{\DR}{\text{DR}}
\newcommand{\DRbar}{{\overline{\text{DR}}}}

\newcommand{\TRe}
{\widetilde{\text{Re}}}
\newcommand{\TCo}
{\widetilde{\text{Co}}}

\newcommand{\CP}{\mathcal{CP}}

\newcommand{\oc}{\textsc{OneCalc}}
\newcommand{\tc}{\textsc{TwoCalc}}

\newcommand{\deltaOL}{\delta^{(1)}}
\newcommand{\deltaTL}{\delta^{(2)}}
\newcommand{\zho}{Z_{\mathcal{H}_1}}
\newcommand{\zht}{Z_{\mathcal{H}_2}}

\newcommand{\nc}{N_c}
\newcommand{\alem}{\alpha_\text{em}}
\newcommand{\als}{\alpha_s}
\newcommand{\alt}{\alpha_t}
\newcommand{\alb}{\alpha_b}
\newcommand{\alq}{\alpha_q}

\newcommand{\cw}{c_\text{w}}
\newcommand{\sw}{s_\text{w}}

\newcommand{\Xt}{X_t}
\newcommand{\XtC}{X_t^*}
\newcommand{\At}{A_t}

\newcommand{\Xb}{X_b}
\newcommand{\XbC}{X_b^*}
\newcommand{\Ab}{A_b}

\newcommand{\Xq}{X_q}

\newcommand{\muC}{\mu^*}

\newcommand{\al}{\alpha}
\newcommand{\bc}{\beta_\text{c}}
\newcommand{\bn}{\beta_\text{n}}
\newcommand{\ca}{c_\alpha}
\newcommand{\sa}{s_\alpha}
\newcommand{\cb}{c_\beta}
\newcommand{\sib}{s_\beta}
\newcommand{\tb}{t_\beta}

\newcommand{\ctb}{c_{2\beta}}

\newcommand{\camb}{c_{\alpha - \beta}}
\newcommand{\samb}{s_{\alpha - \beta}}
\newcommand{\capb}{c_{\alpha + \beta}}
\newcommand{\sapb}{s_{\alpha + \beta}}
\newcommand{\ctamb}{c_{2(\alpha - \beta)}}
\newcommand{\stamb}{s_{2(\alpha - \beta)}}
\newcommand{\ctapb}{c_{2(\alpha + \beta)}}
\newcommand{\stapb}{s_{2(\alpha + \beta)}}

\newcommand{\imag}{\text{i}}

\newcommand{\Mudim}{\mu_\text{D}}
\newcommand{\MudimBar}{\Bar{\mu}_\text{D}}
\newcommand{\EulerGamma}{\gamma_E}
\newcommand{\del}{\varepsilon}
\newcommand{\Div}{\text{Div}}

\newcommand{\gev}{\;\text{GeV}}
\newcommand{\tev}{\;\text{TeV}}

\begin{document}

\thispagestyle{empty}
\def\thefootnote{\fnsymbol{footnote}}

\begin{center}
\begin{flushright}
\begin{footnotesize}
DESY-23-156
\end{footnotesize}
\end{flushright}
\vspace{2em}
\begin{large}
\textbf{Complete electroweak \texorpdfstring{$\order{\nc^2}$}{O(Nc2)}
        two-loop contributions \\[.8em]
        to the Higgs boson masses in the MSSM \\[.8em]
        and aspects of two-loop renormalisation}
\end{large}\\
\vspace{3em}
 {
Henning Bahl$^{1,2}$\footnotetext[0]{bahl@thphys.uni-heidelberg.de, 
daniel.meuser@studium.uni-hamburg.de, georg.weiglein@desy.de
},
Daniel Meuser$^{3,4}$,
and
Georg Weiglein$^{3,4}$
 }\\[2em]
 {\sl $^1$ University of Chicago, Department of Physics and Enrico Fermi Institute,\\[0.2em] 5720 South Ellis Avenue, Chicago, IL 60637 USA}\\[0.2em]
 {\sl $^2$Institut für Theoretische Physik, Universität Heidelberg,\\[0.2em] Philosophenweg 16, 61920 Heidelberg, Germany}\\[0.2em]
 {\sl $^3$    Deutsches Elektronen-Synchrotron DESY, Notkestr.~85, 22607 Hamburg, Germany}\\[0.2em]
 {\sl $^{4}$ II.\  Institut f\"ur  Theoretische  Physik, Universit\"at  Hamburg, Luruper Chaussee 149,\\ 22761 Hamburg, Germany}
\def\thefootnote{\arabic{footnote}}
\setcounter{page}{0}
\setcounter{footnote}{0}
\end{center}
\vspace{2ex}

\renewcommand{\abstractname}{Abstract}
\begin{abstract}
Results for the full electroweak two-loop contributions of $\order{\nc^2}$, where $\nc$ is the colour factor, to the Higgs-boson masses in the MSSM are obtained using a Feynman-diagrammatic approach including the full dependence on the external momentum. These corrections are expected to constitute the dominant part of the two-loop corrections that were still missing up to now. As a consequence of working at $\order{\nc^2}$, the relevant two-loop self-energies decompose into products of one-loop integrals, giving rise to a transparent analytical structure of the self-energies. We compare different renormalisation schemes for $\tan(\beta)$, the ratio of the vacuum expectation values of the two Higgs doublets, and demonstrate under which conditions different renormalisation schemes can be related to each other via a simple reparametrisation. We explicitly show that this is in general not possible for mixed renormalisation schemes due to the presence of evanescent terms. In our numerical analysis, the new corrections are compared with already known two-loop contributions and the experimental uncertainty of the mass of the observed Higgs boson. While smaller than the already known two-loop corrections, the new terms are typically larger in size than the experimental uncertainty. This underlines the relevance of the so-far unknown electroweak two-loop contributions.
\end{abstract}

\renewcommand{\thefootnote}{\arabic{footnote}}

\newpage
\setcounter{tocdepth}{2}
\tableofcontents
\newpage

\section{Introduction}
\label{cha:intro}
In 2012, a scalar particle with a mass of approximately $125 \gev$ has been discovered at the \textit{Large Hadron Collider} (LHC) at the \textit{European Organisation for Nuclear Research} (CERN) \cite{ATLAS:2012yve,CMS:2012qbp,ATLAS:2015yey}. The observed properties of this particle are in agreement with the properties of the Higgs boson predicted by the \textit{Standard Model of Particle Physics} (SM) within the current experimental and theoretical uncertainties \cite{ATLAS:2016neq,ATLAS:2022vkf,CMS:2022dwd}. The combination of the most recent measurements of the ATLAS collaboration yields an observed Higgs boson mass of $M_h = 125.11 \pm 0.11 \gev$ \cite{ATLAS:2023oaq,ATLAS:2023owm}, and the most recent measurement from CMS in the four lepton final state yields the value of $M_h = 125.08 \pm 0.12 \gev$ \cite{CMSMhnew}. While the observation of this particle elucidates the mechanism through which the massive vector bosons and charged fermions obtain their masses, there are still many open questions with respect to the nature of \textit{electroweak symmetry breaking} (EWSB) that are left to be answered.
 
A commonly studied extension of the SM is the \textit{Minimal Supersymmetric extension of the Standard Model} (MSSM) \cite{Nilles:1983ge,Haber:1984rc}, which is able to address several of the open issues of the SM. The Higgs sector of the MSSM  consists of two Higgs doublets with five physical Higgs bosons, three neutral ones and a charged pair. At the tree level, the Higgs boson masses are fully determined by (experimentally known) SM parameters and two additional parameters, one of which is a Higgs boson mass. The remaining MSSM Higgs boson masses can therefore be predicted. In the SM, on the other hand, no such prediction is possible as the Higgs boson mass is an input parameter of the theory.

At the tree-level, the mass of the lightest MSSM Higgs boson is bounded from above by the mass of the $Z$ boson \cite{Inoue:1982ej} ($M_Z \approx 91 \gev$), and it is therefore not in agreement with the observed value. The large size of the quantum corrections shifts the predicted value closer to the observed one, rendering a precision calculation crucial in order to profit from the high experimental accuracy in order to identify the phenomenologically viable parameter space of the model. Predicting the masses of the MSSM Higgs bosons and restricting the parameter space of the theory to match the experimental observations provides an important test of the model. 

Different methods are applied in order to obtain an accurate prediction for the MSSM Higgs boson masses. For a SUSY scale $M_S$ not much larger than the electroweak scale, the calculation of Higgs boson self-energies in terms of \textit{Feynman diagrams} (FD) in a (sufficiently high) fixed order of perturbation theory yields a reliable result \cite{Li:1984tc,Gunion:1989dp,Berger:1989hg,Okada:1990vk,Ellis:1990nz,Haber:1990aw,Barbieri:1991tk,Ellis:1991zd,Brignole:1991pq,Chankowski:1991md,Brignole:1991wp,Brignole:1992uf,Chankowski:1992er,Dabelstein:1994hb,Pierce:1996zz,Hempfling:1993qq,Heinemeyer:1998jw,Heinemeyer:1998kz,Zhang:1998bm,Heinemeyer:1998np,Espinosa:1999zm,Carena:2000dp,Espinosa:2000df,Degrassi:2001yf,Brignole:2001jy,Brignole:2002bz,Dedes:2002dy,Dedes:2003km,Allanach:2004rh,Heinemeyer:2004xw,Frank:2013hba,Hollik:2015ema,Martin:2002wn,Martin:2002iu,Martin:2004kr,Martin:2003qz,Martin:2005qm,Heinemeyer:2010mm,Fritzsche:2011nr,Fritzsche:2013fta,Borowka:2014wla,Degrassi:2014pfa,Carter:2010hi,Borowka:2012yc,Borowka:2015ura,Borowka:2018anu,Pilaftsis:1998dd,Demir:1999hj,Pilaftsis:1999qt,Choi:2000wz,Carena:2000yi,Ibrahim:2000qj,Heinemeyer:2001qd,Carena:2001fw,Ibrahim:2002zk,Ellis:2004fs,Frank:2006yh,Heinemeyer:2007aq,Heinemeyer:2004by,Cao:2006xb,Brignole:2015kva,Arana-Catania:2011rnb,Gomez:2014uha,Hollik:2014wea,Hollik:2014bua,Hahn:2015gaa,Passehr:2017ufr,Goodsell:2019zfs,Domingo:2021kud,R:2021bml}. For a large SUSY scale, the appearance of large logarithms spoils the accuracy of the fixed-order prediction. These large logarithms can be resummed by making use of the \textit{renormalisation group} (RG) within an \textit{effective field theory} (EFT) approach \cite{Barbieri:1990ja,Espinosa:1991fc,Casas:1994us,Haber:1993an,Espinosa:1999zm,Carena:2000dp,Carena:1995bx,Carena:1995wu,Haber:1996fp,Degrassi:2002fi,Martin:2007pg,Hahn:2013ria,Draper:2013oza,Arkani-Hamed:2004ymt,Giudice:2004tc,Carena:2008rt,Binger:2004nn,Bernal:2007uv,Giardino:2011aa,Giudice:2011cg,Bagnaschi:2014rsa,Tamarit:2012ie,Benakli:2013msa,Fox:2005yp,Hall:2009nd,Cabrera:2011bi,Degrassi:2012ry,PardoVega:2015eno,Bagnaschi:2017xid,Harlander:2018yhj,Bagnaschi:2019esc,Bahl:2019wzx,Bahl:2020tuq,Carena:2015uoe,Murphy:2019qpm,Gorbahn:2009pp,Bahl:2018jom,Lee:2015uza,Bahl:2020jaq,BhupalDev:2014bir,Bednyakov:2018cmx,Schienbein:2018fsw,Oredsson:2018yho,Herren:2017uxn,Bagnaschi:2015pwa,Bagnaschi:2015hka,Bahl:2019ago,Cheung:2014hya,Kwasnitza:2021idg}. The hybrid approach combines the FD and EFT methods  and thus yields accurate results also for intermediate values of $M_S$ \cite{Hahn:2013ria,Bahl:2016brp,Bahl:2017aev,Bahl:2018jom,Bahl:2018ykj,Bahl:2019hmm,Bahl:2020tuq,Bahl:2020mjy,Bagnaschi:2018ofa,Sobolev:2020cjh,Athron:2016fuq,Staub:2017jnp,Athron:2017fvs,Kwasnitza:2020wli,Harlander:2019dge}. An overview of the different approaches is given in \citere{Slavich:2020zjv}. The remaining theoretical uncertainties of the prediction for the mass of the SM-like Higgs boson have been estimated to be between $\sim 0.5\gev$ for vanishing stop mixing and $\sim 1 \gev$ for large stop mixing \cite{Bahl:2019hmm,Kwasnitza:2020wli}. This theoretical uncertainty clearly exceeds the experimental error of the measured Higgs boson mass, necessitating the calculation of further higher-order corrections.

In this paper, we focus on two-loop corrections in the Feynman-diagrammatic approach. Two-loop corrections to the neutral Higgs boson masses, in the limit of vanishing external momentum and vanishing electroweak gauge couplings, have been calculated in \citeres{Hempfling:1993qq,Heinemeyer:1998jw,Heinemeyer:1998kz,Zhang:1998bm,Heinemeyer:1998np,Espinosa:1999zm,Carena:2000dp,Espinosa:2000df,Degrassi:2001yf,Brignole:2001jy,Brignole:2002bz,Dedes:2002dy,Dedes:2003km,Allanach:2004rh,Heinemeyer:2004xw,Heinemeyer:2007aq,Hollik:2014wea,Hollik:2014bua,Passehr:2017ufr}. The corresponding two-loop calculations for the mass of the charged Higgs boson were performed in \citeres{Frank:2013hba,Hollik:2015ema}. The effective-potential method allowed for the incorporation of electroweak two-loop effects into the predictions for the Higgs boson masses in the MSSM, still in the limit of vanishing external momentum \cite{Martin:2002iu,Martin:2002wn}. Moreover, the full two-loop contributions involving all diagrams in which ${\als \equiv g_s^2/(4 \pi)}$, ${\alt \equiv h_t^2/(4 \pi)}$, ${\alb \equiv h_b^2/(4 \pi)}$, or ${\al_\tau \equiv h_\tau^2/(4 \pi)}$ appear have been calculated \cite{Martin:2004kr}, where $g_s$ denotes the strong gauge coupling, and $h_t$, $h_b$ and $h_\tau$ are third generation Yukawa couplings. For these diagrams, the full dependence on the external momentum $p^2$ was kept, as well as a non-vanishing fine-structure constant $\alem \equiv e^2/(4 \pi)$. Those numerical results, however, do not include contributions from the first or second generation of fermions, or generation mixing. Furthermore, the results of \citere{Martin:2004kr} were obtained in a pure $\DRbar$ scheme, which greatly simplifies the renormalisation process. The leading two-loop contributions of $\order{\alt \als}$ including the contributions from a non-vanishing external momentum were calculated in a mixed $\OS$-$\DRbar$ scheme in \citere{Borowka:2014wla}. \citere{Degrassi:2014pfa} also incorporated the subleading $\order{\alem \als}$ contributions, working both in a mixed and in a pure $\DRbar$ scheme. \citere{Borowka:2018anu} included all QCD contributions, giving results of $\order{\alq \als}$ and $\order{\alem \als}$, where $\alq$ denotes a product of any two Yukawa couplings. In \citere{Borowka:2018anu}, a mixed $\OS$-$\DRbar$ scheme was used, and complex parameters were taken into account. The leading Yukawa corrections of $\order{(\alt + \alb)^2}$ are given in \citeres{Hollik:2014wea,Hollik:2014bua,Hahn:2015gaa,Passehr:2017ufr} for the case of general complex parameters and vanishing external momentum.

In this paper, we go beyond these works and calculate all electroweak two-loop corrections of $\order{(\alem + \alq)^2 \nc^2}$, where, as above, $\alem$ is the fine-structure constant, $\alq$ denotes any product of the top and bottom Yukawa couplings, and $\nc$ is the number of quark colours in the theory. \footnote{In the context of electroweak precision observables in the SM, contributions of $\order{\nc^2}$ and $\order{\nc^3}$ have been investigated in \citeres{Weiglein:1998jz,AchimDipl,Chen:2020xzx,Chen:2020xot}.} We take into account the full momentum dependence of the self-energies. All mixing contributions between the Higgs bosons, the (would-be) Goldstone bosons and the electroweak gauge bosons are incorporated. The obtained predictions are valid also for the case of large mixing between the lowest-order mass eigenstates. For the first time, we perform a two-loop prediction including pure gauge contributions in combination with an on-shell renormalisation scheme. Based on the enhancement by $\nc^2$ and experiences from calculations in the SM~\cite{Weiglein:1998jz,AchimDipl,Chen:2020xzx,Chen:2020xot}, we expect these contributions to be the dominant electroweak corrections beyond the ones that are already known.

To obtain the desired contributions, we have to perform a complete renormalisation of the MSSM Higgs-gauge sector at the two-loop level under the full inclusion of electroweak effects. This leads to more complicated relations between the two-loop counterterms in comparison to what has been encountered up to now. These relations have to be used in order to obtain a finite result in the general case where all electroweak two-loop contributions entering at $\order{\nc^2}$ are taken into account. From our analysis of the structure of the two-loop self-energies, we can infer under which conditions results in different renormalisation schemes can be related to each other by employing a simple reparametrisation. We will also show that our newly calculated contributions are larger than the experimental uncertainty of the mass of the observed Higgs boson, and hence they should be incorporated in the theoretical predictions as a further improvement of the theoretical precision.

This paper is structured as follows. In \cha{cha:mssm_ren}, we discuss the renormalisation of the Higgs and the gauge sector of the MSSM at the one- and two-loop level. The actual calculation of the $\order{(\alem + \alq)^2 \nc^2}$ corrections is discussed in \cha{cha:calculation}. In \cha{sec:del_cancellation}, we demonstrate under which circumstances calculations within different renormalisation schemes can be related to each other using a simple reparametrisation. In \cha{cha:two-loop}, we investigate how our newly calculated contributions affect the Higgs boson mass prediction in five different MSSM scenarios. We draw our conclusions in \cha{cha:conclusions}. A series of appendices provides additional details of our calculation. Further details can be found in \citere{Meuser:2023}.

\section{Renormalisation of the Higgs and the gauge sector}
\label{cha:mssm_ren}
In order to obtain predictions for the Higgs boson masses in the MSSM at $\order{(\alem + \alq)^2 \nc^2}$, two sectors of the model have to be renormalised. A renormalisation of the quark-squark sector is needed at the one-loop level, while the Higgs-gauge sector (which we treat as a common sector in the following) needs to be renormalised up to the two-loop order.

Here, we focus on the renormalisation of the Higgs-gauge sector, for which $\tan(\beta)$ plays an important role. The renormalisation of the quark-squark sector is discussed in detail in \appx{app:squark_ren}. Expressions for dependent counterterms are given in the present section and in \appx{app:CTs}. Below we first fix the notation for the Higgs and gauge sector in the MSSM, following closely \citere{Frank:2006yh}. We then discuss the relevant parameters of this sector and give the renormalisation transformations for the parameters and fields. From these, we derive the resulting expressions for the renormalised tadpole and self-energy diagrams up to two-loop order. We explain the renormalisation of each independent parameter, usually in the form of a renormalisation condition and a formula for the counterterm. The renormalisation of $\tan(\beta)$ is discussed in \sct{sec:tanbeta_ren}.

\subsection{Higgs and gauge sector at the tree-level}
\label{ssec:Higgs_gauge_tree}

The MSSM Higgs Lagrangian contains, inter alia, the following terms \cite{Frank:2006yh}:
\begin{equation}
\begin{split}
    \mathcal{L}_\text{Higgs} \supset{}& - (m_1^2 + \abs{\mu}^2) \mathcal{H}_1^\dagger \mathcal{H}_1 - (m_2^2 + \abs{\mu}^2) \mathcal{H}_2^\dagger \mathcal{H}_2 + \left( m_{12}^2 \mathcal{H}_1 \cdot \mathcal{H}_2 + \text{h.c.} \right) \\
    & - \tfrac{1}{8} (g^2 + g'^2)(\mathcal{H}_1^\dagger \mathcal{H}_1 - \mathcal{H}_2^\dagger \mathcal{H}_2) - \tfrac{1}{2} g'^2 \abs{\mathcal{H}_1^\dagger \mathcal{H}_2}^2.
\end{split}
\end{equation}
In the first line, we used the SU(2) product $a \cdot b = a_1 b_2 - a_2 b_1$, where $a$ and $b$ are SU(2) doublets. Furthermore, the gauge couplings $g$ and $g'$, and the potentially complex higgsino mass parameter $\mu$ appear. The parameters $m_1^2$, $m_2^2$, and $m_{12}^2$, of which the latter is possibly complex, break supersymmetry softly. The phase of $m_{12}^2$ can be removed by a Peccei-Quinn transformation \cite{Peccei:1977hh,Peccei:1977ur,Dimopoulos:1995kn}. From this point on, we will treat $m_{12}^2$ as a real parameter.

We write the Higgs doublets in terms of component fields:
\begin{subequations}
\begin{align}
    \mathcal{H}_1 ={}&
    \begin{pmatrix}
        v_1 + \tfrac{1}{\sqrt{2}}(\phi_1 - \imag \chi_1) \\
        - \phi_1^-
    \end{pmatrix}, \\
    \mathcal{H}_2 ={}& e^{\imag \xi}
    \begin{pmatrix}
        \phi_2^+ \\
        v_2 + \tfrac{1}{\sqrt{2}}(\phi_2 + \imag \chi_2)
    \end{pmatrix},
\end{align}
\end{subequations}
where $v_1$ and $v_2$ are the vacuum expectation values of the Higgs doublets, and $\xi$ is a phase between the doublets. The doublets have hypercharges $Y_{\mathcal{H}_1} = - 1$ and $Y_{\mathcal{H}_2} = + 1$ \cite{Drees:2004jm}. They couple to down- and up-type (s)fermions, respectively.

In terms of the component fields, the linear and quadratic terms of the Higgs Lagrangian are
\begin{equation}
\label{eqn:Higgs_Lag_gauge}
\begin{split}
    \mathcal{L}^\text{lin.+bil.}_\text{Higgs} ={}& T_{\phi_1} \, \phi_1 + T_{\phi_2} \, \phi_2 + T_{\chi_1} \, \chi_1 + T_{\chi_2} \, \chi_2 \\
    &+ \tfrac{1}{2} (\partial_\mu \phi_i) (\partial^\mu \phi_i) + \tfrac{1}{2} (\partial_\mu \chi_i) (\partial^\mu \chi_i) + (\partial_\mu \phi_i^+) (\partial^\mu \phi_i^-) \\
    &- \tfrac{1}{2}
    \begin{pmatrix}
    \phi_1 & \phi_2 & \chi_1 & \chi_2
    \end{pmatrix}
    \begin{pmatrix}
    \textbf{M}^2_{\phi\phi} & \textbf{M}^2_{\phi\chi} \\
    \textbf{M}^2_{\chi\phi} & \textbf{M}^2_{\chi\chi}
    \end{pmatrix}
    \begin{pmatrix}
    \phi_1 \\ \phi_2 \\ \chi_1 \\ \chi_2
    \end{pmatrix} \\
    &-
    \begin{pmatrix}
    \phi_1^+ & \phi_2^+
    \end{pmatrix}
    \textbf{M}^2_{\phi^-\phi^+}
    \begin{pmatrix}
    \phi_1^- \\ \phi_2^-
    \end{pmatrix}.
\end{split}
\end{equation}
The mass (sub-)matrices, whose entries are given in \citere{Frank:2006yh}, fulfil the following relations:
\begin{subequations}
\label{eqs:Higgs_mass_trace}
\begin{align}
    \big( \textbf{M}^2_{\phi\phi} \big)^\text{T} ={}& \textbf{M}^2_{\phi\phi}, \\
    \big( \textbf{M}^2_{\phi\chi} \big)^\text{T} ={}& \textbf{M}^2_{\chi\phi} = -\textbf{M}^2_{\phi\chi}, \\
    \big( \textbf{M}^2_{\chi\chi} \big)^\text{T} ={}& \textbf{M}^2_{\chi\chi}, \\
    \big( \textbf{M}^2_{\phi^-\phi^+} \big)^\dagger ={}& \textbf{M}^2_{\phi^-\phi^+}.
\end{align}
\end{subequations}

In the Higgs--gauge sector, we now have eight independent real parameters: $g$, $g'$, $v_1$, $v_2$, $m_1^2$, $m_2^2$, $m_{12}^2$, and $\xi$. It should be noted that we do not consider $\mu$ to be part of the Higgs--gauge but rather the chargino--neutralino sector; we address its renormalisation in \sct{ssec:DRbarMUEAq}. This set of parameters is, however, not the most convenient to work with. To obtain input parameters which can be linked to physical observables more easily, we replace the gauge couplings and vacuum expectation values (VEVs) by the elementary charge $e$, the gauge boson masses $M_Z$ and $M_W$, and the VEV ratio $\tan(\beta) = \tb$:
\begin{subequations}
\begin{align}
    e ={}& g \cw = g' \sw, \\
    M_Z^2 ={}& \tfrac{1}{2} (g^2 + g'^2)(v_1^2 + v_2^2), \\
    M_W^2 ={}& \tfrac{1}{2} g'^2(v_1^2 + v_2^2) = M_Z^2 \cw^2, \\
    \tb ={}& \frac{v_2}{v_1}.
\end{align}
\end{subequations}
$\cw$ and $\sw$ are the cosine and sine of the weak-mixing angle $\theta_\text{w}$, respectively.

With these definitions, the mass matrices introduced in \eqn{eqn:Higgs_Lag_gauge} fulfil
\begin{subequations}
\begin{align}
    \Tr \textbf{M}^2_{\phi\phi} ={}& \Tr\textbf{M}^2_{\chi\chi} + M_Z^2, \\
    \Tr\textbf{M}^2_{\phi^-\phi^+} ={}& \Tr\textbf{M}^2_{\chi\chi} + M_W^2.
\end{align}
\end{subequations}

To replace the four remaining (unphysical) input parameters, we first rotate the Higgs component fields into their mass eigenstates:
\begin{subequations}
\begin{align}
    \begin{pmatrix}
    h \\ H \\ A \\ G 
    \end{pmatrix} ={}& 
    \begin{pmatrix}
    -\sa & \ca & 0 & 0 \\
    \ca & \sa & 0 & 0 \\
    0 & 0 & - s_{\bn} & c_{\bn} \\
    0 & 0 & c_{\bn} & s_{\bn}
    \end{pmatrix}
    \begin{pmatrix}
    \phi_1 \\ \phi_2 \\ \chi_1 \\ \chi_2 
    \end{pmatrix}, \\
    \begin{pmatrix}
    H^\pm \\ G^\pm
    \end{pmatrix} ={}&
    \begin{pmatrix}
    - s_{\bc} & c_{\bc} \\
    c_{\bc} & s_{\bc}
    \end{pmatrix}
    \begin{pmatrix}
    \phi_1^\pm \\ \phi_2^\pm
    \end{pmatrix},
\end{align}
\end{subequations}
where $c_x = \cos(x)$ and $s_x = \sin(x)$ for $x \in \{\al, \bn, \bc\}$. We use the same notation for all linear combinations of these angles. We call the $\{\phi_1,\phi_2,\chi_1,\chi_2\}$-basis the gauge eigenstates and the $\{h, H, A, G\}$-basis the tree-level mass eigenstates; we use the same terms for the charged sector. The fields $G$ and $G^\pm$ are the unphysical (would-be) Goldstone fields. Expressed in terms of mass eigenstates, the linear and quadratic terms of the Higgs Lagrangian read
\begin{equation}
\label{eqn:Higgs_Lag_mass}
\begin{split}
    \mathcal{L}^\text{lin.+bil.}_\text{Higgs} ={}& T_h \, h + T_H \, H + T_A \, A + T_G \, G \\
    &+ \tfrac{1}{2} (\partial_\mu h) (\partial^\mu h) + \tfrac{1}{2} (\partial_\mu H) (\partial^\mu H) + \tfrac{1}{2} (\partial_\mu A) (\partial^\mu A) \\
    &+ \tfrac{1}{2} (\partial_\mu G) (\partial^\mu G) + (\partial_\mu H^+) (\partial^\mu H^-) + (\partial_\mu G^+) (\partial^\mu G^-) \\
    &- \tfrac{1}{2}
    \begin{pmatrix}
    h & H & A & G
    \end{pmatrix}
    \begin{pmatrix}
    m_h^2 & m_{hH}^2 & m_{hA}^2 & m_{hG}^2 \\
    m_{hH}^2 & m_H^2 & m_{HA}^2 & m_{HG}^2 \\
    m_{hA}^2 & m_{HA}^2 & m_A^2 & m_{AG}^2 \\
    m_{hG}^2 & m_{HG}^2 & m_{AG}^2 & m_G^2
    \end{pmatrix}
    \begin{pmatrix}
    h \\ H \\ A \\ G
    \end{pmatrix} \\
    & -
    \begin{pmatrix}
    H^+ & G^+
    \end{pmatrix}
    \begin{pmatrix}
    m_{H^\pm}^2 & m_{H^-G^+}^2 \\
    m_{G^-H^+}^2 & m_{G^\pm}^2
    \end{pmatrix}
    \begin{pmatrix}
    H^- \\ G^-
    \end{pmatrix}.
\end{split}
\end{equation}
It is important to note that we use a different convention for labelling the off-diagonal entries of the charged Higgs boson mass matrix than \citere{Frank:2006yh}, which leads to differences also in the respective counterterms. Instead, our charged counterterms agree with the expressions given in \citere{Fritzsche:2013fta}.

To complete our choice of physical input parameters, we choose the tadpole coefficients $T_h$, $T_H$ and $T_A$, as well as one of the masses $m_A^2$/$m_{H^\pm}^2$. If all MSSM parameters are real, we refer to the model as the rMSSM. In this case, the $\CP$-odd pseudoscalar $A$ does not mix with the $\CP$-even scalars $h$ and $H$, and we use $m_A^2$ as an input parameter. In the cMSSM, the MSSM with complex parameters, $A$ mixes with the $\CP$-even states through loop corrections, and the $\CP$ eigenstate is not a mass eigenstate anymore. Instead, we use the charged mass $m_{H^\pm}^2$ as input in this case.

To sum up, in the Higgs-gauge sector, we use the input parameters
\begin{equation}
\label{eqn:input_Higgs_gauge}
    e, M_Z^2, M_W^2, m_A^2/m_{H^\pm}^2, T_h, T_H, T_A, \tb.
\end{equation}
The remaining mass parameters and $T_G$ can be expressed in terms of these input parameters and the mixing angles $\al$, $\bn$ and $\bc$. These relations are needed to derive counterterm expressions and can be found in \citere{Frank:2006yh}. As stated above, we use a different convention for the off-diagonal charged mass counterterms; our $m_{H^-G^+}^2$ is denoted by $m_{G^-H^+}^2$ in \citere{Frank:2006yh} and vice versa.

The trace of a matrix is invariant under a unitary transformation, so \eqs{eqs:Higgs_mass_trace} can be rewritten by the masses defined in \eqn{eqn:Higgs_Lag_mass}:
\begin{subequations}
\label{eqs:Higgs_mass_rel}
\begin{align}
    m_h^2 + m_H^2 ={}& m_A^2 + m_G^2 + M_Z^2, \\
    m_{H^\pm}^2 + m_{G^\pm}^2 ={}& m_A^2 + m_G^2 + M_W^2.
\end{align}
\end{subequations} 
These relations hold before and after applying the minimisation conditions for the Higgs potential.

At tree-level, and without including a contribution from the gauge-fixing part of the Lagrangian (see below), $m_G^2$ and $m_{G^\pm}^2$ vanish and we get the familiar relations
\begin{subequations}
\begin{align}
    m_h^2 + m_H^2 ={}& m_A^2 + M_Z^2, \\
    m_{H^\pm}^2 ={}& m_A^2 + M_W^2.
\end{align}
\end{subequations}

The tree-level masses of the $\CP$-even fields are given by
\begin{equation}
\label{eqn:CP-even_tree-level_masses}
    m_{h/H}^2 = \tfrac{1}{2} \left(m_A^2 + M_Z^2 \mp \sqrt{(m_A^2 + M_Z^2)^2 - 4 m_A^2 M_Z^2 \ctb^2} \right),
\end{equation}
where $m_h^2 \leq m_H^2$. The tadpoles, the phase $\xi$, the off-diagonal mass terms, and the $m_G^2$ and $m_{G^\pm}^2$ entries of the mass matrices vanish at tree-level. The mixing angles at the minimum are
\begin{equation}
    \bn = \bc = \beta, \quad 0 < \beta < \tfrac{\pi}{2},
\end{equation}
and
\begin{equation}
    \al = \arctan \left[ \frac{- (m_A^2 + M_Z^2)\sib \cb}{M_Z^2 \cb^2 + m_A^2 \sib^2 - m_h^2} \right], \quad - \tfrac{\pi}{2} < \al < 0.
\end{equation}

The bilinear MSSM Lagrangian for electroweak gauge bosons is identical to its SM counterpart. It reads
\begin{equation}
\begin{aligned}
    \mathcal{L}_\text{gauge}^\text{bil.} ={}& -\tfrac{1}{2} \left( \partial_\mu A_\nu \partial^\mu A^\nu - \partial_\mu A_\nu \partial^\nu A^\mu \right) \\
    & -\tfrac{1}{2} \left( \partial_\mu Z_\nu \partial^\mu Z^\nu - \partial_\mu Z_\nu \partial^\nu Z^\mu \right) + \tfrac{1}{2}M_Z^2 Z_\mu Z^\mu \\
    & - \left( \partial_\mu W^+_\nu \partial^\mu W^{-\nu} - \partial_\mu W^+_\nu \partial^\nu W^{-\mu} \right) + M_W^2 W^+_\mu W^{-\mu}.
\end{aligned}
\end{equation}

There are also mixing terms between gauge and Higgs bosons:
\begin{equation}
\label{eqn:gauge_Higgs_mix}
    \mathcal{L}_\text{Higgs-gauge}^\text{bil.} = M_Z (\cb \chi_1 + \sib \chi_2)\partial^\mu Z_\mu + \left( \imag M_W (\cb \phi^+_1 + \sib \phi^+_2) \partial^\mu W^-_\mu + \text{h.c.} \right).
\end{equation}

In the 't Hooft-gauge, these mixing terms are exactly cancelled by the gauge-fixing terms at tree level. At higher orders, they generate counterterms which enter the renormalisation of the scalar--vector self-energies. In the following we adopt a prescription where the gauge-fixing part of the Lagrangian does not contribute to the renormalisation.

\subsection{Renormalisation transformations}

All parameters are renormalised via
\begin{equation}
\label{eqn:ren_trafo}
    p \to p + \delta p = p + \deltaOL p + \deltaTL p.
\end{equation}
This means in particular that $\tb \to \tb + \delta \tb$, which is also written in this way in e.g.\ \citeres{Fritzsche:2011nr,Fritzsche:2013fta}, but not in \citeres{Frank:2006yh,Borowka:2015ura}. It should be noted that the mixing angles $\al$, $\bn$ and $\bc$ are not renormalised. Only after the renormalisation transformation we set $\bn = \bc = \beta$. For the elementary charge, we write $e \to e + \delta e \equiv (1 + \delta Z_e) e$. All mass parameters in \eqn{eqn:Higgs_Lag_mass} are also renormalised in the form of \eqn{eqn:ren_trafo}. Since only the parameters given in \eqn{eqn:input_Higgs_gauge} are independent in the Higgs and gauge sector, most mass counterterms will be dependent quantities.

We renormalise the fields by
\begin{subequations}
\begin{align}
    \mathcal{H}_i \to{}& \sqrt{1 + \delta Z_{\mathcal{H}_i}} \, \mathcal{H}_i, \\
    \begin{pmatrix}
    h \\ H \\ A \\ G
    \end{pmatrix}
    \to{}&
    \begin{pmatrix}
    \sqrt{1 + \delta Z_{hh}} & \frac{1}{2}\delta Z_{hH} & 0 & 0 \\
    \frac{1}{2}\delta Z_{hH} & \sqrt{1 + \delta Z_{HH}} & 0 & 0 \\
    0 & 0 & \sqrt{1 + \delta Z_{AA}} & \frac{1}{2}\delta Z_{AG} \\
    0 & 0 & \frac{1}{2}\delta Z_{AG} & \sqrt{1 + \delta Z_{GG}}
    \end{pmatrix}
    \begin{pmatrix}
    h \\ H \\ A \\ G
    \end{pmatrix}, \\
    \begin{pmatrix}
    H^- \\ G^-
    \end{pmatrix}
    \to{}& 
    \begin{pmatrix}
    \sqrt{1 + \delta Z_{H^-H^+}} & \frac{1}{2} \delta Z_{H^-G^+} \\
    \frac{1}{2} \delta Z_{G^-H^+} & \sqrt{1 + \delta Z_{G^-G^+}}
    \end{pmatrix}
    \begin{pmatrix}
    H^- \\ G^-
    \end{pmatrix}, \\
    \begin{pmatrix}
    H^+ \\ G^+
    \end{pmatrix}
    \to{}& 
    \begin{pmatrix}
    \sqrt{1 + \delta Z_{H^-H^+}} & \frac{1}{2} \delta Z_{G^-H^+} \\
    \frac{1}{2} \delta Z_{H^-G^+} & \sqrt{1 + \delta Z_{G^-G^+}}
    \end{pmatrix}
    \begin{pmatrix}
    H^+ \\ G^+
    \end{pmatrix}, \\
    \begin{pmatrix}
    A_\mu \\ Z_\mu
    \end{pmatrix}
    \to{}&
    \begin{pmatrix}
    \sqrt{1 + \delta Z_{\gamma \gamma}} & \frac{1}{2} \delta Z_{\gamma Z} \\
    \frac{1}{2} \delta Z_{Z \gamma} & \sqrt{1 + \delta Z_{ZZ}}
    \end{pmatrix}
    \begin{pmatrix}
    A_\mu \\ Z_\mu
    \end{pmatrix}, \\
    W^\pm_\mu \to{}& \sqrt{1 + \delta Z_{WW}} \, W^\pm_\mu,
\end{align}
\end{subequations}
where $\delta Z = \deltaOL Z + \deltaTL Z$ as for the parameter renormalisation. All scalar field renormalisation constants are fixed by $\delta \zho$ and $\delta \zht$. As the MSSM Higgs sector is $\CP$-conserving at tree level, this means in particular that the mixing field renormalisation constants between the $\CP$-even and $\CP$-odd fields vanish at all orders. The counterterms for the masses in \eqn{eqn:Higgs_Lag_mass} are determined by the counterterms for our chosen input parameters. All relations between the one- and two-loop counterterms are collected in \appx{app:CTs}.

Applying the renormalisation transformation to \eqs{eqs:Higgs_mass_rel}, we find the relations
\begin{subequations}
\allowdisplaybreaks
\begin{align}
    \delta^{(n)} m_h^2 + \delta^{(n)} m_H^2 ={}& \delta^{(n)} m_A^2 + \delta^{(n)} m_G^2 + \delta^{(n)} M_Z^2, \\
    \label{eqn:mHpm_mA0_rel}
    \delta^{(n)} m_{H^\pm}^2 + \delta^{(n)} m_{G^\pm}^2 ={}& \delta^{(n)} m_A^2 + \delta^{(n)} m_G^2 + \delta^{(n)} M_W^2 .
\end{align}
\end{subequations}
These relations between the independent and the dependent counterterms hold to all orders. In the present work, the counterterms $\delta^{(n)} m_h^2$, $\delta^{(n)} m_H^2$, $\delta^{(n)} m_{G^\pm}^2$ and $\delta^{(n)} m_G^2$ are always dependent quantities while the gauge boson mass counterterms, $\delta^{(n)} M_Z^2$ and $\delta^{(n)} M_W^2$, are always defined in an on-shell scheme. Depending on the scenario, either $\delta^{(n)} m_A^2$ or $\delta^{(n)} m_{H^\pm}^2$ is defined on-shell as well, while the other one becomes a dependent counterterm.

\subsection{One-loop renormalisation conditions and counterterms}
\label{ssec:higgs_gauge_1L_ren}

In this section, we will discuss the one-loop renormalisation of all parameters and fields that are relevant for our $\order{(\alem + \alq)^2 \nc^2}$ calculation. While some of these counterterms only matter for the two-loop part of our work, most are relevant already for a one-loop prediction.

Here, we work with the renormalised one-point, $\hat \Gamma^{(1)}_i$, and two-point vertex functions, $\hat \Sigma^{(1)}_{ij}$. Their definition in terms of the unrenormalised vertex functions, $\Gamma^{(1)}_i$ and $\Sigma^{(1)}_{ij}$, can be found in \cha{ssec:tad_se_ren_1L}.

The tadpole counterterms are chosen such that the renormalised one-point vertex functions vanish:
\begin{subequations}
\begin{align}
    &\hat \Gamma^{(1)}_i \overset{!}{=} 0, \quad i \in \{h,H,A\}, \\
    \Rightarrow &\deltaOL T_i = - \Gamma^{(1)}_i.
\end{align}
\end{subequations}

Due to the relation
\begin{equation}
\label{eqn:TG_and_TA}
    T_G = -\tan(\beta - \beta_n) T_A,
\end{equation}
the counterterm $\deltaOL T_G$ vanishes:
\begin{equation}
    \deltaOL T_G = 0.
\end{equation}

Since the unrenormalised vertex function $\Gamma^{(1)}_G$ also vanishes, all renormalised Higgs one-point vertex functions can be set to zero simultaneously. This ensures that the vacuum expectation values in our calculation receive no shifts from loop corrections \cite{Peskin:1995ev,Denner:1991kt}.

We use the input masses $M_Z^2$, $M_W^2$, and $m_A^2$ (in the rMSSM) or $m_{H^\pm}^2$ (in the cMSSM). They are determined by expanding the pole equation to one-loop order and by identifying the squared physical mass $M_i^2$ with  the real part of the complex pole,
\begin{equation}
    M_i^2 - m_i^2 + \Re \hat \Sigma^{(1)}_{ii}(M_i^2) \overset{!}{=} 0.
\end{equation}

In an $\OS$ scheme, $M_i^2 = m_i^2$, and we find
\begin{subequations}
\begin{alignat}{2}
    \deltaOL M_Z^2 ={}& \Re \Sigma^{T,(1)}_{ZZ}(M_Z^2), \\
    \deltaOL M_W^2 ={}& \Re \Sigma^{T,(1)}_{W^-W^+}(M_W^2), \\
    \deltaOL m_A^2 ={}& \Re \Sigma^{(1)}_{AA}(m_A^2) & \text{(in the rMSSM)}, \\
    \deltaOL m_{H^\pm}^2 ={}& \Re \Sigma^{(1)}_{H^-H^+}(m_{H^\pm}^2) & \quad \text{(in the cMSSM)}.
\end{alignat}
\end{subequations}

Taking the one-loop version of \eqn{eqn:mHpm_mA0_rel}, we get
\begin{equation}
    \deltaOL m_{H^\pm}^2 + \deltaOL m_{G^\pm}^2 = \deltaOL m_A^2 + \deltaOL m_G^2 + \deltaOL M_W^2.
\end{equation}
The neutral and charged Goldstone mass counterterms are identical at the one-loop level, see \appx{app:CTs}. The dependent mass counterterm is therefore given by
\begin{subequations}
\begin{alignat}{2}
    \deltaOL m_{H^\pm}^2 ={}& \deltaOL m_A^2 + \deltaOL M_W^2 & \text{(in the rMSSM)}, \\
    \deltaOL m_A^2 ={}& \deltaOL m_{H^\pm}^2 - \deltaOL M_W^2 & \quad \text{(in the cMSSM)}.
\end{alignat}
\end{subequations}

For the field renormalisation constants, different approaches are used in the Higgs and the gauge sector. In the Higgs sector, we renormalise the fields in a $\DRbar$ scheme. It is most convenient to determine the doublet field counterterms from the $\CP$-even, diagonal self-energies at vanishing mixing angle $\al$:
\begin{subequations}
\begin{align}
    \deltaOL Z_{\mathcal{H}_1}^\DR ={}& - \Big[ \partial\Sigma^{(1)}_{HH} \big|_{\al = 0} \Big]_\Div, \\
    \deltaOL Z_{\mathcal{H}_2}^\DR ={}& - \Big[ \partial\Sigma^{(1)}_{hh} \big|_{\al = 0} \Big]_\Div.
\end{align}
\end{subequations}
The `$\Div$' operator performs a series expansion in $\del$, where the space--time dimension $D$ is given by $D = 4 - 2\del$, and keeps only the part proportional to the divergence $\del^{-1}$. As mentioned above, all Higgs field renormalisation constants are fixed by this choice for the doublet field counterterms.

In the gauge sector, the field counterterms are determined from on-shell conditions. The off-diagonal counterterms $\delta Z_{Z \gamma}$ and $\delta Z_{\gamma Z}$ are chosen such that the mixing self-energy $\hat \Sigma^T_{\gamma Z}$ vanishes at the two on-shell momenta $p^2 = 0$ and $p^2 = M_Z^2$:
\begin{subequations}
\begin{align}
\label{eqn:dZZA1_ren_cond}
    &\hat \Sigma^{T,(1)}_{\gamma Z}(0) \overset{!}{=} 0, \\
\label{eqn:dZZA1_expr}
    \Rightarrow{}& \deltaOL Z_{Z \gamma} = \frac{2}{M_Z^2} \Sigma^{T,(1)}_{\gamma Z}(0) \overset{\order{\nc}}{=} 0,
\end{align}
\end{subequations}
and
\begin{subequations}
\begin{align}
    &\hat \Sigma^{T,(1)}_{\gamma Z}(M_Z^2) \overset{!}{=} 0, \\
    \Rightarrow{}& \deltaOL Z_{\gamma Z} = -\frac{2}{M_Z^2} \Sigma^{T,(1)}_{\gamma Z}(M_Z^2).
\end{align}
\end{subequations}

Whenever we set the symbol $\order{\nc}$ over an equal sign, the identity holds in our calculation at $\order{\nc}$ but not necessarily in a more inclusive one.

The diagonal field counterterms, on the other hand, are used to set the residues of the propagators to unity. Expanding to one-loop order, we arrive at
\begin{subequations}
\begin{align}
    & \partial \hat \Sigma^{T,(1)}_{ZZ}(M_Z^2) \overset{!}{=} 0, \\
    \Rightarrow{}& \deltaOL Z_{ZZ} = - \partial \Sigma^{T,(1)}_{ZZ}(M_Z^2),
\end{align}
\end{subequations}
where $\partial \Sigma(p^2) \equiv \partial/\partial p^2 \Sigma(p^2)$.

For the $W$ boson, we analogously find
\begin{equation}
    \deltaOL Z_{WW} = - \partial \Sigma^{T,(1)}_{W^-W^+}(M_W^2).
\end{equation}

For the photon, no mass parameter in the Lagrangian is associated with the photon and so there is also no counterterm to be generated from the renormalisation transformation. This poses no problem, however, as the transverse part of the photon self-energy, at one-loop order, vanishes at zero momentum due to a Slavnov-Taylor identity \cite{Denner:1991kt}, and so the propagator pole is not shifted away from zero by loop corrections:
\begin{equation}
    \Sigma^{T,(1)}_{\gamma\gamma}(0) = 0.
\end{equation}
Because of \eqn{eqn:gaga_1L_ren}, this also means
\begin{equation}
\label{eqn:gaga_1L_Ward_ren}
    \hat \Sigma^{T,(1)}_{\gamma\gamma}(0) = 0.
\end{equation}
Consequently, we require
\begin{subequations}
\begin{align}
\label{eqn:dZAA1_ren_cond}
    &\partial \hat \Sigma^{T,(1)}_{\gamma \gamma}(0) \overset{!}{=} 0, \\
    \Rightarrow{}& \deltaOL Z_{\gamma \gamma} = - \partial \Sigma^{T,(1)}_{\gamma \gamma}(0).
\end{align}
\end{subequations}
We define the one-loop vacuum polarisation by
\begin{subequations}
\begin{align}
    \Pi^{(1)}_{\gamma \gamma}(p^2) \equiv{}& \frac{\Sigma^{T,(1)}_{\gamma \gamma}(p^2)}{p^2} \quad \text{for $p^2 \neq 0$}, \\
    \Pi^{(1)}_{\gamma \gamma}(0) \equiv{}& \partial \Sigma^{T,(1)}_{\gamma \gamma}(0),
\end{align}
\end{subequations}
so
\begin{equation}
    \deltaOL Z_{\gamma \gamma} = - \Pi^{(1)}_{\gamma \gamma}(0).
\end{equation}

When evaluating the vacuum polarisation $\Pi^{(1)}_{\gamma \gamma}$ at zero momentum, we can no longer treat the first two generations of quarks as massless since this would lead to infrared divergences. Instead, we split the contributions from quarks and squarks to the vacuum polarisation into parts stemming from light and heavy particles:
\begin{equation}
    \Pi^{(1)}_{\gamma \gamma}(0) = \Pi^{(1),\text{light}}_{\gamma \gamma}(0) + \Pi^{(1),\text{heavy}}_{\gamma \gamma}(0).
\end{equation}
The light part includes contributions from the five light quarks whereas the heavy part contains the squark and top contributions; the heavy part can be calculated perturbatively. In accordance with \citeres{Bharucha:2012nx,Degrassi:2014sxa}, we rewrite the light part of the vacuum polarisation as
\begin{equation}
\begin{split}
    \Pi^{(1),\text{light}}_{\gamma \gamma}(0) ={}& \Pi^{(1),\text{light}}_{\gamma \gamma}(0) - \Re \Pi^{(1),\text{light}}_{\gamma \gamma}(M_Z^2) + \Re \Pi^{(1),\text{light}}_{\gamma \gamma}(M_Z^2) \\
    \equiv{}& \Delta \alem(M_Z^2)  + \Re \Pi^{(1),\text{light}}_{\gamma \gamma}(M_Z^2),
\end{split}
\end{equation}
with
\begin{equation}
\begin{split}
    \Delta \alem(M_Z^2) ={}& \Delta \alpha_{\text{lep}}(M_Z^2) + \Delta \alpha^{(5)}_{\text{had}}(M_Z^2).
\end{split}
\end{equation}
The numerical values that we use for $\Delta \alpha_{\text{lep}}(M_Z^2)$ and $\Delta \alpha^{(5)}_{\text{had}}(M_Z^2)$ are given in \eqn{eqn:input_pars}. $\Delta \alpha_{\text{lep}}(M_Z^2)$ is calculated perturbatively, and $\Delta \alpha^{(5)}_{\text{had}}(M_Z^2)$ is extracted from experimental input via a dispersion relation. As the leptonic contributions to the running of the fine-structure constant are sizeable, we include them in our definition of $\Delta \alem(M_Z^2)$ although they are formally not of $\order{\nc}$.

With this, our expression for the photon field counterterm is modified to
\begin{equation}
    \deltaOL Z_{\gamma \gamma} = - \Pi^{(1),\text{heavy}}_{\gamma \gamma}(0) - \Re \Pi^{(1),\text{light}}_{\gamma \gamma}(M_Z^2) - \Delta \alem(M_Z^2).
\end{equation}

In the second term, we can now safely set the quark masses of the first two generations to zero without encountering infrared divergences.

The elementary charge is renormalised such that all corrections to the $e e \gamma$-vertex (and, by charge universality, to any $ff\gamma$-vertex) vanish for external on-shell particles in the Thomson limit. With this renormalisation condition, we get the relation
\begin{equation}
\label{eqn:el_ren_cond}
    Z_e \left( \sqrt{Z_{\gamma \gamma}} - \frac{\sw + \delta \sw}{\cw + \delta \cw} \frac{\delta Z_{Z \gamma}}{2} \right) = 1,
\end{equation}
which holds to all orders \cite{Bauberger:1997zz,Freitas:2002ja,Awramik:2002vu,Dittmaier:2021loa}. Expanding up to one-loop order, the elementary charge counterterm is fully determined by gauge field counterterms:
\begin{equation}
    \deltaOL Z_e \equiv \frac{\deltaOL e}{e} = \frac{1}{2}\left( \frac{\sw}{\cw} \deltaOL Z_{Z \gamma} - \deltaOL Z_{\gamma \gamma} \right).
\end{equation}
The sign difference with respect to \citeres{Freitas:2002ja,Awramik:2002vu} stems from a different convention in the SU(2) term of the gauge-covariant derivative, which is often found between the SM and the MSSM. While \eqn{eqn:el_ren_cond} corresponds to the common MSSM choice, in the SM often a convention is used that is obtained by exchanging the minus sign with a plus sign.

The weak mixing angle is not an independent parameter but fixed by the electroweak vector boson mass counterterms via the relation
\begin{subequations}
\begin{align}
    \cw^2 ={}& \frac{M_W^2}{M_Z^2}, \quad \sw^2 = 1 - \cw^2, \\
    \Rightarrow \deltaOL \sw ={}& \frac{1}{2} \frac{\cw^2}{\sw} \left( \frac{\deltaOL M_Z^2}{M_Z^2} - \frac{\deltaOL M_W^2}{M_W^2} \right).
\end{align}
\end{subequations}

Lastly, we introduce the auxiliary renormalisation constants related to the ubiquitous factor $e/(\sw M_W)$:
\begin{equation}
    \deltaOL Z_\text{w} \equiv \deltaOL Z_e - \frac{\deltaOL M_W^2}{2 M_W^2} - \frac{\deltaOL \sw}{\sw}.
\end{equation}

The only parameter that we have not renormalised at the one-loop level so far is $\tb$. Its renormalisation is discussed in \sct{sec:tanbeta_ren}.

\subsection{Renormalisation at the two-loop level}
\label{ssec:higgs_gauge_2L_ren}

For the two-loop renormalisation, the relations between the renormalised one- and two-point functions ($\hat\Gamma_i^{(2)}$ and $\hat\Sigma_{ij}^{(2)}$) and their unrenormalised counterparts ($\Gamma_i^{(2)}$ and $\Sigma_{ij}^{(2)}$) are needed. They are given in \cha{ssec:tad_se_ren_2L}.

Similarly to the one-loop level, we choose the tadpole counterterms such that the one-point vertex functions vanish:
\begin{subequations}
\begin{align}
    &\hat \Gamma^{(2)}_i \overset{!}{=} 0, \quad i \in \{h,H,A\}, \\
    \Rightarrow &\deltaTL T_h = - \Gamma^{(2)}_h - \tfrac{1}{2} \deltaOL Z_{hh} \deltaOL T_h - \tfrac{1}{2} \deltaOL Z_{hH} \deltaOL T_H, \\
    &\deltaTL T_H = - \Gamma^{(2)}_H - \tfrac{1}{2} \deltaOL Z_{HH} \deltaOL T_H - \tfrac{1}{2} \deltaOL Z_{hH} \deltaOL T_h, \\
    &\deltaTL T_A = - \Gamma^{(2)}_A - \tfrac{1}{2} \deltaOL Z_{AA} \deltaOL T_A - \tfrac{1}{2} \deltaOL Z_{AG} \deltaOL T_G.
\end{align}
\end{subequations}

The field renormalisation constants which appear explicitly on the right-hand side cancel with the ones from the sub-loop renormalisation of the $\Gamma^{(2)}_i$. As a consequence, the two-loop tadpole counterterms are independent of any field renormalisation.

From \eqn{eqn:TG_and_TA}, we obtain the dependent $\deltaTL T_G$ counterterm:
\begin{equation}
    \deltaTL T_G = -\cb^2 \deltaOL \tb \deltaOL T_A.
\end{equation}
Using this counterterm, the remaining one-point vertex function $\hat \Gamma^{(2)}_G$ vanishes as well.

The masses of the electroweak vector bosons and the Higgs bosons are renormalised in an $\OS$ scheme as before. At the two-loop level, however, mixing effects have to be taken into account. For the case of $2\times 2$ mixing, this can be done using the effective self-energy,
\begin{equation}
    \hat \Sigma_{ii}^\text{eff}(p^2) \equiv \hat \Sigma_{ii}(p^2) - \frac{\hat \Sigma_{ij}(p^2) \hat \Sigma_{ji}(p^2)}{p^2 - m_j^2 + \hat \Sigma_{jj}(p^2)},
\end{equation}
with $i\neq j$. The loop-corrected propagator can be written in terms of the effective self-energy as (for more details see e.g.\ \citere{Fuchs:2016swt})
\begin{equation}
    \hat \Delta_{ii}(p^2) = \frac{1}{p^2 - m_i^2 + \hat \Sigma_{ii}^\text{eff}(p^2)}.
\end{equation}
The following effective self-energies enter in our results:
\begin{subequations}
\label{eqn:gaga_2L_eff}
\begin{align}
    \hat \Sigma^{T,(2),\text{eff}}_{\gamma\gamma}(p^2) ={}& \hat \Sigma^{T,(2)}_{\gamma\gamma}(p^2) - \frac{\big( \hat \Sigma^{T,(1)}_{\gamma Z}(p^2) \big)^2}{p^2 - M_Z^2}, 
    \label{eqn:gaga_eff_se} \\
    \hat \Sigma^{T,(2),\text{eff}}_{ZZ}(p^2) ={}& \hat \Sigma^{T,(2)}_{ZZ}(p^2) - \frac{\big( \hat \Sigma^{T,(1)}_{\gamma Z}(p^2) \big)^2}{p^2}, \\
    \hat \Sigma^{T,(2),\text{eff}}_{W^- W^+}(p^2) ={}& \hat \Sigma^{T,(2)}_{W^- W^+}(p^2), \\
    \hat \Sigma^{(2),\text{eff}}_{AA}(p^2) ={}& \hat \Sigma^{(2)}_{AA}(p^2) - \frac{\big( \hat \Sigma^{(1)}_{AG}(p^2) \big)^2}{p^2 - \xi_Z M_Z^2} + \xi_Z p^2 \frac{\hat \Sigma^{L,(1)}_{AZ}(p^2) \hat \Sigma^{L,(1)}_{ZA}(p^2)}{p^2 - \xi_Z M_Z^2}, \\
    \begin{split}
    \hat \Sigma^{(2),\text{eff}}_{H^- H^+}(p^2) ={}& \hat \Sigma^{(2)}_{H^- H^+}(p^2) - \frac{\hat \Sigma^{(1)}_{H^- G^+}(p^2) \hat \Sigma^{(1)}_{G^- H^+}(p^2)}{p^2 - \xi_W M_W^2} \\
    &+ \xi_W p^2 \frac{\hat \Sigma^{L,(1)}_{H^- W^+}(p^2) \hat \Sigma^{L,(1)}_{W^- H^+}(p^2)}{p^2 - \xi_W M_W^2}.
    \end{split}
\end{align}
\end{subequations}

It should be noted that these expressions have already been expanded up to the two-loop level. The effective Higgs self-energies depend explicitly on the gauge parameters $\xi_Z$ and $\xi_W$. This dependence vanishes once we go on-shell:
\begin{subequations}
\begin{align}
    \hat \Sigma^{(2),\text{eff}}_{AA}(m_A^2) ={}& \hat \Sigma^{(2)}_{AA}(m_A^2) - \frac{\big( \hat \Sigma^{(1)}_{AG}(m_A^2) \big)^2}{m_A^2}, \\
    \hat \Sigma^{(2),\text{eff}}_{H^- H^+}(m_{H^\pm}^2) ={}& \hat \Sigma^{(2)}_{H^- H^+}(m_{H^\pm}^2) - \frac{\hat \Sigma^{(1)}_{H^- G^+}(m_{H^\pm}^2) \hat \Sigma^{(1)}_{G^- H^+}(m_{H^\pm}^2)}{m_{H^\pm}^2}.
\end{align}
\end{subequations}
Here, we have used the on-shell Slavnov-Taylor identities given in \cha{app:ST_id}. To determine the two-loop mass counterterms, we expand the pole equation 
\begin{equation}
    \eval{p^2 - m_i^2 + \hat \Sigma_{ii}^\text{eff}(p^2)}_{p^2 = \mathcal{M}_i^2} = 0,
\end{equation}
up to the two-loop order. We arrive at
\begin{subequations}
\allowdisplaybreaks
\begin{align}
\begin{split}
    \deltaTL M_Z^2 ={}& \Re \Sigma^{T,(2)}_{ZZ}(M_Z^2) - \Re\{ \deltaOL Z_{ZZ} \} \deltaOL M_Z^2 + \tfrac{1}{4} \Re\{ (\deltaOL Z_{\gamma Z})^2 \} M_Z^2 \\
    &+ \Im\{ \hat \Sigma^{T,(1)}_{ZZ}(M_Z^2) \} \Im\{ \partial \hat \Sigma^{T,(1)}_{ZZ}(M_Z^2) \} + \frac{\big( \Im \hat \Sigma^{T,(1)}_{\gamma Z}(M_Z^2) \big)^2}{M_Z^2},
\end{split} \\
\begin{split}
    \deltaTL M_W^2 ={}& \Re \Sigma^{T,(2)}_{W^- W^+}(M_W^2) - \Re\{ \deltaOL Z_{WW} \} \deltaOL M_W^2 \\
    &+ \Im\{ \hat \Sigma^{T,(1)}_{W^- W^+}(M_W^2) \} \Im\{ \partial \hat \Sigma^{T,(1)}_{W^- W^+}(M_W^2) \},
\end{split} \\
\begin{split}
    \deltaTL m_A^2 ={}& \Re \Sigma^{(2)}_{AA}(m_A^2) - \deltaOL Z_{AA} \deltaOL m_A^2 - \deltaOL Z_{AG} \deltaOL m_{AG}^2 \\
    & + \tfrac{1}{4} (\deltaOL Z_{AG})^2 m_A^2 + \Im\{ \hat \Sigma^{(1)}_{AA}(m_A^2) \} \Im\{ \partial \hat \Sigma^{(1)}_{AA}(m_A^2) \} \\
    &- \Re \frac{\big( \hat \Sigma^{(1)}_{AG}(m_A^2) \big)^2}{m_A^2} \quad (\text{in the rMSSM}),
\end{split} \\
\begin{split}
    \deltaTL m_{H^\pm}^2 ={}& \Re \Sigma^{(2)}_{H^- H^+}(m_{H^\pm}^2) - \deltaOL Z_{H^- H^+} \deltaOL m_{H^\pm}^2 - \tfrac{1}{2} \deltaOL Z_{H^- G^+} \deltaOL m_{G^- H^+}^2 \\
    &- \tfrac{1}{2} \deltaOL Z_{G^- H^+} \deltaOL m_{H^- G^+}^2 + \tfrac{1}{4} \deltaOL Z_{H^- G^+} \deltaOL Z_{G^- H^+} m_{H^\pm}^2 \\
    &+ \Im\{ \hat \Sigma^{(1)}_{H^- H^+}(m_{H^\pm}^2) \} \Im\{ \partial \hat \Sigma^{(1)}_{H^- H^+}(m_{H^\pm}^2) \} \\
    &- \Re \frac{\hat \Sigma^{(1)}_{H^- G^+}(m_{H^\pm}^2) \hat \Sigma^{(1)}_{G^- H^+}(m_{H^\pm}^2)}{m_{H^\pm}^2} \quad (\text{in the cMSSM}).
\end{split}
\end{align}
\end{subequations}
The last terms in the expressions for the $Z$, $A$ and $H^\pm$ mass counterterm stem from the mixing contribution in the effective self-energy. As indicated, we use two different input parameters for the rMSSM and the cMSSM also at the two-loop level. In all mass counterterms, the diagonal one-loop field renormalisation constants drop out. To get a relation between the Higgs boson mass counterterms, we take the two-loop version of \eqn{eqn:mHpm_mA0_rel}:
\begin{equation}
    \deltaTL m_{H^\pm}^2 + \deltaTL m_{G^\pm}^2 = \deltaTL m_A^2 + \deltaTL m_G^2 + \deltaTL M_W^2.
\end{equation}
At the two-loop level, the neutral and charged Goldstone mass counterterms do not agree with each other anymore. Instead, they fulfil
\begin{equation}
    \deltaTL m_{G^\pm}^2 - \deltaTL m_G^2 = \cb^4 M_W^2 \big( \deltaOL \tb \big)^2,
\end{equation}
see \appx{app:CTs}.
From there,
\begin{equation}
\label{eqn:MH_MA_MW_2L_rel}
    \deltaTL m_{H^\pm}^2 - \deltaTL m_A^2 - \deltaTL M_W^2 + \cb^4 M_W^2 \big( \deltaOL \tb \big)^2 = 0
\end{equation}
follows directly. All previous Feynman-diagrammatic two-loop calculations for the Higgs boson mass did not require the $(\deltaOL \tb)^2$ term as they were focusing on QCD corrections \cite{Borowka:2014wla,Degrassi:2014pfa,Borowka:2018anu,Heinemeyer:2007aq} or on pure Yukawa corrections \cite{Hollik:2014wea,Hollik:2014bua,Hahn:2015gaa,Passehr:2017ufr}. In both of these cases, the $(\deltaOL \tb)^2$ term does not contribute, and to the best of our knowledge this term has also not been mentioned in the 
literature so far. In our calculation, however, this term is needed in order to render all scalar two-loop self-energies finite.

Depending on the scenario, the dependent mass counterterm is given by
\begin{subequations}
\begin{alignat}{2}
    \deltaTL m_{H^\pm}^2 ={}& \deltaTL m_A^2 + \deltaTL M_W^2 - \cb^4 M_W^2 \big( \deltaOL \tb \big)^2 \quad &&\text{(in the rMSSM)}, \\
    \deltaTL m_A^2 ={}& \deltaTL m_{H^\pm}^2 - \deltaTL M_W^2 + \cb^4 M_W^2 \big( \deltaOL \tb \big)^2 \quad &&\text{(in the cMSSM)}.
\end{alignat}
\end{subequations}

The two-loop Higgs field counterterms are again defined in a $\DRbar$ scheme. First, we define the $\DRbar$ counterterms via\footnote{The following expressions technically hold for a definition in the $\DR$ scheme. We explain the relation between the $\DR$ and $\DRbar$ schemes in \appx{app:reg_ren}.}
\begin{subequations}
\begin{align}
    \deltaTL Z_{\mathcal{H}_1}^\DRbar ={}& - \Big[ \partial\Sigma^{(2)}_{HH} \big|_{\al = 0} \Big]_\Div, \\
    \deltaTL Z_{\mathcal{H}_2}^\DRbar ={}& - \Big[ \partial\Sigma^{(2)}_{hh} \big|_{\al = 0} \Big]_\Div,
\end{align}
\end{subequations}
where,as above, $\al$ denotes the $\CP$-even Higgs mixing angle.

The two-loop weak-mixing angle counterterm is already determined by the counterterms of the electroweak gauge boson masses:
\begin{equation}
\begin{split}
    \deltaTL \sw = \frac{\cw^2}{2\sw} \Bigg[ &\frac{\deltaTL M_Z^2}{M_Z^2} - \frac{\deltaTL M_W^2}{M_W^2} - \Bigg( \frac{\deltaOL M_Z^2}{M_Z^2} \Bigg)^2 + \frac{\deltaOL M_W^2}{M_W^2} \frac{\deltaOL M_Z^2}{M_Z^2} - \Bigg( \frac{\deltaOL \sw}{\cw} \Bigg)^2 \Bigg].
\end{split}
\end{equation}

As at the one-loop level, we again fix the elementary charge via the electromagnetic vertices in the Thomson limit. This means that \eqn{eqn:el_ren_cond} holds again. Expanding this relation up to two-loop order, we get
\begin{equation}
    \deltaTL Z_e = -\frac{1}{2} \deltaTL Z_{\gamma \gamma} + \frac{\sw}{2 \cw} \deltaTL Z_{Z \gamma} + \big( \deltaOL Z_e \big)^2 + \frac{1}{8} \big( \deltaOL Z_{\gamma \gamma} \big)^2 + \frac{1}{2 \cw^3} \deltaOL Z_{Z \gamma} \, \deltaOL \sw,
\end{equation}
in agreement with \citere{Hessenberger:2018xzo}.

The off-diagonal field renormalisation constants are chosen such that the renormalised mixing self-energies vanish on-shell:
\begin{subequations}
\begin{align}
\begin{split}
    &\hat \Sigma^{T,(2)}_{\gamma Z}(0) \overset{!}{=} 0 \\
    \Rightarrow{}& \deltaTL Z_{Z \gamma} = \frac{2}{M_Z^2} \Sigma^{T,(2)}_{\gamma Z}(0) - \frac{1}{2} \deltaOL Z_{Z \gamma} \deltaOL Z_{ZZ} - \deltaOL Z_{Z \gamma} \frac{\deltaOL M_Z^2}{M_Z^2},
\end{split} \\
\begin{split}
    &\hat \Sigma^{T,(2)}_{\gamma Z}(M_Z^2) \overset{!}{=} 0 \\
    \Rightarrow{}& \deltaTL Z_{\gamma Z} = -\frac{2}{M_Z^2} \Sigma^{T,(1)}_{\gamma Z}(M_Z^2) - \frac{1}{2} \deltaOL Z_{\gamma Z} \deltaOL Z_{\gamma \gamma} + \deltaOL Z_{Z \gamma} \frac{\deltaOL M_Z^2}{M_Z^2}.
\end{split}
\end{align}
\end{subequations}

The renormalisation constant $\deltaTL Z_{\gamma Z}$ is not needed in our calculation but was included for the sake of completeness.  The unrenormalised transverse part of the $\gamma Z$ self-energy vanishes at zero momentum also at the two-loop order for the class of $\order{\nc^2}$ contributions that is considered here. This implies
\begin{equation}
    \deltaTL Z_{Z \gamma} \overset{\order{\nc^2}}{=} 0.
\end{equation}

The diagonal photon field counterterm, on the other hand, is again used to set the residue of the propagator to unity. The derivation proceeds in analogous fashion to the one-loop case. The transverse part of the unrenormalised $\gamma\gamma$ two-loop self-energy vanishes on-shell
\begin{equation}
    \Sigma^{T,(2)}_{\gamma\gamma}(0) \overset{\order{\nc^2}}{=} 0 ,
\end{equation}
and therefore, with \eqn{eqn:dZZA1_expr}
\begin{equation}
    \hat \Sigma^{T,(2)}_{\gamma\gamma}(0) \overset{\order{\nc^2}}{=} 0.
\end{equation}

We can thus impose a renormalisation condition on the derivative of the self-energy to fix the residue of the propagator:
\begin{subequations}
\begin{align}
    &\partial \hat \Sigma^{T,(2)}_{\gamma \gamma}(0) = 0, \\
    \Rightarrow &\deltaTL Z_{\gamma \gamma} = - \partial \Sigma^{T,(2)}_{\gamma \gamma}(0) - \frac{1}{4} \left( \deltaOL Z_{Z \gamma} \right)^2.
\end{align}
\end{subequations}

We also give a two-loop version of the auxiliary counterterm $\delta Z_\text{w}$:
\begin{equation}
\begin{split}
    \deltaTL Z_\text{w} \equiv{}& \frac{1}{2} \left( \deltaOL Z_\text{w} \right)^2 + \deltaTL Z_e - \frac{1}{2} \left( \deltaOL Z_e \right)^2 - \frac{\deltaTL M_W^2}{2 M_W^2} + \left( \frac{\deltaOL M_W^2}{2 M_W^2} \right)^2 \\
    &- \frac{\deltaTL \sw}{\sw} + \frac{1}{2} \left( \frac{\deltaOL \sw}{\sw} \right)^2.
\end{split}
\end{equation}

\subsection{Renormalisation of \texorpdfstring{$\tan(\beta)$}{tan(beta)}}
\label{sec:tanbeta_ren}

In this section, we discuss the renormalisation of the MSSM parameter $\tb$. Both the precise definition and the numerical value of this parameter have a large impact on the prediction of MSSM observables, in particular the mass of the SM-like Higgs boson \cite{Freitas:2002pe}. The parameter $\tb$ appears in the calculation of the $\CP$-even Higgs boson masses already at the tree-level. Thus, for a prediction at two-loop order, expressions for the $\tb$ counterterms at two-loop order are required.

In this section, we will discuss two renormalisation schemes for $\tb$: the $\DRbar$ scheme and an $\OS$ definition via the decay $A \to \tau^- \tau^+$. The $\DRbar$ choice is popular since $\tb$ has no natural physical observable to which it is directly related, and a renormalisation via minimal subtraction 
often simplifies a calculation, cf.\ \citeres{Freitas:2002pe,Freitas:2002um,Frank:2006yh,Baro:2008bg,Hollik:2014bua,Hollik:2014wea,Borowka:2015ura}. Furthermore, the $\DRbar$ definition is process-independent and provides numerically stable results in the sense of renormalisation scale dependence \cite{Freitas:2002pe}. It does, however, lead to a gauge-dependent definition of $\tb$ in an $R_\xi$ gauge at the two-loop level. As an alternative, we investigate an $\OS$ definition of $\tb$ in terms of the decay width $\Gamma(A \to \tau^- \tau^+).$\footnote{In a $\CP$-violating scenario, one could use $\Gamma(H^- \to \tau^- \bar \nu_\tau)$ instead.} We choose this particular decay width as it has a relatively clean signature and in the region of larger $\tb$ involves a sizeable coupling to the leptons \cite{Baro:2008bg}. As a renormalisation condition, we require the square of the amplitude $A \to \tau^- \tau^+$ to not receive any higher-order corrections and from this determine an $\OS$ $\tb$ counterterm. This definition is gauge-independent, numerically stable and, of course, process-dependent \cite{Freitas:2002pe,Freitas:2002um}.

Before we discuss the different renormalisation schemes, we introduce the one- and two-loop counterterms for $\sib$ and $\cb$:
\begin{subequations}
\begin{align}
    \deltaOL \sib ={}& \cb^3 \deltaOL \tb, \\
    \deltaOL \cb ={}& - \sib \cb^2 \deltaOL \tb, \\
    \deltaTL \sib ={}& \cb^3 \deltaTL \tb - \tfrac{3}{2} \cb^4 \sib \big( \deltaOL \tb \big)^2, \\
    \deltaTL \cb ={}& - \sib \cb^2 \deltaTL \tb - \frac{1}{2} \cb^3 \big( \cb^2 - 2 \sib^2 \big) \big( \deltaOL \tb \big)^2.
\end{align}
\end{subequations}
The renormalisation of $\tb$ is closely tied to the renormalisation of the vacuum expectation values and the Higgs fields. We write the renormalisation transformation for the VEVs in two equivalent ways:
\begin{equation}
\begin{split}
    v_i \to{}& v_i + \delta v_i \\
    ={}& Z_{\mathcal{H}_i} (v_i + \delta \bar v_i).
\end{split}
\end{equation}
At the one-loop level, the $\tb$ counterterm reads
\begin{equation}
\label{eqn:dTB1fromdv}
    \deltaOL \tb = \tb \Big( \frac{\deltaOL \bar v_2}{v_2} - \frac{\deltaOL \bar v_1}{v_1} + \frac{1}{2} \big[ \deltaOL \zht - \deltaOL \zho \big] \Big).
\end{equation}

The one-loop relation
\begin{equation}
\label{eqn:dv1div}
    \Big[ \frac{\deltaOL \bar v_1}{v_1} \Big]_\Div = \Big[ \frac{\deltaOL \bar v_2}{v_2} \Big]_\Div
\end{equation}
was noted in \citeres{Dabelstein:1994hb,Chankowski:1992er}.

\subsubsection{\texorpdfstring{$\DRbar$}{DRbar} renormalisation via the \texorpdfstring{$AZ$}{AZ} transition}

If $\tb$ is renormalised in the $\DRbar$-scheme, its counterterm consists of divergent terms only. Taking the divergent part of \eqn{eqn:dTB1fromdv} and using the one-loop relation in \eqn{eqn:dv1div}, we find
\begin{equation}
\label{eqn:dTB1DRbar}
    \deltaOL \tb^\DRbar = \frac{\tb}{2} \big( \deltaOL \zht^\DRbar - \deltaOL \zho^\DRbar \big).
\end{equation}
This relation is often used to determine $\deltaOL \tb^\DRbar$ in schemes with $\DRbar$ field renormalisation \cite{Passehr:2014xwu}. At the two-loop level, \eqn{eqn:dv1div} does not hold in general, and another approach has to be taken.

We choose here to determine the $\tb$ counterterm by demanding the finiteness of the $AZ$ mixing self-energy:
\begin{subequations}
\begin{align}
    \big[ \hat \Sigma^{L,(1)}_{AZ} \big]_\Div ={}& 0, \\
    \Rightarrow \deltaOL \tb^\DRbar ={}& \frac{1}{\imag \cb^2 M_Z} \big[ \Sigma^{L,(1)}_{AZ} \big]_\Div - \frac{\deltaOL Z_{AG}^\DR}{2\cb^2}.
\end{align}
\end{subequations}
This agrees with the expression given in \eqn{eqn:dTB1DRbar}. This prescription allows us to determine a counterterm for $\tb$ without having to consider the renormalisation of the VEVs.

The $\DRbar$ renormalisation via $AZ$ transitions can easily be extended to the two-loop level:
\begin{subequations}
\begin{align}
    \big[ \hat \Sigma^{L,(2)}_{AZ} \big]_\Div ={}& 0, \\
\begin{split}
    \Rightarrow \deltaTL \tb^\DRbar ={}& \frac{1}{\imag \cb^2 M_Z} \big[ \Sigma^{L,(2)}_{AZ} \big]_\Div - \frac{\deltaTL Z_{AG}^\DRbar}{2 \cb^2} \\
    &+ \cb \sib \big( \deltaOL \tb^\DRbar \big)^2 - \tfrac{1}{2} \deltaOL \tb^\DRbar \deltaOL Z_{AA}^\DRbar \\
    &- \frac{1}{2} \Big( \deltaOL \tb^\DRbar + \frac{\deltaOL Z_{AG}^\DRbar}{2 \cb^2} \Big) \Big[ \frac{\deltaOL M_Z^2}{M_Z^2} + \deltaOL Z_{ZZ} \Big]_\Div.
\end{split}
\end{align}
\end{subequations}

Of course, we could also use the $H^-W^+$ self-energy to determine an expression for $\delta \tb$. Since, in this section, we are only interested in extracting divergences to define $\tb$ in a minimal subtraction scheme, the charged self-energies would yield the same result. We checked the finiteness of the charged self-energy as a validation.

\subsubsection{\texorpdfstring{$\OS$}{OS} renormalisation via the decay \texorpdfstring{$A \to \tau^- \tau^+$}{Atautau}}
\label{ssec:tanbeta_OS_ren}

In this section, we present an $\OS$ scheme which is defined via the Higgs decay process $A \to \tau^- \tau^+$ \cite{Freitas:2002pe,Freitas:2002um,Baro:2008bg}. This approach yields a gauge-independent definition of $\tb$ due to its direct relation to an observable (the partial decay width $\Gamma(A \to \tau^- \tau^+)$). Furthermore, this method provides a numerically stable prediction due to the smallness of loop corrections to the decay \cite{Freitas:2002um}.

This definition, however, comes with its own drawbacks as well. First of all, it is process- and flavour-dependent and as such it is somewhat inconvenient; any decay into fermions or even an entirely different observable could be used to define $\tb$ instead. Secondly, if such a decay were observed and the corresponding partial decay width were measured, the extraction of an experimental value for $\tb$ would require the calculation of the respective three-point vertex to the desired order. Beyond the one-loop level, this can become quite tedious \cite{Freitas:2002pe,Freitas:2002um}. It should be noted in this context that the advantage of the $\DRbar$ renormalisation of being simple and process-independent applies only to intermediate steps of the calculation. Ultimately, within a quantum field theory, one is interested in predictions for relations between physical observables. For this purpose the relation of $\tb$ to a physical observable is needed in spite of the fact that up to now no experimental input for such an observable exists (with the exception of the mass of the SM-like Higgs boson, see also the discussion in \citeres{Bahl:2019hmm,Bahl:2022kzs}). We note that for the class of contributions considered in this paper the evaluation of the decay width at the two-loop level is largely simplified by the fact that no virtual corrections to the three-point vertex exist.

For our on-shell definition of $\tb$, we impose the renormalisation condition that the decay width $\Gamma(A \to \tau^- \tau^+$) receives no quantum corrections. This is equivalent to demanding that the absolute value of the physical three-point amplitude $\Gamma_{A\tau\tau}^\text{ph}$ receives no loop corrections. We focus here on a $\CP$-conserving scenario, in which the $\CP$-odd Higgs boson only mixes with the neutral (would-be) Goldstone boson $G$ and the longitudinal part of the $Z$ boson.

Our starting point is the physical vertex amplitude
\begin{equation}
    \hat \Gamma_{A\tau\tau}^\text{ph} = \sqrt{\hat Z_A} \big( \hat \Gamma_{A\tau\tau} + \hat Z_{AG} \hat \Gamma_{G\tau\tau} + \text{$AZ$ mixing} \big).
\end{equation}
It takes into account mixing effects with unphysical states as well as the correct normalisation of the $\mathcal{S}$-matrix by including finite wave-function normalisation factors. The contribution from the unphysical states is gauge-dependent for each term separately, but in the sum the dependence drops out. We have shown this explicitly at the one-loop level utilising the Slavnov-Taylor identities presented in \appx{app:ST_id}. As a specific choice, we work here in the Landau gauge, $\xi_Z = 0$:\footnote{Due to gauge independence we can of course work in any arbitrary $R_\xi$ gauge; the Landau gauge is the most convenient for the following discussion as it sets the $AZ$ mixing to 0.}
\begin{equation}
\label{eqn:Atautau_phys}
    \hat \Gamma_{A\tau\tau}^\text{ph} = \sqrt{\hat Z_A} \big( \hat \Gamma_{A\tau\tau} + \hat Z_{AG} \hat \Gamma_{G\tau\tau} \big) \big|_{\xi_Z = 0}.
\end{equation}

Comparing this with the notation used in \citeres{Fuchs:2016swt,Bahl:2021rts}, some remarks are in order. In \citere{Fuchs:2016swt}, the mixing of tree-level mass eigenstates into loop-corrected mass eigenstates has been discussed. In the case of $\CP$ violation, the three tree-level eigenstates $h$, $H$, and $A$ mix into three loop-corrected eigenstates $h_1$, $h_2$, and $h_3$. We only consider here the case of $\CP$ conservation, so no mixing between the $\CP$-even and the $\CP$-odd states takes place. There is, however, still mixing between the $\CP$-odd states $A$ and $G$, which is only well-defined when taking into account contributions from the longitudinal degrees of the $Z$ vector boson as well. To this end, we employ the formalism established in \citere{Fuchs:2016swt}.

Therein, the diagonal and off-diagonal wave function normalisation factors are defined as
\begin{subequations}
\begin{align}
    \hat Z_A ={}&  \Bigg[ 1 + \partial \hat \Sigma_{AA}(p^2) - \frac{\partial}{\partial p^2} \frac{\big( \hat \Sigma_{AG}(p^2) \big)^2}{p^2 - \xi_Z M_Z^2 + \hat \Sigma_{GG}(p^2)} \Bigg]^{-1} \Bigg|_{p^2 = \mathcal{M}_A^2}, \\
    \hat Z_{AG} ={}& - \frac{\hat \Sigma_{AG}(\mathcal{M}_A^2)}{\mathcal{M}_A^2 - \xi_Z M_Z^2 + \hat \Sigma_{GG}(\mathcal{M}_A^2)}.
\end{align}
\end{subequations}

As expected, for the case of a full on-shell renormalisation,\footnote{Proper on-shell renormalisation determines diagonal field counterterms such that the corresponding propagator has unit residue and the off-diagonal field counterterms such that mixing contributions vanish on the mass-shell.} one finds $\hat Z_A = 1$ and $\hat Z_{AG} = 0$.

Now we only need expressions for the vertex functions in \eqn{eqn:Atautau_phys}. We derive them from the Lagrangian
\begin{equation}
    \mathcal{L}_{\chi\tau\tau + Z\tau\tau}= \frac{\imag e m_\tau}{2 \sw M_W \cb} \overline{\tau} \gamma_5 \tau \left( s_{\bn} A - c_{\bn} G \right) - \frac{e}{\sw \cw} \overline{\tau} \gamma^\mu \left[ g_L^\tau P_L - g_R^\tau P_R \right] \tau Z_\mu,
\end{equation}
which contains all interactions of bosons with $\tau$ leptons relevant to us, via a renormalisation transformation. We have introduced the abbreviations $g_L^\tau = T_{3L}^\tau(1 - 4 T_{3L}^\tau Q_\tau \sw^2)$ and $g_R^\tau = 4 (T_{3L}^\tau)^2 Q_\tau \sw^2 $.

As a renormalisation condition, we require that the absolute square of the physical amplitude should not receive any higher order corrections:
\begin{equation}
\begin{split}
    \big| \hat \Gamma_{A\tau\tau}^\text{ph} \big|^2 ={}& \big| \Gamma_{A\tau\tau}^{(0)} + \hat \Gamma_{A\tau\tau}^\text{ph,(1)} + \hat \Gamma_{A\tau\tau}^\text{ph,(2)} + \cdots \big|^2 \\
    ={}& \big| \Gamma_{A\tau\tau}^{(0)} \big|^2 \big(1 + 2 \Re \tilde \Gamma_{A\tau\tau}^\text{ph,(1)} + 2 \Re \tilde \Gamma_{A\tau\tau}^\text{ph,(2)} + \big| \tilde \Gamma_{A\tau\tau}^\text{ph,(1)} \big|^2 + \cdots \big) \\
    \overset{!}{=}{}& \big| \Gamma_{A\tau\tau}^{(0)} \big|^2,
\end{split}
\end{equation}
where we defined $\tilde \Gamma_{A\tau\tau}^\text{ph,(i)}$ via
\begin{equation}
    \hat \Gamma_{A\tau\tau}^\text{ph,(i)} \equiv \tilde \Gamma_{A\tau\tau}^\text{ph,(i)} \Gamma_{A\tau\tau}^{(0)}.
\end{equation}

From this, we obtain both the one-loop and the two-loop renormalisation condition
\begin{subequations}
\allowdisplaybreaks
\begin{align}
    \label{eqn:dTB1OS_cond} 
    \Re \tilde \Gamma_{A\tau\tau}^\text{ph,(1)} ={}& 0,  \\
    \label{eqn:dTB2OS_cond}
    \Re \tilde \Gamma_{A\tau\tau}^\text{ph,(2)} ={}& - \frac{1}{2} \Im^2 \tilde \Gamma_{A\tau\tau}^\text{ph,(1)}.
\end{align}
\end{subequations}

\paragraph*{One-loop order}

Expanding \eqn{eqn:Atautau_phys} to the one-loop order, we obtain
\begin{equation}
    \hat \Gamma_{A\tau\tau}^\text{ph,(1)} = \hat \Gamma_{A\tau\tau}^{(1)} - \frac{1}{2} \partial \hat \Sigma^{(1)}_{AA}(m_A^2) \Gamma_{A\tau\tau}^{(0)} - \frac{\hat \Sigma^{(1)}_{AG}(m_A^2)}{m_A^2} \Gamma_{G\tau\tau}^{(0)}.
\end{equation}

The first term is the renormalised one-loop vertex, which in our case is just the vertex counterterm, as no loop contributions exist at $\order{\nc}$. We obtain the counterterm by applying the renormalisation transformation presented in \sct{ssec:Higgs_gauge_tree} to the tree-level vertex:
\begin{equation}
\begin{split}
    \hat \Gamma_{A\tau\tau}^{(1)} \overset{\order{\nc}}{=}{}& \deltaOL \Gamma_{A\tau\tau} \overset{\order{\nc}}{=} \Big( -\frac{\deltaOL \cb}{\cb} + \deltaOL Z_\text{w} + \frac{1}{2} \deltaOL Z_{AA} - \frac{1}{2 \tb} \deltaOL Z_{AG} \Big) \Gamma_{A\tau\tau}^{(0)}.
\end{split}
\end{equation}

The $G\tau\tau$ tree-level vertex is simply
\begin{equation}
    \Gamma_{G\tau\tau}^{(0)} = -\frac{1}{\tb} \Gamma_{A\tau\tau}^{(0)}.
\end{equation}

All field renormalisation constants drop out in the physical amplitude, as one would expect. This allows us to define $\deltaOL \tb$ independently of the renormalisation conditions for the fields. The $\tb$ counterterm appears in both the vertex counterterm (through $\deltaOL \cb$) and the renormalised $AG$ self-energy (through the mass counterterm $\deltaOL m_{AG}^2$).

Solving \eqn{eqn:dTB1OS_cond} for $\deltaOL \tb$ leads to
\begin{equation}
\label{eqn:deltaOL_tanbeta}
    \frac{\deltaOL \tb^\OS}{\tb} = - \deltaOL Z_\text{w} + \frac{1}{2} \Re \partial \Sigma^{(1)}_{AA}(m_A^2) - \frac{\Re \Sigma^{(1)}_{AG}(m_A^2) - \deltaOL m_{AG}^2}{\tb m_A^2} \Big|_{\deltaOL \tb = 0}.
\end{equation}

The last term can be rewritten by noting
\begin{equation}
\begin{split}
    \Re \Sigma^{(1)}_{AG}(m_A^2) - \deltaOL m_{AG}^2 \big|_{\deltaOL \tb = 0} ={}& \Re \Sigma^{(1)}_{AG}(m_A^2) - \deltaOL m_{AG}^2 - \deltaOL \tb \cb^2 m_A^2 \\
    ={}& -\frac{m_A^2}{M_Z} \Im \Sigma^{L,(1)}_{AZ}(m_A^2),
\end{split}
\end{equation}
where we used \eqn{eqn:AGZ_ST_identity} from the first to the second line. With this, we can write
\begin{equation}
    \frac{\deltaOL \tb^\OS}{\tb} = - \deltaOL Z_\text{w} + \frac{1}{2} \Re \partial \Sigma^{(1)}_{AA}(m_A^2) + \frac{1}{\tb} \frac{\Im \Sigma^{L,(1)}_{AZ}(m_A^2)}{M_Z}.
\end{equation}

\paragraph*{Two-loop order}

The physical two-loop vertex is obtained by expanding \eqn{eqn:Atautau_phys} up to the two-loop order:
\begin{equation}
\begin{split}
    \hat \Gamma_{A\tau\tau}^\text{ph,(2)} ={}& \hat \Gamma_{A\tau\tau}^{(2)} - \frac{1}{2} \partial \hat \Sigma^{(1)}_{AA}(m_A^2) \hat \Gamma_{A\tau\tau}^{(1)} - \frac{\hat \Sigma^{(1)}_{AG}(m_A^2)}{m_A^2} \hat \Gamma_{G\tau\tau}^{(1)} \\
    &- \frac{1}{2} \Bigg[ \partial \hat \Sigma^{(2)}_{AA}(m_A^2) - \frac{3}{4} \big( \partial \hat \Sigma^{(1)}_{AA}(m_A^2) \big)^2 - \frac{\partial}{\partial p^2} \frac{\big( \hat \Sigma^{(1)}_{AG}(p^2) \big)^2}{p^2} \Big|_{p^2 = m_A^2} \Bigg] \Gamma_{A\tau\tau}^{(0)} \\
    &- \Bigg[ \frac{\hat \Sigma^{(2)}_{AG}(m_A^2)}{m_A^2} - \frac{\hat \Sigma^{(1)}_{AG}(m_A^2) \hat \Sigma^{(1)}_{GG}(m_A^2)}{m_A^4} - \frac{1}{2} \partial \hat \Sigma^{(1)}_{AA}(m_A^2) \frac{\hat \Sigma^{(1)}_{AG}(m_A^2)}{m_A^2} \Bigg] \Gamma_{G\tau\tau}^{(0)} \\
    &+ \frac{\imag}{2} \Im \hat \Sigma^{(1)}_{AA}(m_A^2) \partial^2 \hat \Sigma^{(1)}_{AA}(m_A^2) \Gamma_{A\tau\tau}^{(0)} \\
    &+ \frac{\imag \, \Im \hat \Sigma^{(1)}_{AA}(m_A^2)}{m_A^2} \Bigg[ \partial \hat \Sigma^{(1)}_{AG}(m_A^2) - \frac{\hat \Sigma^{(1)}_{AG}(m_A^2)}{m_A^2} \Bigg] \Gamma_{G\tau\tau}^{(0)}.
\end{split}
\end{equation}

The terms in the last line appear because we defined the wave function normalisation constants at the complex rather than at the real pole. When taking the real part of the physical amplitude, the last line will produce products of imaginary parts. As the real part of the two-loop vertex and the imaginary part of the one-loop vertex appear in the two-loop renormalisation condition, we give their explicit expressions here:
\begin{subequations}
\label{eqn:GammaAtautau2L}
\allowdisplaybreaks
\begin{align}
\begin{split}
    \Re \hat \Gamma_{A\tau\tau}^\text{ph,(2)} ={}& \hat \Gamma_{A\tau\tau}^{(2)} - \frac{1}{2} \Re \partial \hat \Sigma^{(1)}_{AA}(m_A^2) \hat \Gamma_{A\tau\tau}^{(1)} - \frac{\Re \hat \Sigma^{(1)}_{AG}(m_A^2)}{m_A^2} \hat \Gamma_{G\tau\tau}^{(1)} \\
    &- \frac{1}{2} \Re \Bigg\{ \partial \hat \Sigma^{(2)}_{AA}(m_A^2) - \frac{3}{4} \big( \partial \hat \Sigma^{(1)}_{AA}(m_A^2) \big)^2 - \frac{\partial}{\partial p^2} \frac{\big( \hat \Sigma^{(1)}_{AG}(p^2) \big)^2}{p^2}\Big|_{p^2 = m_A^2} \Bigg\} \Gamma_{A\tau\tau}^{(0)} \\
    &- \Re \Bigg\{ \frac{\hat \Sigma^{(2)}_{AG}(m_A^2)}{m_A^2} - \frac{\hat \Sigma^{(1)}_{AG}(m_A^2) \hat \Sigma^{(1)}_{GG}(m_A^2)}{m_A^4} - \frac{\partial \hat \Sigma^{(1)}_{AA}(m_A^2)}{2} \frac{\hat \Sigma^{(1)}_{AG}(m_A^2)}{m_A^2} \Bigg\} \Gamma_{G\tau\tau}^{(0)} \\
    &- \frac{1}{2} \Im \hat \Sigma^{(1)}_{AA}(m_A^2) \Im \partial^2 \hat \Sigma^{(1)}_{AA}(m_A^2) \Gamma_{A\tau\tau}^{(0)}  \\
    &- \frac{\Im \hat \Sigma^{(1)}_{AA}(m_A^2)}{m_A^2} \Bigg[ \Im \partial \hat \Sigma^{(1)}_{AG}(m_A^2) - \frac{\Im \hat \Sigma^{(1)}_{AG}(m_A^2)}{m_A^2} \Bigg] \Gamma_{G\tau\tau}^{(0)},
\end{split} \\
    \Im \hat \Gamma_{A\tau\tau}^\text{ph,(1)} ={}& - \frac{1}{2} \partial \Im \hat \Sigma^{(1)}_{AA}(m_A^2) \Gamma_{A\tau\tau}^{(0)} - \frac{\Im \hat \Sigma^{(1)}_{AG}(m_A^2)}{m_A^2} \Gamma_{G\tau\tau}^{(0)}.
\end{align}
\end{subequations}
    
In our calculation, the two-loop vertex again just contains the counterterm
\begin{equation}
\begin{split}
    \hat \Gamma_{A\tau\tau}^{(2)} \overset{\order{\nc^2}}{=} \deltaTL \Gamma_{A\tau\tau} \overset{\order{\nc^2}}{=}{}& \Bigg[ -\frac{\deltaTL \cb}{\cb} + \Big( \frac{\deltaOL \cb}{\cb} \Big)^2 + \deltaTL Z_\text{w} - \deltaOL Z_\text{w} \frac{\deltaOL \cb}{\cb} \\
    & + \frac{1}{2} \deltaTL Z_{AA} - \frac{1}{8} \big( \deltaOL Z_{AA} \big)^2 + \frac{1}{2} \deltaOL Z_{AA} \Big( \deltaOL Z_\text{w} - \frac{\deltaOL \cb}{\cb} \Big) \\
    & - \frac{1}{2 \tb} \deltaTL Z_{AG} - \frac{1}{2 \tb} \deltaOL Z_{AG} \Big( \deltaOL Z_\text{w} - \frac{\deltaOL \cb}{\cb} \Big) \Bigg] \Gamma_{A\tau\tau}^{(0)}.
\end{split}
\end{equation}

At the two-loop level, also the one-loop $G\tau\tau$ vertex appears:
\begin{equation}
    \hat \Gamma_{G\tau\tau}^{(1)} \overset{\order{\nc}}{=} \deltaOL \Gamma_{G\tau\tau} \overset{\order{\nc}}{=} \Big( -\frac{\deltaOL \cb}{\cb} + \deltaOL Z_\text{w} + \frac{1}{2} \deltaOL Z_{GG} - \frac{\tb}{2} \deltaOL Z_{AG} \Big) \Gamma_{G\tau\tau}^{(0)}.
\end{equation}

Before we give an explicit expression for the two-loop $\tb$ counterterm, we introduce the symbol
\begin{equation}
    \tilde \Sigma^{(2)}(p^2) = \Sigma^{(2)}(p^2) \Big|_{\deltaOL Z = 0},
\end{equation}
which denotes an unrenormalised two-loop self-energy where the one-loop field counterterms in the sub-loop diagrams have been set to 0. This means in particular
\begin{subequations}
\begin{align}
    \Sigma^{(2)}_{AA} ={}& \tilde \Sigma^{(2)}_{AA} + \deltaOL Z_{AA} \Sigma^{(1)}_{AA} + \deltaOL Z_{AG} \Sigma^{(1)}_{AG}, \\
    \Sigma^{(2)}_{AG} ={}& \tilde \Sigma^{(2)}_{AG} + \tfrac{1}{2} \big( \deltaOL Z_{AA} + \deltaOL Z_{GG} \big) \Sigma^{(1)}_{AG} + \tfrac{1}{2} \deltaOL Z_{AG} \big( \Sigma^{(1)}_{AA} + \Sigma^{(1)}_{GG} \big).
\end{align}
\end{subequations}

We now insert \eqs{eqn:GammaAtautau2L} into \eqn{eqn:dTB2OS_cond} and solve for $\deltaTL \tb$. As in the one-loop case, the $\tb$ counterterm appears as a contribution to $\deltaTL \cb$ in the vertex counterterm and to the mass counterterm $\deltaTL m_{AG}^2$ in the renormalised two-loop $AG$ self-energy.

Putting everything together, we obtain
\begin{equation}
\label{eqn:deltaTL_tanbeta}
\begin{split}
    \frac{\deltaTL \tb^\OS}{\tb} ={}& -\frac{1}{2} \cb^2 \big( \cb^2 - 2 \sib^2 \big) \big( \deltaOL \tb \big)^2 - \Big( \frac{\deltaOL \cb}{\cb} \Big)^2 - \deltaTL Z_\text{w} + \deltaOL Z_\text{w} \frac{\deltaOL \cb}{\cb} \\
    &+ \frac{1}{2} \Re \partial \Sigma^{(1)}_{AA}(m_A^2) \Big( -\frac{\deltaOL \cb}{\cb} + \deltaOL Z_\text{w} \Big) \\
    &- \frac{\Re \Sigma^{(1)}_{AG}(m_A^2) - \deltaOL m_{AG}^2}{\tb m_A^2} \Big( -\frac{\deltaOL \cb}{\cb} + \deltaOL Z_\text{w} \Big) \\
    &+ \frac{1}{2} \Re \Bigg\{ \partial \tilde \Sigma^{(2)}_{AA}(m_A^2) - \frac{3}{4} \big( \partial \Sigma^{(1)}_{AA}(m_A^2) \big)^2 \\
    &- \frac{2 \big( \Sigma^{(1)}_{AG}(m_A^2) - \deltaOL m_{AG}^2 \big) \partial \Sigma^{(1)}_{AG}(m_A^2)}{m_A^2} + \frac{\big( \Sigma^{(1)}_{AG}(m_A^2) - \deltaOL m_{AG}^2 \big)^2}{m_A^4} \Bigg\} \\
    &- \Re \Bigg\{ \frac{\tilde \Sigma^{(2)}_{AG}(m_A^2) - \deltaTL m_{AG}^2}{\tb m_A^2}\Big|_{\deltaTL \tb = 0} - \frac{\partial \Sigma^{(1)}_{AA}(m_A^2)}{2} \frac{\Sigma^{(1)}_{AG}(m_A^2) - \deltaOL m_{AG}^2}{\tb m_A^2} \\
    &- \frac{\big( \Sigma^{(1)}_{AG}(m_A^2) - \deltaOL m_{AG}^2 \big) \big( \Sigma^{(1)}_{GG}(m_A^2) - \deltaOL m_G^2 \big)}{\tb m_A^4} \Bigg\} \\
    &+ \frac{1}{2} \Im \Sigma^{(1)}_{AA}(m_A^2) \Im \partial^2 \Sigma^{(1)}_{AA}(m_A^2) \\
    &- \frac{\Im \Sigma^{(1)}_{AA}(m_A^2)}{\tb m_A^2} \Bigg[ \Im \partial \Sigma^{(1)}_{AG}(m_A^2) - \frac{\Im \Sigma^{(1)}_{AG}(m_A^2)}{m_A^2} \Bigg] \\
    &- \frac{1}{2} \Im^2 \Big\{ \frac{1}{2} \partial \Sigma^{(1)}_{AA}(m_A^2) - \frac{\Sigma^{(1)}_{AG}(m_A^2)}{\tb m_A^2} \Big\} \\
    &- \frac{1}{2} \deltaOL Z_{AG} \frac{\Re \Sigma^{(1)}_{AA}(m_A^2) - \deltaOL m_A^2}{\tb m_A^2}.
\end{split}
\end{equation}

Again, any field renormalisation constant drops out in the final expression for the two-loop $\tb$ counterterm. In order to assure this feature, however, we have to make use of the fact that $m_A^2$ has been defined as an on-shell quantity, as can be seen from the last term. In a $\CP$-violating scenario, we would thus have to use the decay of a charged Higgs boson into $\tau$ and $\nu_\tau$ together with a charged on-shell mass.

\section{Calculation of electroweak \texorpdfstring{$\order{\nc^2}$}{O(Nc2)} terms to the Higgs boson masses}
\label{cha:calculation}
In this paper, we calculate for the first time the complete electroweak two-loop corrections of $\order{(\alem + \alq)^2 \nc^2}$ to the Higgs boson masses in the MSSM. The setup of our calculation is general enough to allow for both real and complex input parameters.\footnote{So far, we have determined the on-shell $\tb$ counterterms from the $A \to \tau^- \tau^+$ decay for a scenario with $\CP$-symmetry conservation. The modification for the definition of an on-shell $\tb$ in a $\CP$-violating scenario via the decay $H^+ \to \tau^+ \nu_\tau$ is straightforward.} We refer to these scenarios as the rMSSM and the cMSSM, respectively. The masses of the $\CP$-even Higgs bosons $h$ and $H$ obtain contributions from particle mixing at the two-loop order and beyond in both scenarios. The presence of non-vanishing phases in the cMSSM gives rise to non-vanishing $\CP$-violating self-energies and thus leads to mixing with the $\CP$-odd Higgs boson $A$. Hence, a $2 \times 2$ propagator matrix occurs for the $\CP$-even Higgs bosons in the rMSSM and a $3 \times 3$ matrix for the neutral Higgs bosons in the cMSSM. In most scenarios, the difference between the tree-level masses $m_H^2$ and $m_A^2$ is rather small, leading potentially to large mixing effects in the cMSSM.

\begin{figure}
	\centering
	\includegraphics[width = \textwidth]{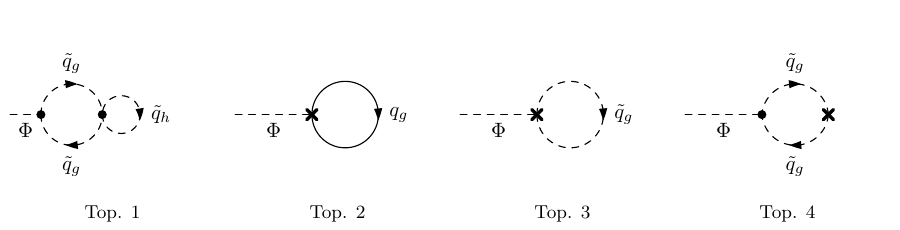}
	\caption{Topologies of the two-loop tadpole diagrams. $\Phi = h, H, A, G$; $g$ and $h$ are flavour indices. The cross denotes the insertion of a one-loop counterterm. Counterterm topologies which have not been listed vanish for the considered class of contributions.}
	\label{fig:tadpoles}
\end{figure}

\begin{figure}
	\centering
	\includegraphics[width = \textwidth]{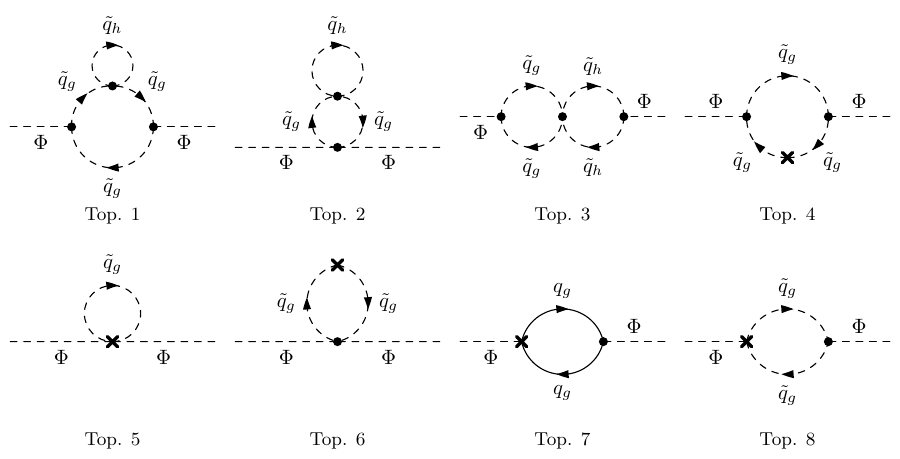}
	\caption{Topologies of the neutral two-loop self-energy diagrams. $\Phi = h, H, A, G$; $g$ and $h$ are flavour indices. The cross denotes the insertion of a one-loop counterterm. Counterterm topologies which have not been listed vanish for the considered class of contributions.}
	\label{fig:neutral_self-energies}
\end{figure}

We fully take into account the electroweak and Yukawa two-loop contributions for non-vanishing external momenta in a mixed $\OS$-$\DRbar$ scheme within the considered class of contributions. In our calculation, we allow for the general case of complex parameters in the MSSM, and we take into account flavour- and generation-mixing Feynman diagrams. We note, however, that we use a unit CKM matrix as a simplifying approximation. This restriction could be lifted as a possible extension of our results. Since we focus on the contributions of $\order{(\alem + \alq)^2 \nc^2}$, the calculated diagrams do not contain internal leptons, Higgs and gauge bosons as well as their respective supersymmetric partners.

\begin{figure}
	\centering
	\includegraphics[width = \textwidth]{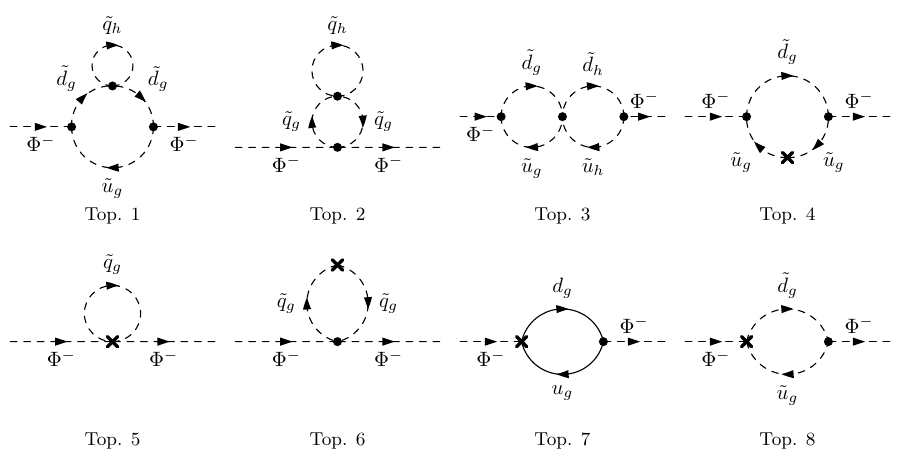}
	\caption{Topologies of the charged two-loop self-energy diagrams. $\Phi^- = H^-, G^-$; $g$ and $h$ are flavour indices. The cross denotes the insertion of a one-loop counterterm. Counterterm topologies which have not been listed vanish for the considered class of contributions.}
	\label{fig:charged_self-energies}
\end{figure}

The relevant tadpole diagrams are shown in \fig{fig:tadpoles}, while the neutral and charged self-energies consist of the diagrams shown in \figs{fig:neutral_self-energies} and \ref{fig:charged_self-energies}, respectively. The calculated two-loop vector boson self-energies and scalar--vector mixing self-energies have the same topological structure. All of these topologies lead to diagrams which can be expressed as products of one-loop integrals.

When discussing two-loop self-energies, we distinguish between so-called ``genuine'' diagrams with two independent loop momenta, see e.g.\ topologies 1--3 in \fig{fig:neutral_self-energies}, and ``sub-loop'' diagrams with a one-loop counterterm insertion, see e.g.\ topologies 4--8 in \fig{fig:neutral_self-energies}.

All genuine two-loop diagrams have two loop momenta that are integrated over. In our case, no internal propagator depends on both loop momenta, and our two-loop diagrams decompose into mere products of one-loop integrals. The sub-loop diagrams have a very similar form as they are products of a one-loop integral and a one-loop counterterm.

\begin{figure}
	\centering
	\includegraphics[width = .3\textwidth]{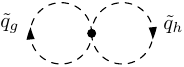}
	\caption{The basic two-loop vacuum diagram $V_{\tilde q_g \tilde q_h}$ with the four-squark vertex. Here, $g$ and $h$ are flavour indices.}
	\label{fig:2L_vacuum}
\end{figure}

The genuine diagrams all contain a four-squark interaction vertex. To better understand their overall structure in terms of colour factors and coupling constants, we investigate the vacuum diagram shown in \fig{fig:2L_vacuum}. All genuine tadpole and neutral self-energy diagrams can be obtained from this diagram by simply adding external legs to the vacuum bubble. As such, all genuine diagrams will have the same colour structure as the vacuum diagram.

The colours of the internal squark propagators in \fig{fig:2L_vacuum} can have all possible values, and hence they need to be summed over; we label them by the indices $a$ and $b$. Keeping the colour sum explicit, the vacuum bubble has the structure
\begin{equation}
\begin{split}
    V_{\tilde q_g \tilde q_h} = \sum_{a,b = 1}^{\nc} \Big[ &\als \frac{1}{2} \Big( \delta_{ab} \delta_{ab} - \frac{1}{\nc} \delta_{aa} \delta_{bb} \Big) S_{gh} + \als \frac{1}{2} \Big( \delta_{aa} \delta_{bb} - \frac{1}{\nc} \delta_{ab} \delta_{ab} \Big) S'_{gh} \\
    &+ \alem \delta_{ab} G_{gh} + \alem G'_{gh} + \alq \delta_{ab} Y_{gh} + \alq Y'_{gh} \Big],
\end{split}
\end{equation}
where the coefficients $S_{gh}$, $S'_{gh}$, $G_{gh}$, $G'_{gh}$, $Y_{gh}$, and $Y'_{gh}$ contain entries of the squark-mixing matrices and other numerical factors, but no coupling constants or colour factors. After performing the colour sums, we find
\begin{equation}
    V_{\tilde q_g \tilde q_h} = \als \nc C_F S'_{gh} + \alem (\nc G_{gh} + \nc^2 G'_{gh}) + \alq (\nc Y_{gh} + \nc^2 Y'_{gh}),
\end{equation}
where we introduced the Casimir operator $C_F$ of the fundamental representation,
\begin{equation}
\label{eqn:cf}
    C_F = \frac{\nc^2 - 1}{2 \nc}.
\end{equation}

Depending on which squark flavours appear in the loops, the coefficients have the following properties:
\begin{itemize}
    \item If the squarks have the same flavour, $g = h$: \\
    $S_{gg} = S'_{gg}$, $G_{gg} = G'_{gg}$ and $Y_{gg} = Y'_{gg}$. All coefficients are non-vanishing.
    \item If the squarks have different flavours but are of the same generation: \\
    $S'_{gh} = 0$, $G_{gh} \neq G'_{gh}$, and $Y'_{gh} = 0$.
    \item If the squarks stem from different generations but are of the same type: \\
    $S'_{gh} = 0$, $G_{gh} = 0$ and $Y_{gh} = 0$.
    \item If the squarks stem from different generations and are of a different type: \\
    $S'_{gh} = 0$, $G_{gh} = 0$, and $Y_{gh} = Y'_{gh} = 0$.
\end{itemize}
These relations are read off the four-squark-vertex Feynman rules, see e.g.\ \citere{Drees:2004jm}.

The contributions parameterised by $S_{gh}$, $G_{gh}$ and $Y_{gh}$ are irrelevant to us as they do not produce terms of $\order{\nc^2}$. The term with the coefficient $S'_{gh}$ is proportional to $\nc C_F$, and hence formally contains a factor $\nc^2$, as one sees from \eqn{eqn:cf}. Other genuine two-loop diagrams also contribute at $\order{\als \nc C_F}$. This type of diagrams does not decompose into a simple product of one-loop integrals. Since the complete two-loop QCD contributions have already been calculated in \citeres{Heinemeyer:2007aq,Borowka:2014wla,Degrassi:2014pfa,Borowka:2018anu}, we set
\begin{equation}
    \als \equiv 0
\end{equation}
in the diagrams that are evaluated in this work. We give estimates for the size of the left-out QCD corrections in the scenarios discussed in \cha{cha:two-loop}.

With this restriction, all diagrams appearing in our calculation contain only quarks and squarks as internal particles. We are interested in the $\order{\nc^2}$ parts of these two-loop diagrams, i.e.\ the contributions parameterised by the coefficients $G'_{gh}$ and $Y'_{gh}$. The genuine two-loop self-energy diagrams in \fig{fig:neutral_self-energies} are obtained from the vacuum bubble in \fig{fig:2L_vacuum} by adding two cubic Higgs-squark-squark vertices or a single quartic Higgs-Higgs-squark-squark vertex; the self-energies are hence of $\order{(\alem + \alq)^2 \nc^2}$.

We treat the first and second generation quarks as massless. This leaves us with the three non-vanishing couplings
\begin{equation}
    \alem, \alt, \alb.
\end{equation}
These are sufficient to describe any four-squark vertex, as well as any Higgs-quark or Higgs-squark vertex. The coupling structure of the one- and two-loop corrections to the Higgs boson pole masses is then typically of the form
\begin{subequations}
\allowdisplaybreaks
\begin{align}
\begin{split}
    \Delta^{(1)}M_{h/H}^2 \sim{} \mathcal{O}\bigg(&\nc \alt (m_t + \mu + \At)^2 + \nc \alb (m_b + \mu + \Ab)^2 \\
    &+ \nc \alem \big[ m_t (m_t + \mu + \At) + m_b (m_b + \mu + \Ab) + M_Z^2 \big]\bigg),
\end{split} \\
\begin{split}
\label{eqn:two_loop_contr}
    \Delta^{(2)}M_{h/H}^2 \sim{} \mathcal{O}\bigg(&\nc^2 \alt^2 (m_t + \mu + \At)^2 + \nc^2 \alb^2 (m_b + \mu + \Ab)^2 \\
    &+ \nc^2 \alem \big[ \sqrt{\alt} (m_t + \mu + \At) + \sqrt{\alb} (m_b + \mu + \Ab) \big]^2 \\
    &+ \nc^2 \alem^2 \big[ m_t (m_t + \mu + \At) + m_b (m_b + \mu + \Ab) + M_Z^2 \big]\bigg),
\end{split}
\end{align}
\end{subequations}
where the plus signs in the brackets and parentheses are used to indicate possible combinations but do not imply that the terms always occur in this exact form.

While the one-loop result was already fully known, for the two-loop contribution only the $\order{\alt^2}$ and $\order{\alb^2}$ part (first line of \eqn{eqn:two_loop_contr}) had so far been evaluated. The second and third lines of \eqn{eqn:two_loop_contr} vanish in the gaugeless limit and are calculated for the first time in the present work.

We stress again that two-loop terms of $\order{\nc}$ are not included in our calculation. Two-loop diagrams with one internal squark and an additional internal Higgs boson are also of this order and therefore would have to be included in a full discussion of $\order{\nc}$ contributions.

For the generation of the loop amplitudes, we use \textsc{FeynArts 3.11} \cite{Kublbeck:1990xc,Eck:1992ms,Hahn:2000kx} employing the \texttt{MSSMCT} \cite{Fritzsche:2013fta} model file. For the case of a single (s)quark generation, 4080 genuine two-loop diagrams and 1242 sub-loop diagrams have been calculated, amounting to a total of 5322 diagrams. When taking into account all three generations of matter, we have 36720 genuine and 3726 sub-loop diagrams, for a combined total number of 40446 diagrams at the particle level.

The sub-loop diagrams have a simple one-loop topology and hence they can be reduced using the package \textsc{FormCalc 9.9} \cite{Hahn:1998yk,Hahn:2016ebn}. The genuine two-loop diagrams, on the other hand, are reduced with \tc{} \cite{Weiglein:1993hd,Weiglein:1995qs}. Before they can be processed, the genuine diagrams have to be adapted to the conventions used in \tc{}. The sub-loop diagrams can also be reduced using the one-loop version of \tc{}, which is called \oc{} \cite{Weiglein:1993hd,Weiglein:1995qs}. We find full agreement between the \oc{} and \textsc{FormCalc} results for the sub-loop diagrams.

We combine the unrenormalised two-loop self-energies, which consist of the genuine two-loop diagrams as well as the sub-loop renormalisation diagrams, and the counterterms using the expressions specified in \sct{ssec:higgs_gauge_2L_ren}. This yields the renormalised self-energies, which are finite. Each one-loop integral is thereby written as
\begin{equation}
\label{eqn:int_ser}
    L = L^\text{div} \frac{1}{\del} + L^\text{fin} + L^\del \del + \order{\del^2},
\end{equation}
where $L$ is either an $A_0$, a $B_0$, or a $B_0'$ loop function. Similarly, we expand the one-loop counterterms as
\begin{equation}
\label{eqn:ct_ser}
    \deltaOL c = \deltaOL c^\text{div} \frac{1}{\del} + \deltaOL c^\text{fin} + \deltaOL c^\del \del + \order{\del^2},
\end{equation}
where $c$ is a parameter or field. One-loop counterterms appear in the sub-loop part of the unrenormalised two-loop self-energies as well as in the two-loop counterterms. Keeping the expansion coefficients of one-loop integrals and counterterms as symbols speeds up the expansion in $\del$ significantly. The coefficients can later be evaluated numerically.

We expand each two-loop self-energy up to $\order{\del^0}$:
\begin{equation}
    \Sigma^{(2)} = \Sigma^{(2),\text{ddiv}} \frac{1}{\del^2} + \Sigma^{(2),\text{div}} \frac{1}{\del} + \Sigma^{(2),\text{fin}} + \order{\del}.
\end{equation}
The one-loop integrals and counterterms enter the self-energy coefficients via
\begin{subequations}
\begin{align}
    \Sigma^{(2),\text{ddiv}} \supset{}& L^\text{div}, \deltaOL c^\text{div}, \\
    \Sigma^{(2),\text{div}} \supset{}& L^\text{div}, \deltaOL c^\text{div}, L^\text{fin}, \deltaOL c^\text{fin}, \\
    \Sigma^{(2),\text{fin}} \supset{}& L^\text{div}, \deltaOL c^\text{div}, L^\text{fin}, \deltaOL c^\text{fin}, L^\del, \deltaOL c^\del.
\end{align}    
\end{subequations}

The renormalised self-energies are UV-finite, so
\begin{subequations}
\begin{align}
    \hat \Sigma^{(2),\text{ddiv}} ={}& 0, \\
    \hat \Sigma^{(2),\text{div}} ={}& 0.
\end{align}
\end{subequations}
We have numerically checked the finiteness of all renormalised two-loop tadpoles, the neutral and charged two-loop Higgs self-energies, and the scalar-vector mixing two-loop self-energies $A\gamma$, $AZ$, $G\gamma$ and $H^-W^+$.

If all parameters which need to be renormalised at the two-loop level are defined in an $\OS$ scheme, the $\order{\del}$ parts $L^\del$ and $\deltaOL c^\del$ will drop out in the determination of $\hat \Sigma^{(2),\text{fin}}$. We will analyse this issue in the following section.

\section{Investigation of contributions from \texorpdfstring{$\order{\del}$}{O(del)} parts of one-loop terms}
\label{sec:del_cancellation}
In order to analyse to what extent $\order{\del}$ parts of loop integrals and counterterms (the so-called evanescent terms) contribute in a renormalised self-energy, we have to study the structure of the unrenormalised two-loop self-energies first. We note in this context that we make use here of the property of the considered set of two-loop contributions to fully decompose into products of one-loop integrals and counterterms. This allows a very transparent treatment of the occurrence of evanescent terms in the two-loop expressions. While for general two-loop calculations the appearances of evanescent terms may be somewhat more difficult to trace, our results for this complete sub-set of two-loop contributions are well-suited for drawing general conclusions for two-loop calculations.

If the results for the considered diagrams are expressed in an analytic form where the one-loop integrals are kept as symbols, there is some freedom involved in the choice of the resulting expression. This is due to reduction formulae relating different loop integrals to one another, for example
\begin{equation}
    B_0(0, m^2, m^2) = (1 - \del) \frac{A_0(m^2)}{m^2}.
    \label{eqn:B0toA0}
\end{equation}

Inserting \eqn{eqn:int_ser} into \eqn{eqn:B0toA0} allows us to relate the coefficients of $A_0$ and $B_0$ in the expansion with respect to $\del$ to each other:
\begin{subequations}
\begin{align}
    B_0^\text{div}(0, m^2, m^2) &= \frac{A_0^\text{div}(m^2)}{m^2}, \\
    B_0^\text{fin}(0, m^2, m^2) &= \frac{A_0^\text{fin}(m^2) - A_0^\text{div}(m^2)}{m^2}, \\
    B_0^\del(0, m^2, m^2) &= \frac{A_0^\del(m^2) - A_0^\text{fin}(m^2)}{m^2}.
\end{align}
\end{subequations}

We can also invert these relations:
\begin{subequations}
\begin{align}
    A_0^\text{div}(m^2) &= m^2 B_0^\text{div}(0, m^2, m^2), \\
    A_0^\text{fin}(m^2) &= m^2 \left( B_0^\text{fin}(0, m^2, m^2) + B_0^\text{div}(0, m^2, m^2) \right), \\
    A_0^\del(m^2) &= m^2 \left( B_0^\del(0, m^2, m^2) + B_0^\text{fin}(0, m^2, m^2) + B_0^\text{div}(0, m^2, m^2) \right).
\end{align}
\end{subequations}
This implies that if a cancellation of $\order{\del}$ parts of loop integrals occurs for a particular choice of the base integrals, it will also occur for the other possible choices.

As mentioned before, all two-loop diagrams appearing in our calculation are products of one-loop integrals and counterterms. This allows us to cast the unrenormalised two-loop self-energy (including subloop renormalisation contributions) in the following still quite general form:
\begin{equation}
\label{eqn:two-loop_SE_unr}
    \Sigma^{(2)} = \sum_i L_i M_i + \sum_jN_j \delta c_j.
\end{equation}
Here, the $L_i$, $M_i$, and $N_j$ are one-loop integrals. The $\delta c_j$ are one-loop counterterms. The first term contains all genuine diagrams, the second one the diagrams with sub-loop renormalisation. In the following, we leave the summation over $i$ and $j$ implicit.

To see how the aforementioned cancellation can take place, we first expand the functions and counterterms according to \eqs{eqn:int_ser} and (\ref{eqn:ct_ser}):
\begin{equation}
\begin{split}
    \Sigma^{(2)} ={}& \frac{L_i^\text{div} M_i^\text{div} + N_j^\text{div} \delta c_j^\text{div}}{\del^2} + \frac{L_i^\text{div} M_i^\text{fin} + L_i^\text{fin} M_i^\text{div} + N_j^\text{div} \delta c_j^\text{fin} + N_j^\text{fin} \delta c_j^\text{div}}{\del} \\
    &+ L_i^\text{div} M_i^\del + L_i^\text{fin} M_i^\text{fin} + L_i^\del M_i^\text{div} + N_j^\text{div} \delta c_j^\del + N_j^\text{fin} \delta c_j^\text{fin} + N_j^\del \delta c_j^\text{div} \\
    &+ \order{\del}.
\end{split}
\end{equation}

For the next step, we derive a relation between the genuine contributions and the contributions involving sub-loop renormalisation. Every genuine two-loop diagram comes with a number of sub-loop renormalisation diagrams that are associated with it. The sum of a genuine diagram and its sub-loop renormalisation diagrams is free from non-local divergences, i.e.\ terms of the form $\log(p^2) \del^{-1}$. These terms have to cancel after the process of sub-loop renormalisation in a renormalisable theory \cite{Schwartz:2014sze}.

We obtain the sub-loop renormalisation diagrams by shrinking one of the loops of the genuine two-loop diagram to a single point and inserting a one-loop counterterm at this point. Let us demonstrate this for the squark topologies in \fig{fig:neutral_self-energies}: Topology 1 leads to the sub-loop renormalisation topology 4 when shrinking the upper loop, and to topology 5 when shrinking the lower one. Topology 2 will similarly lead to the sub-loop renormalisation topologies 5 and 6; topology 3 leads to two diagrams of topology 8. We see that two genuine diagrams of a different topology can lead to the same sub-loop renormalisation diagram. Therefore, the following relations are understood to hold only when all genuine and all sub-loop renormalisation diagrams are taken into account.

To derive the required relations, as an example, let us first consider a simple two-loop ``test'' self-energy
\begin{equation}
    \Sigma^{(2)}_\text{test} = A_0 B_0 + A_0 \delta c_B + B_0 \delta c_A.
\end{equation}
The first term corresponds to a genuine diagram (like topology 2 in \fig{fig:neutral_self-energies}, but without any couplings), and the second and third terms stem from sub-loop renormalisation diagrams. The divergent parts of the counterterms are given by the divergence of the loop which was shrunk but with an additional minus sign:
\begin{subequations}
\begin{align}
    \delta c_A^\text{div} ={}& -A_0^\text{div}, \\
    \delta c_B^\text{div} ={}& -B_0^\text{div}.
\end{align}
\end{subequations}
This implies the relations
\begin{subequations}
\begin{align}
    A_0^\text{div} \delta c_B^\text{div} + B_0^\text{div} \delta c_A^\text{div}
    ={}& - 2 A_0^\text{div} B_0^\text{div}, \\
    A_0^\text{fin} \delta c_B^\text{div} + B_0^\text{fin} \delta c_A^\text{div}
    ={}& - A_0^\text{fin} B_0^\text{div} - B_0^\text{fin} A_0^\text{div}, \\
    A_0^\del \delta c_B^\text{div} + B_0^\del \delta c_A^\text{div}
    ={}& - A_0^\del B_0^\text{div} - B_0^\del A_0^\text{div}.
\end{align}
\end{subequations}

This derivation can be straightforwardly extended to other products of one-loop integrals. Consequently, in the notation of \eqn{eqn:two-loop_SE_unr}, this means
\begin{subequations}
\begin{align}
    N_j^\text{div} \delta c_j^\text{div} ={}& - 2 L_i^\text{div} M_i^\text{div}, \\
    N_j^\text{fin} \delta c_j^\text{div} ={}& - L_i^\text{fin} M_i^\text{div} - L_i^\text{div} M_i^\text{fin}, \\
    N_j^\del \delta c_j^\text{div} ={}& - L_i^\del M_i^\text{div} - L_i^\text{div} M_i^\del,
\end{align}
\end{subequations}
where now all sub-loop renormalisation diagrams are included on the left-hand side and all genuine diagrams contribute to the right-hand side. These relations hold independently of the renormalisation scheme as only the divergent (and therefore scheme-independent) parts of the one-loop counterterms appear. Inserting all three relations into the expression for the unrenormalised self-energy leaves us with
\begin{equation}
\label{eqn:del_part} 
    \Sigma^{(2)} = -\frac{L_i^\text{div} M_i^\text{div}}{\del^2} + \frac{N_j^\text{div} \delta c_j^\text{fin}}{\del}
    + L_i^\text{fin} M_i^\text{fin} + N_j^\text{div} \delta c_j^\del + N_j^\text{fin} \delta c_j^\text{fin} + \order{\del}.
\end{equation}
We have numerically verified that, in our calculation, the $\order{\del^{-1}}$ part is indeed only generated by the finite parts of the one-loop counterterms. Similarly, all $\order{\del}$ parts of loop integrals in the final result stem from the one-loop counterterms. It becomes clear that some, if not all, two-loop counterterms have to include finite pieces in order to cancel the $\delta c_j^\del$-terms in the renormalised self-energy, as subtracting only the divergences---as is done in a $\DRbar$ scheme---will leave the finite part unaltered. In a pure $\DRbar$ scheme, however, all $\order{\del}$ parts cancel after the sub-loop renormalisation already.

We demonstrate these findings with the example of the two-loop $AA$ self-energy and show under which circumstances the $\order{\del}$ part of the one-loop counterterm $\deltaOL M_W^2$ cancels after renormalisation. To simplify the analysis, we restrict ourselves to a $\DRbar$ renormalisation of $\tb$, which in this case is needed only up to the one-loop order.

The renormalised two-loop $AA$ self-energy is given in \eqn{eqn:A0A0_2L_ren}. Genuine two-loop counterterms as well as products of one-loop counterterms appear. The counterterm products can be neglected in our discussion as $\deltaOL M_W^2$ plays no role at the one-loop level if $\tb$ is renormalised in a $\DRbar$ scheme. The $W$ mass counterterm appears, however, in the sub-loop renormalisation part of $\Sigma^{(2)}_{AA}$ in a product with one-loop integrals. Therefore, it plays a role in the determination of the two-loop counterterms. The relevant terms in the renormalised self-energy are
\begin{equation}
\begin{split}
    \hat \Sigma_{AA}^{(2)}(p^2) ={}& \Sigma_{AA}^{(2)}(p^2) + \deltaTL Z_{AA} (p^2 - m_A^2)  - \deltaTL m_A^2 + \text{terms without $\deltaOL M_W^2$} \\
    ={}& N^\text{div}(p^2) (\deltaOL M_W^2)^\del + \deltaTL Z_{AA} (p^2 - m_A^2)  - \deltaTL m_A^2 \\
    &+ \text{terms without ($\deltaOL M_W^2)^\del$}.
\end{split}
\end{equation}
As before, $N^\text{div}(p^2)$ is the divergent part of the loop integrals which are multiplied with $\deltaOL M_W^2$ in the sub-loop renormalisation diagrams. It has to be a polynomial of degree one in $p^2$, so we can write $N^\text{div}(p^2) = \nu_1 p^2 + \nu_0$.

When defining the mass $m_A^2$ in an $\OS$ scheme, it contains the $\order{\del}$ part of $\deltaOL M_W^2$ as well, since
\begin{equation}
\begin{split}
    \deltaTL m_A^2 ={}& \Re \Sigma_{AA}^{(2)}(m_A^2) + \text{terms without $\deltaOL M_W^2$} \\
    ={}& \underbrace{N^\text{div}(m_A^2)}_{= \nu_1 m_A^2 + \nu_0} (\deltaOL M_W^2)^\del + \text{terms without ($\deltaOL M_W^2)^\del$}.
\end{split}
\end{equation}

Inserting this back into the renormalised self-energy, we arrive at
\begin{equation}
\begin{split}
    \hat \Sigma_{AA}^{(2)}(p^2) ={}& \underbrace{\left( N^\text{div}(p^2) - N^\text{div}(m_A^2) \right)}_{= \nu_1 (p^2 - m_A^2)} (\deltaOL M_W^2)^\del + \deltaTL Z_{AA} (p^2 - m_A^2) \\
    &+ \text{terms without ($\deltaOL M_W^2)^\del$}.
\end{split}
\end{equation}

We can see that the on-shell self-energy $\hat \Sigma_{AA}^{(2)}(m_A^2)$ is free from ($\deltaOL M_W^2)^\del$. To obtain this property also for off-shell momenta, we need to use an on-shell renormalisation for the field counterterm $\deltaTL Z_{AA}$ as well:
\begin{equation}
\begin{split}
    \deltaTL Z_{AA} ={}& - \partial \Sigma_{AA}^{(2)}(m_A^2) + \text{terms without $\deltaOL M_W^2$} \\
    ={}& - \underbrace{\pdv{N^\text{div}(p^2)}{p^2}}_{= \nu_1} (\deltaOL M_W^2)^\del + \text{terms without ($\deltaOL M_W^2)^\del$}.
\end{split}
\end{equation}

With $\deltaTL m_A^2$ and $\deltaTL Z_{AA}$ defined in an on-shell scheme, we find
\begin{equation}
\begin{split}
    \hat \Sigma_{AA}^{(2)}(p^2) \overset{\OS}{=}{}& \nu_1 (p^2 - m_A^2) (\deltaOL M_W^2)^\del - \nu_1 (\deltaOL M_W^2)^\del (p^2 - m_A^2) \\
    &+ \text{terms without ($\deltaOL M_W^2)^\del$} \\
    ={}& \text{terms without ($\deltaOL M_W^2)^\del$}.
\end{split}
\end{equation}

The same logic applies to all other one-loop counterterms as well. Thus, in a full $\OS$ renormalisation, all $\order{\del}$ parts of loop integrals will drop out for arbitrary momenta.

Alternatively, we could have used a full $\DRbar$ renormalisation for all one- and two-loop counterterms. In this case, \eqn{eqn:del_part} takes the form
\begin{equation}
    \Sigma^{(2)} \overset{\DRbar}{=} -\frac{L_i^\text{div} M_i^\text{div}}{\del^2} + L_i^\text{fin} M_i^\text{fin} + \order{\del}.
\end{equation}
The two-loop counterterms will remove the $\order{\del^2}$ divergence, and the renormalised self-energy reads
\begin{equation}
    \hat \Sigma^{(2)} \overset{\DRbar}{=} L_i^\text{fin} M_i^\text{fin} + \order{\del}.
\end{equation}
Again, the renormalised self-energy is free from $\order{\del}$ terms of loop integrals and counterterms.

It becomes clear that, at the two-loop level, the $\order{\del}$ parts of loop integrals and counterterms contribute only in a mixed renormalisation scheme where at least one one-loop counterterm is defined in an on-shell scheme and at least one two-loop counterterm is defined in a minimal subtraction scheme. To demonstrate this, let us assume that we need two counterterms $\delta c_1$ and $\delta c_2$ to renormalise the two-loop self-energy. $\delta c_1$ is only needed at the one-loop level and defined in the $\OS$ scheme, $\delta c_2$ is needed at the one- and two-loop level and defined in the $\DRbar$ scheme. Then
\begin{equation}
    \Sigma^{(2)} \overset{\text{MIX}}{=} -\frac{L_i^\text{div} M_i^\text{div}}{\del^2} + \frac{N_1^\text{div} \delta c_1^\text{fin}}{\del}
    + L_i^\text{fin} M_i^\text{fin} + N_1^\text{div} \delta c_1^\del + N_1^\text{fin} \delta c_1^\text{fin} + \order{\del}.
\end{equation}
The $\DRbar$ $\delta c_2$ counterterm will then only remove the divergences and we find that evanescent terms remain in the renormalised self-energy,
\begin{equation}
    \hat \Sigma^{(2)} \overset{\text{MIX}}{=} L_i^\text{fin} M_i^\text{fin} + N_1^\text{div} \delta c_1^\del + N_1^\text{fin} \delta c_1^\text{fin} + \order{\del}.
\end{equation}

This finding is important regarding the comparison between results that have been obtained in different renormalisation schemes. In schemes where the $\order{\del}$ terms of the counterterms cancel out, it is possible to carry out the renormalisation in a chosen scheme and then do a finite reparameterisation into a different scheme. This is often done, for instance, in order to convert a result that has been obtained using the on-shell definition of the top quark mass into the corresponding result where the $\MSbar$ definition of the top-quark mass is employed. However, the application of this familiar procedure is not possible if, in one of the two schemes, the $\order{\del}$ parts of the counterterms contribute.

The possible occurrence of $\order{\del}$ parts of counterterms in two-loop results for the masses of the Higgs bosons in the MSSM was already noted in \citeres{Borowka:2015ura,Degrassi:2014pfa}. There, a comparison of two calculations was carried out, one of which employed a mixed renormalisation scheme \cite{Borowka:2014wla} while the other one used a full $\DRbar$ renormalisation \cite{Degrassi:2014pfa}. In \citere{Degrassi:2014pfa}, a finite reparameterisation from the $\DRbar$ scheme to an $\OS$ scheme was subsequently performed for the top quark mass and the stop squark masses. Based on this finite reparameterisation a significant disagreement between the two results was pointed out in \citere{Degrassi:2014pfa}. This disagreement was traced back to the occurrence of $(\deltaOL m_t)^\del$ terms in the result of \citere{Borowka:2014wla}, which are not generated in the scheme of \citere{Degrassi:2014pfa}.

Our discussion above clearly demonstrates the complications arising in mixed renormalisation schemes. In this case, it is essential to take into account the possible appearance of evanescent terms if comparing calculations in different renormalisation schemes. Alternatively, the occurrence of evanescent terms in mixed renormalisation schemes can be avoided by modifying the renormalisation prescription --- e.g., by starting with a complete $\DRbar$ scheme and then reparameterising (partly) to the OS scheme.

We emphasise that in a prediction for the relation between physical observables all evanescent terms will drop out, while this is not necessarily the case in relations between physical observables and parameters that have been defined in a minimal subtraction scheme at the two-loop level and beyond. Since in mixed renormalisation schemes (without modification of the renormalisation prescription) terms like $(\deltaOL m_t)^\del$ contribute in the latter relations, while they will eventually drop out in relations between physical observables, the incomplete cancellation of evanescent terms may lead to numerical instabilities or gauge-dependent effects which should be seen as a theoretical limitation of such relations between parameters in a minimal subtraction scheme and physical observables. This also motivated us to use an OS renormalisation for $\tan\beta$ at the two-loop level. Finally, we stress that evanescent terms do not only appear in BSM theories but can also arise in SM calculations if parameters like $\alem$, $\als$ or the weak mixing angle are renormalised in the $\MSbar$ scheme, while some or all of the masses are defined in the on-shell scheme.

\section{Numerical analysis of the full electroweak \texorpdfstring{$\order{\nc^2}$}{O(Nc2)} two-loop results}
\label{cha:two-loop}
We now turn to the numerical results of our two-loop prediction for the MSSM Higgs boson masses. Our main emphasis lies on the size of our newly calculated contributions relative to the experimental uncertainty of the mass of the detected Higgs boson at $M_h = 125.11 \pm 0.11 \gev$ \cite{ATLAS:2023oaq,ATLAS:2023owm}. In our numerical discussion for the different scenarios, it will therefore not be our goal to include all known one-loop\cite{Chankowski:1991md,Brignole:1991wp,Brignole:1992uf,Chankowski:1992er,Dabelstein:1994hb,Pierce:1996zz,Pilaftsis:1998dd,Demir:1999hj,Pilaftsis:1999qt,Choi:2000wz,Carena:2000yi,Ibrahim:2000qj,Heinemeyer:2001qd,Carena:2001fw,Ibrahim:2002zk,Ellis:2004fs,Frank:2006yh} and two-loop contributions \cite{Hempfling:1993qq,Heinemeyer:1998jw,Heinemeyer:1998kz,Zhang:1998bm,Heinemeyer:1998np,Espinosa:1999zm,Carena:2000dp,Espinosa:2000df,Degrassi:2001yf,Brignole:2001jy,Brignole:2002bz,Dedes:2002dy,Dedes:2003km,Allanach:2004rh,Heinemeyer:2004xw,Martin:2002wn,Martin:2004kr,Heinemeyer:2007aq,Borowka:2014wla,Degrassi:2014pfa,Borowka:2018anu,Hollik:2014wea,Hollik:2014bua,Hahn:2015gaa,Passehr:2017ufr} to the MSSM Higgs boson mass and to perform a resummation of large logarithms \cite{Barbieri:1990ja,Espinosa:1991fc,Casas:1994us,Haber:1993an,Espinosa:1999zm,Carena:2000dp,Carena:1995bx,Carena:1995wu,Haber:1996fp,Degrassi:2002fi,Martin:2007pg,Hahn:2013ria,Draper:2013oza,Arkani-Hamed:2004ymt,Giudice:2004tc,Carena:2008rt,Binger:2004nn,Bernal:2007uv,Giardino:2011aa,Giudice:2011cg,Bagnaschi:2014rsa,Tamarit:2012ie,Benakli:2013msa,Fox:2005yp,Hall:2009nd,Cabrera:2011bi,Degrassi:2012ry,PardoVega:2015eno,Bagnaschi:2017xid,Harlander:2018yhj,Bagnaschi:2019esc,Bahl:2019wzx,Bahl:2020tuq,Carena:2015uoe,Murphy:2019qpm,Gorbahn:2009pp,Bahl:2018jom,Lee:2015uza,Bahl:2020jaq,BhupalDev:2014bir,Bednyakov:2018cmx,Schienbein:2018fsw,Oredsson:2018yho,Herren:2017uxn,Bagnaschi:2015pwa,Bagnaschi:2015hka,Bahl:2019ago,Cheung:2014hya,Kwasnitza:2021idg}. Instead, we take into account all one-loop contributions of $\order{\nc}$ and the full two-loop contributions of $\order{\nc^2}$ and concentrate on an analysis of the relative shifts induced by these corrections instead of the absolute predictions for the Higgs boson masses.\footnote{A consistent combination of the newly calculated $\order{\nc^2}$ corrections with the other known two-loop corrections and a resummation of large logarithms requires a significant effort (i.e., the avoidance of double-counting and a consistent parameter definition). This is clearly beyond the scope of the current paper, which is focused on the $\order{\nc^2}$ corrections itself. We leave this for future work.} As we explained in the previous sections, we neglect the quark masses of the first and second generation. Since they do not belong to the class of $\order{\nc^2}$ contributions, we note that we also do not include the numerically sizeable two-loop QCD corrections that have been calculated previously \cite{Heinemeyer:2007aq,Borowka:2014wla,Degrassi:2014pfa,Borowka:2018anu}. The couplings relevant to us are $\alem$, $\alt$, and $\alb$. We go beyond \citeres{Hollik:2014wea,Hollik:2014bua,Hahn:2015gaa,Passehr:2017ufr} in including the dependence on the external momentum also for the Yukawa terms of $\order{\nc^2}$.

The MSSM in its most general $R$-parity conserving form---with complex parameters and taking into account all possible mixing contributions between the sfermions---has 124 input parameters compared to the 19 parameters in the SM \cite{Haber:1997if,Drees:2004jm}. Despite having worked out the renormalisation for the most general case of complex parameters in \cha{cha:mssm_ren}, we restrict ourselves to $\CP$-conserving scenarios in the following analyses. As explained in \appx{app:squark_ren}, we also assume flavour diagonal squark mass matrices and a unit CKM matrix. This already greatly reduces the number of MSSM parameters entering our calculation.

In the Higgs-gauge sector, the most important parameters for a Higgs boson mass prediction are the mass of the $\CP$-odd Higgs boson, $m_A$, and the VEV ratio $\tb$. They fully determine the tree-level masses of the $\CP$-even Higgs bosons, see \eqn{eqn:CP-even_tree-level_masses}. Starting from the one-loop level, parameters from the squark sector also enter the mass prediction through self-energy diagrams containing squarks in the loops. These parameters appear in the squark mass matrices given in \eqs{eqs:squark_mass_matrices_params}, namely the squark mass parameters $M_{\tilde q_g}$, $M_{\tilde u_g}$, and $M_{\tilde d_g}$ for each generation $g$, the trilinear couplings $A_t$ and $A_b$,\footnote{The trilinear couplings $A_u$, $A_c$, $A_d$, and $A_s$ do not appear in our calculation since we neglect the corresponding quark masses.} and the higgsino mass parameter $\mu$. All of these parameters have a non-vanishing mass dimension and, with the exception of $\mu$, break supersymmetry softly. The trilinear couplings and $\mu$ determine the off-diagonal elements of the squark mass matrices and are hence responsible for the strength of the squark mass mixing. In our analysis, we typically set these parameters to one single common value, the SUSY scale $M_S$.

We investigate the calculated corrections for four different MSSM scenarios:
\begin{itemize}
    \item{The dependence of the light Higgs boson mass $M_h$ on the SUSY scale $M_S$ for $\tb = 15$ and $A_q = 0$.}
    \item{The dependence of the light Higgs boson mass $M_h$ on the SUSY scale $M_S$ for $\tb = 15$ and $A_q = - 2 M_S$.}
    \item{The dependence of the light Higgs boson mass $M_h$ on the trilinear coupling $A_q$ for $\tb = 15$ and $M_S = 1.5 \tev$.}
    \item{The dependence of the $\CP$-even Higgs boson masses on the trilinear coupling $A_q$ in the $M_h^{125}$ benchmark scenario \cite{Bagnaschi:2018ofa} for $\tb = 5$ and $m_A = 90 \gev$.}
\end{itemize}

In order to estimate the size of the individual corrections contributing to the $\order{\nc^2}$ prediction, we perform calculations for different values of the coupling constants in each scenario. For the full prediction, we use the parameter values given below. Additionally, we make predictions for sufficiently small values for the electric charge and the bottom mass, the former allowing us to take the gaugeless limit numerically. By appropriately adding and subtracting the different predictions, we can then separate the Yukawa contributions from the gauge contributions. The limit of vanishing bottom mass allows us to separate the dominant top contributions from the smaller bottom and the top-bottom-mixing contributions. The details are given in \appx{app:PlotGen}.

Our considered scenarios respect the $\CP$ symmetry and hence only the $\CP$-even Higgs bosons $h$ and $H$ mix with each other. We therefore use an on-shell definition for the mass of the $\CP$-odd Higgs boson $A$. Furthermore, we use the on-shell renormalisation for $\tb$ which we explained in detail in \sct{ssec:tanbeta_OS_ren}. For the quark--squark sector we choose a mixed on-shell--$\DRbar$ renormalisation: We define the stop masses on-shell, while the trilinear stop coupling $A_t$ is renormalised in the $\DRbar$ scheme. In the first three scenarios, we define the lighter sbottom mass on-shell, whereas in the last scenario, the heavier sbottom mass is renormalised on-shell.\footnote{These specific choices yielded the best numerical stability.}

The squared Higgs boson masses are calculated by finding the pole of the inverse propagator matrix. Except for scenario 4, we work at the strict two-loop order implying that (see e.g.\ \citere{Bahl:2018ykj})
\begin{subequations}
\label{eqs:fixed-order_Higgs_masses}
\begin{align}
\begin{split}
    M_h^2 ={}& m_h^2 - \Re \hat \Sigma_{hh}^{(1)}(m_h^2) - \Re \hat \Sigma_{hh}^{(2)}(m_h^2) \\
    &+ \Re \Big\{ \hat \Sigma_{hh}^{(1)}(m_h^2) \partial \hat \Sigma_{hh}^{(1)}(m_h^2) \Big\} - \Re \frac{\big( \hat \Sigma_{hH}^{(1)}(m_h^2) \big)^2}{m_H^2 - m_h^2},
\end{split} \\
\begin{split}
    M_H^2 ={}& m_H^2 - \Re \hat \Sigma_{HH}^{(1)}(m_H^2) - \Re \hat \Sigma_{HH}^{(2)}(m_H^2) \\
    &+ \Re \Big\{ \hat \Sigma_{HH}^{(1)}(m_H^2) \partial \hat \Sigma_{HH}^{(1)}(m_H^2) \Big\} + \Re \frac{\big( \hat \Sigma_{hH}^{(1)}(m_H^2) \big)^2}{m_H^2 - m_h^2}.
\end{split}
\end{align}
\end{subequations}
In scenario 4, in which mixing effects are important, we determine the pole numerically via a fixed-point iteration.

It is important to note that the fixed-order approach via the self-energies specified above gives corrections to the squared Higgs boson masses since the mass parameters appearing in the Lagrangian are of mass dimension two, see \eqn{eqn:Higgs_Lag_mass}. In order to be able to compare our results with the experimental value for the Higgs boson mass, which is of mass dimension one, we have to take the square root of the above expressions. This will naturally mix different contributions and orders of perturbation theory. For our analysis we therefore calculate:
\begin{itemize}
    \item{The Higgs boson mass, $M_{h_i}$. It contains the full tree-level as well as one-loop and two-loop contributions of order $\order{\nc}$ and $\order{\nc^2}$, respectively. It is obtained by simply taking the square root of the squared Higgs boson mass, see \eqs{eqs:Mh2plots}. This calculation gives an estimate of the overall value of the Higgs boson mass (where, as explained above, numerically sizeable contributions that are not of $\order{\nc}$ or $\order{\nc^2}$ have not been incorporated).}
    \item{Two-loop contributions of $\order{\nc^2}$ to the Higgs boson mass, $\Delta^{(2)} M_{h_i}$. They are calculated by subtracting different predictions for the Higgs boson mass, see \eqs{eqs:DeltaMhplots}. These calculations allow us to compare the size of our newly calculated contributions to the experimental uncertainty of the observed Higgs boson mass.}
\end{itemize}

For our parameters, unless explicitly stated otherwise, we use the following values:
\begin{equation}
\label{eqn:input_pars}
\begin{aligned}
    \alem^{-1} ={}& 137.035999084 \text{\cite{ParticleDataGroup:2020ssz}},
    & \als ={}& 0, \\
    \Delta \alpha_\text{lep}(M_Z^2) ={}& 0.031497687 \text{\cite{Steinhauser:1998rq}},
    & \Delta \alpha^{(5)}_\text{had}(M_Z^2) ={}& 0.02766 \text{\cite{ParticleDataGroup:2020ssz}}, \\
    M_Z ={}& 91.1876 \gev \text{\cite{ParticleDataGroup:2020ssz}},
    & M_W ={}& 80.379 \gev \text{\cite{ParticleDataGroup:2020ssz}}, \\
    m_u ={}& 0,
    & m_c ={}& 0, \\
    m_t ={}& M_t = 172.76 \gev \text{\cite{ParticleDataGroup:2020ssz}},
    & m_d ={}& 0, \\
    m_s ={}& 0,
    & m_b ={}& 4.18 \gev \text{\cite{ParticleDataGroup:2020ssz}}, \\
    \tb ={}& \tb^\OS = 15,
    & m_A ={}& M_A = M_S, \\
    M_{\tilde q_g}^2 ={}& M_S^2,
    & M_{\tilde u_g}^2 ={}& M_S^2, \\
    M_{\tilde d_g}^2 ={}& M_S^2,
    & \mu ={}& M_S, \\
    A_t ={}& A_q,
    & A_b ={}& A_q.
\end{aligned}
\end{equation}
The trilinear couplings of the first and second generation do not need to be specified since only the product $m_q A_q$ enters the calculation, and the associated quark masses vanish in our approximation. The given top quark mass is defined as the pole mass, and all plots in the following analyses are based on predictions employing the on-shell scheme for the top quark mass.

\subsection{Scenario 1: The dependence of \texorpdfstring{$M_h$}{Mh} on the scale \texorpdfstring{$M_S$}{MS} for \texorpdfstring{$A_q = 0$}{Aq=0}}
\label{sec:scenario1}

\begin{figure}
    \centering
    \includegraphics[width = .6 \textwidth]{"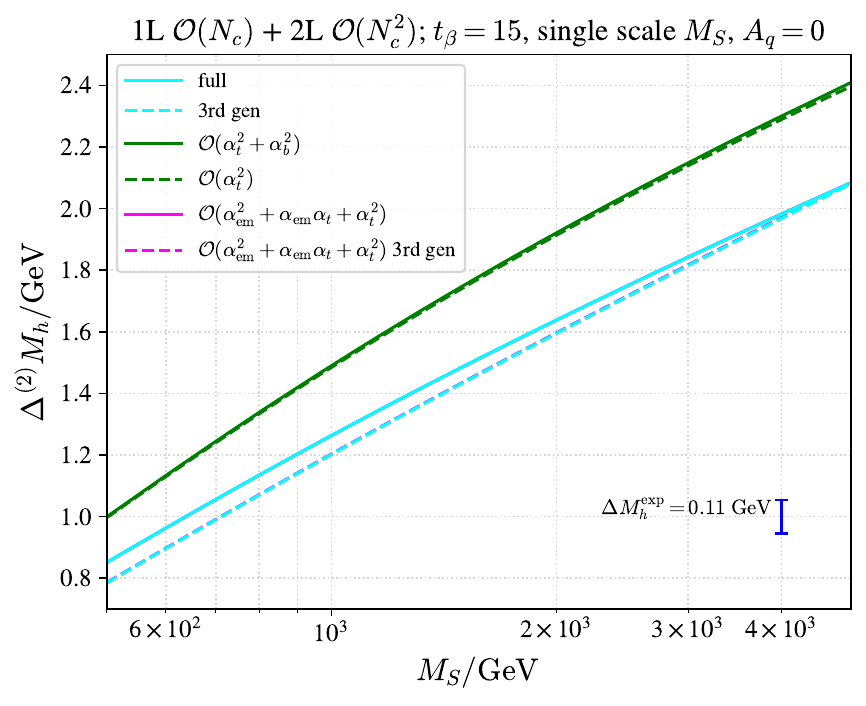"}
    \caption{Two-loop contributions to the light Higgs boson mass $M_h$ for $A_q = 0$. The solid cyan curve includes all two-loop contributions at $\order{\nc^2}$, the dashed cyan curve uses only contributions from the third generation. The green curves give the two-loop Yukawa contributions at $\order{\nc^2}$. The magenta curves do not contain contributions proportional to the bottom Yukawa coupling. They lie under the cyan curves. In blue, we give the experimental uncertainty for the Higgs boson mass.}
    \label{fig:Aq0_big}
\end{figure}

For our first scenario, we investigate how the mass of the light $\CP$-even Higgs boson, $h$, depends on the SUSY scale $M_S$. We set the VEV ratio $\tb = 15$ and we assume the third generation trilinear couplings, $A_t$ and $A_b$, to vanish. The remaining soft SUSY-breaking parameters, the squark masses and $\mu$, are set to the same value $M_S$. Furthermore, we set the $\CP$-odd mass $m_A$ to $M_S$ as well.

In \fig{fig:Aq0_big}, we show the shifts in $M_h$ induced by the $\order{\nc^2}$ two-loop contributions in this scenario. The pure Yukawa corrections, shown in green, are dominated by the top and stop contributions of $\order{\alt^2}$. The bottom and sbottom contributions are negligible in this scenario. The cyan curves contain the full electroweak two-loop contributions of $\order{\nc^2}$. We also made a prediction in the limit $m_b \to 0$, which is supposed to be shown in magenta. Due to the aforementioned smallness of the $\order{\alb^2}$ terms, the magenta curves are not distinguishable from the cyan ones and hence lie behind them.

The green curves represent the contributions of our calculation that were already known. The cyan curves additionally contain pure gauge ($\order{\alem^2}$) and mixed gauge-Yukawa ($\order{\alem \alq}$) contributions that were calculated for the first time in this paper. Independently of the value chosen for $M_S$, these terms lower the $\order{\nc^2}$ two-loop corrections by approximately $15\%$ of the pure Yukawa contributions. This reduction is even larger when regarding only the contributions of the third generation of quarks and squarks (the dashed cyan curve). Our additional contributions lead to a shift of the Higgs boson mass of $0.15 \gev$ for smaller values of $M_S$ and more than $0.3 \gev$ for larger values. The new contributions shift the mass of the light Higgs boson by an amount that is larger than the current experimental uncertainty. The remaining uncertainty translates into an uncertainty on the SUSY parameters when using the measured Higgs mass as an input for constraining the SUSY parameters. We also note that additional uncertainties arise not only from additional higher-order corrections (e.g., electroweak two-loop contributions not enhanced by $\order{\nc^2}$ or three-loop corrections beyond the known logarithmic contributions) but also from the uncertainties of the SM parameters used as an input for our calculation (see e.g.\ Ref.~\cite{Slavich:2020zjv} for more details). To bring the theoretical uncertainties to the same level as the current experimental one, additional higher-order corrections beyond the ones obtained in the present work will be required.

\subsection{Scenario 2: The dependence of \texorpdfstring{$M_h$}{Mh} on the scale \texorpdfstring{$M_S$}{MS} for \texorpdfstring{$A_q = -2 M_S$}{Aq=-2MS}}
\label{sec:scenario2}

\begin{figure}
    \centering
    \includegraphics[width = .6 \textwidth]{"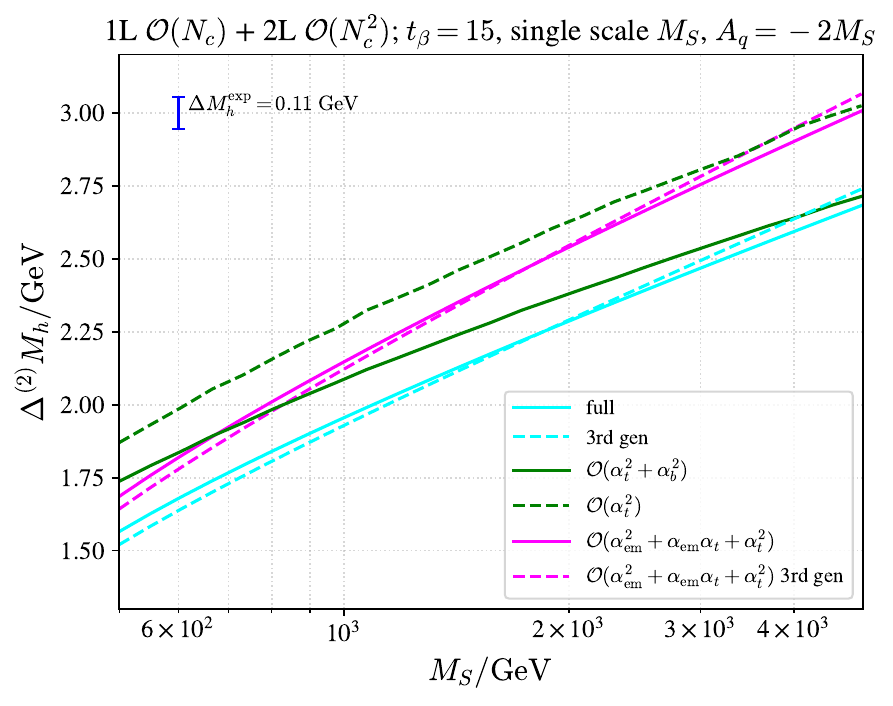"}
    \caption{Two-loop contributions to the light Higgs boson mass $M_h$ for $A_q = -2 M_S$. The solid cyan curve includes all two-loop contributions at $\order{\nc^2}$, the dashed cyan curve uses only contributions from the third generation. The green curves give the two-loop Yukawa contributions at $\order{\nc^2}$. The magenta curves do not contain contributions proportional to the bottom Yukawa coupling. In blue, we give the experimental uncertainty band for the Higgs boson mass.}
    \label{fig:Aqn2_big}
\end{figure}

\begin{figure}
    \centering
    \includegraphics[width = .6\textwidth]{"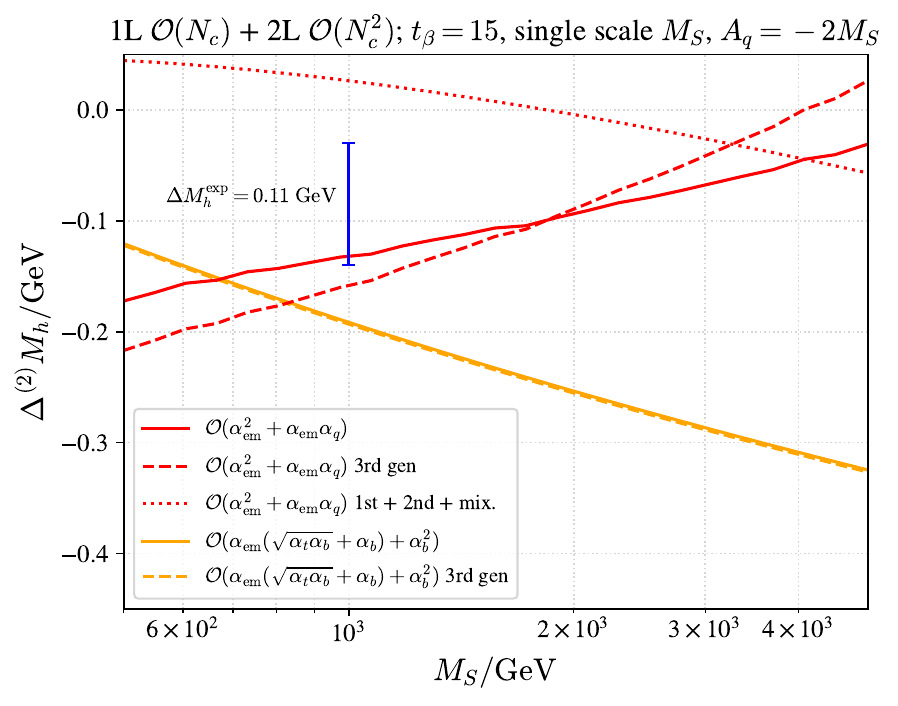"}
    \caption{Subleading two-loop contributions to the light Higgs boson mass $M_h$ for $A_q = -2 M_S$. The red curves correspond to the contributions depending on the fine-structure constant $\alem$. The orange curves give the contributions which vanish in the limit $m_b \to 0$. In blue, we give the experimental uncertainty for the Higgs boson mass.}
    \label{fig:Aqn2_sml}
\end{figure}

The second scenario uses the same parameters as the first one, the only difference is that the trilinear couplings are now set to $A_q = - 2 M_S$. The value of the stop-mixing parameter $X_t$ is therefore close to the one for which the maximal value of $M_h$ is obtained (for the case of negative values of $X_t$) \cite{Carena:1999xa,Carena:2002qg,Bagnaschi:2018ofa}.

We display the shifts in $M_h$ induced by the $\order{\nc^2}$ two-loop corrections in \fig{fig:Aqn2_big}. Regarding the pure Yukawa corrections (green curves), we see that the bottom and sbottom contributions now make up a considerable part of the full Yukawa contribution. The full two-loop correction (solid cyan curve) is again dominated by the Yukawa terms; the combined gauge and the mixed gauge-Yukawa contributions corresponds to a shift of 0.03--$0.17 \gev$ for the Higgs boson mass prediction. The bottom and sbottom contributions, on the other hand, give a shift of 0.12--$0.32 \gev$.

In \fig{fig:Aqn2_sml}, we show the subleading contributions to our Higgs mass prediction in the second scenario. They have been obtained by appropriate subtraction of the curves from \fig{fig:Aqn2_big}. We see that a cancellation takes place between the contributions from the quarks and squarks of the third generation (dashed red curve) and the combined first, second, and generation-mixing contributions (dotted red curve). Nevertheless, our newly obtained corrections (solid red curve) are comparable to the experimental uncertainty for not too large values of $M_S$. Also, the bottom corrections (orange curves) have a sizeable impact.

\subsection{Scenario 3: The dependence of \texorpdfstring{$M_h$}{Mh} on the trilinear coupling \texorpdfstring{$A_q$}{Aq}}
\label{sec:scenario3}

\begin{figure}
    \centering
    \includegraphics[width = 0.6 \textwidth]{"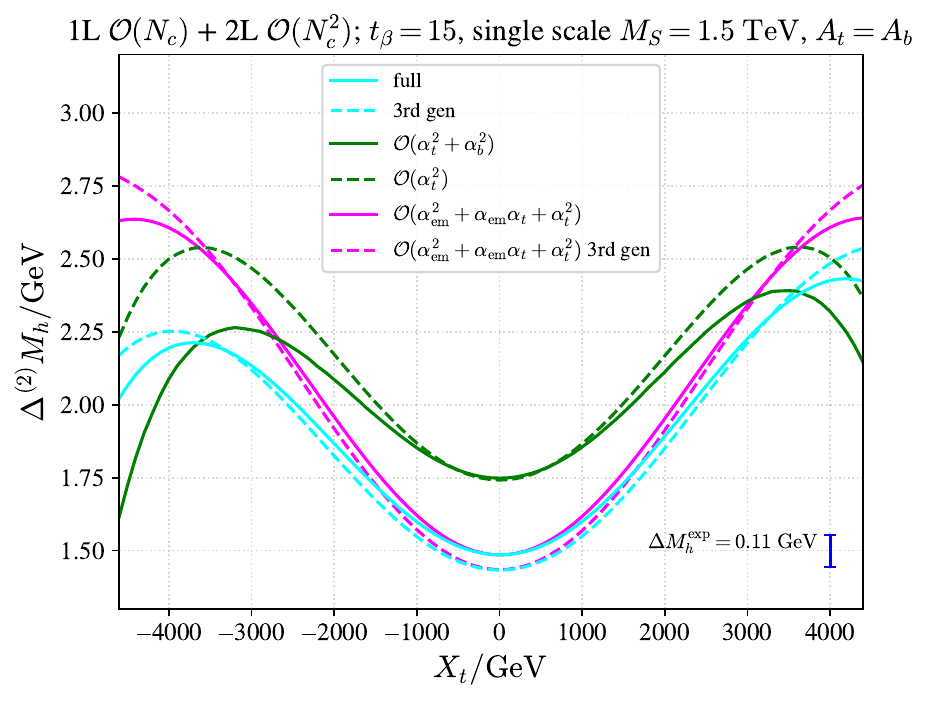"}
    \caption{Two-loop contributions to the light Higgs boson mass $M_h$. We plot the corrections against the stop-mixing parameter $X_t = A_t - \mu/\tb$. The solid cyan curve includes all two-loop contributions at $\order{\nc^2}$, the dashed cyan curve uses only contributions from the third generation. The green curves give the two-loop Yukawa contributions at $\order{\nc^2}$. The magenta curves do not contain contributions proportional to the bottom Yukawa coupling. In blue, we give the experimental uncertainty for the Higgs boson mass.}
    \label{fig:Aqvar_big}
\end{figure}

\begin{figure}
    \centering
    \includegraphics[width = .6\textwidth]{"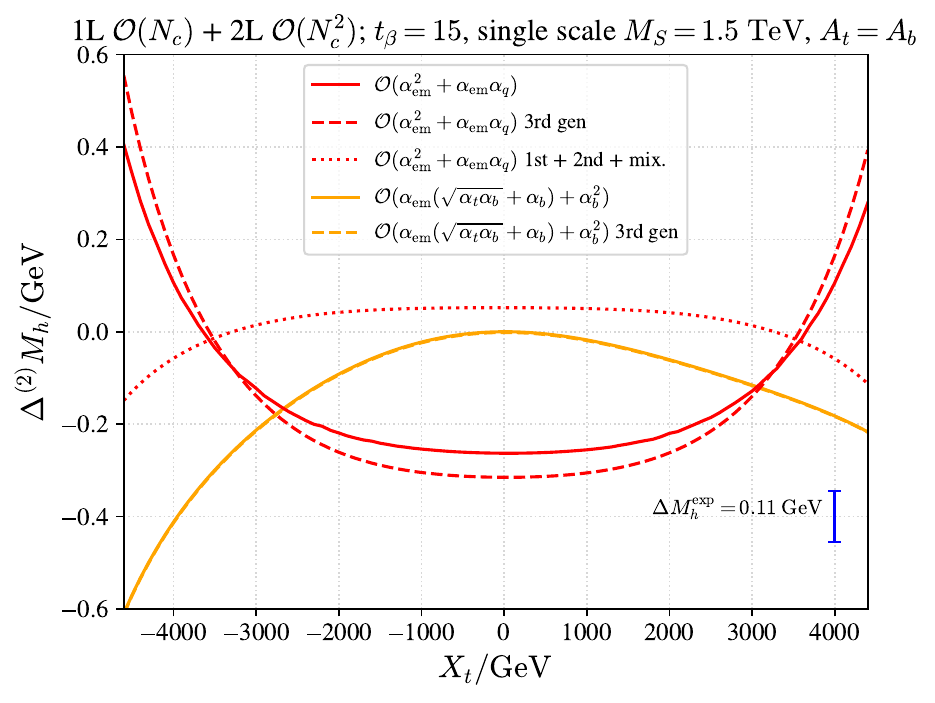"}
    \caption{Subleading two-loop contributions to the light Higgs boson mass $M_h$. We plot the corrections against the stop-mixing parameter $X_t = A_t - \mu/\tb$. The red curves give the contributions depending on the fine-structure constant $\alem$. The orange curves give the contributions which vanish in the limit $m_b \to 0$. In blue, we give the experimental uncertainty for the Higgs boson mass.}
    \label{fig:Aqvar_sml}
\end{figure}

In our third scenario, we analyse how our prediction for the mass of the light $\CP$-even Higgs boson depends on the trilinear couplings $A_t$ and $A_b$. To this end, we use a single value for the trilinear couplings and set $A_t = A_q = A_b$. All other SUSY-breaking parameters, the higgsino mass parameter $\mu$, and the mass of the $\CP$-odd Higgs boson $m_A$ are set to the fixed value $M_S = 1.5 \tev$. Plots for this scenario are shown as a function of the stop-mixing parameter $X_t = A_t - \mu/\tb = A_q - \mu/\tb$.

In \fig{fig:Aqvar_big}, we show the shifts induced by the $\order{\nc^2}$ contributions in the third scenario. Here, we can see that the two-loop corrections shift the Higgs boson mass by 1.4--$2.8 \gev$, depending on the contributions chosen. Again, the Yukawa contributions make up the largest part of the two-loop contributions. The pure top contributions of  $\order{\alt^2}$ (dashed green curve) are very symmetric with respect to their dependence on $X_t$. When including also the bottom contributions (solid green curve), which are symmetric with respect to $X_b = A_b - \mu \tb$, the curve loses its $X_t$ symmetry; for negative values of $X_t$, the pure bottom corrections are larger than for positive values for the same value of $|X_t|$ (see also \fig{fig:Aqvar_sml}). The bottom contributions lower the prediction by roughly $10\%$, as we can see from comparing the magenta curves with the cyan ones, or the dashed green one with the solid green curve.

The additional inclusion of gauge contributions (cyan curves) leaves the maximal value for the corrections largely unaffected (solid cyan vs.\ solid green). They, however, shift the position of the maximum to larger values of $|X_t| \approx 4 \tev$. The minimum remains at $X_t \approx 0$, but the gauge contributions lower it by around $0.25 \gev$ in comparison to the pure Yukawa corrections.

In \fig{fig:Aqvar_sml}, we show the subleading contributions in our third scenario. We can clearly see the aforementioned, strong asymmetry of the bottom corrections with respect to $X_t$ (orange curves). The corrections proportional to the gauge couplings are dominated by contributions from the third-generation squarks. In the whole range of $|X_t| < 2 \tev$, the gauge corrections lower the Higgs boson mass by more than $0.2 \gev$ and hence exceed the experimental uncertainty. In the same range, the gauge corrections which stem from the first- and second-generation squarks and generation mixing increase the Higgs boson mass by $\sim 0.05 \gev$.

\subsection{Scenario 4: The Higgs boson masses in the \texorpdfstring{$M_h^{125}$}{Mh125} scenario with strong mixing}
\label{sec:scenario5}

In the three scenarios discussed so far, we have exclusively used the fixed-order method (see \eqs{eqs:fixed-order_Higgs_masses}) 
to predict the MSSM Higgs boson masses. This allowed us to cleanly separate the different contributions entering the prediction and, hence, we could compare the size of our newly calculated gauge and gauge-Yukawa-mixing terms of $\order{(\alem^2 + \alem \alq)\nc^2}$ against the already known pure Yukawa terms of $\order{(\alt^2 + \alb^2)\nc^2}$. Such an approach yields a reliable prediction only when the difference between the tree-level masses $m_h$ and $m_H$ is sufficiently large. Until now, the tree-level mass split was sizeable enough for the fixed-order method to work.

In this fourth scenario, we now want to investigate a case where we can no longer predict the Higgs boson masses in a strict perturbative approach. Our scenario of choice is a slightly modified version of the ``$M_h^{125}$ scenario'' as it has been defined in \citere{Bagnaschi:2018ofa}. For the Standard Model parameters, we use the values given in \eqs{eqn:input_pars}. For the MSSM parameters, we set
\begin{equation}
\begin{aligned}
    \tb ={}& 5,
    & m_A ={}& 90 \gev, \\
    M_{\tilde q_1}^2 ={}& M_{\tilde u_1}^2 = M_{\tilde d_1}^2 = (2 \tev)^2,
    & M_{\tilde q_2}^2 ={}& M_{\tilde u_2}^2 = M_{\tilde d_2}^2 = (2 \tev)^2, \\
    M_{\tilde q_3}^2 ={}& M_{\tilde u_3}^2 = M_{\tilde d_3}^2 = (1.5 \tev)^2,
    & \mu ={}& 1 \tev, \\
    A_t ={}& A_q,
    & A_b ={}& A_q. 
\end{aligned}
\end{equation}

The ``$M_h^{125}$ scenario'' was designed such that with the theoretical prediction at that time one obtained a Higgs boson mass that was compatible with the experimental value within the theoretical uncertainties over a wide range of $m_A$ and $\tb$. In this original version, the stop-mixing parameter $X_t$ is fixed and hence determines the trilinear coupling $A_q$.

\begin{figure}[t]
    \centering
    \includegraphics[width = .7\textwidth]{"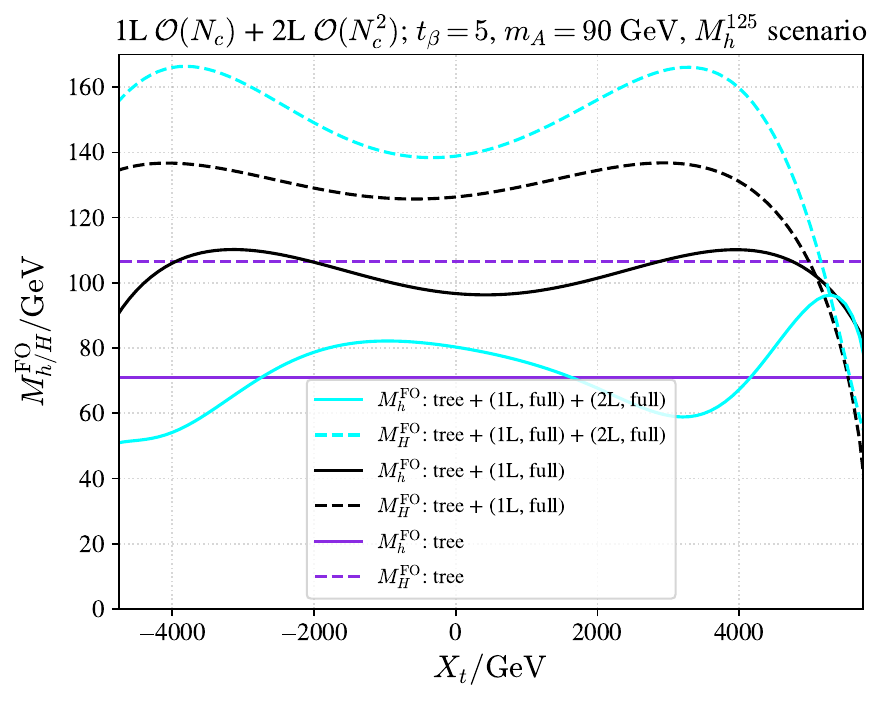"}
    \caption{The dependence of the $\CP$-even Higgs boson masses, $M_h$ and $M_H$, on the trilinear coupling $A_q$ in an illustrative scenario that is not phenomenologically viable (see text). We plot the masses against the stop-mixing parameter $X_t = A_t - \mu/\tb$. The cyan curves include the tree-level prediction as well as the one-loop contributions of $\order{\nc}$ and the two-loop contributions of $\order{\nc^2}$. The black curves show the prediction the Higgs boson masses omitting the two-loop contributions. The purple lines indicate the tree-level predictions. All curves have been generated by including all considered contributions in a strict fixed-order approach (see \eqs{eqs:fixed-order_Higgs_masses}).}
    \label{fig:MH125_FO_overview}
\end{figure}

We pursue here a different approach; we impose values for $\tb$ and $m_A$, allowing us to investigate the dependence of the Higgs boson masses on the trilinear coupling $A_q$. We emphasise that a scenario with two light $\CP$-even states and a similarly light $\CP$-odd state is excluded by experimental measurements and searches. This scenario is, nevertheless, useful to showcase some interesting features of our newly calculated contributions. These features will be present in a similar manner for the mixing between the nearly mass-degenerate two heavy neutral Higgs bosons $H$ and $A$ in a $\CP$-violating scenario.

For our choice of parameters, the $\CP$-even tree-level masses are $m_h \approx 71 \gev$ and $m_H \approx 107 \gev$, shown by the purple lines in \fig{fig:MH125_FO_overview}. The difference between these tree-level values is small enough for large resonance effects to spoil the perturbative ansatz, as we can see from the loop-corrected masses that are obtained using the fixed-order method according to \eqs{eqs:fixed-order_Higgs_masses} (black and cyan curves in \fig{fig:MH125_FO_overview}). The one-loop corrections shift $M_h$ by up to $40 \gev$, $M_H$ is increased by up to $30 \gev$. The two-loop corrections, on the other hand, lower $M_h$ by more than $50 \gev$ around $X_t \approx 3 \tev$; the two-loop prediction for $M_H$ is more than $30 \gev$ larger than the one-loop result for the same value of $X_t$. As a result, in this scenario the perturbative series is no longer well-behaved if the pole masses are determined by a strict fixed-order approach. We therefore determine the pole masses numerically using a fixed-point iteration incorporating the momentum dependence of the $\order{(\alt^2 + \alb^2)\nc^2}$ Yukawa terms, which has not been available before.

\begin{figure}
    \centering
    \begin{subfigure}{.6\textwidth}
        \centering
        \includegraphics[width = \textwidth]{"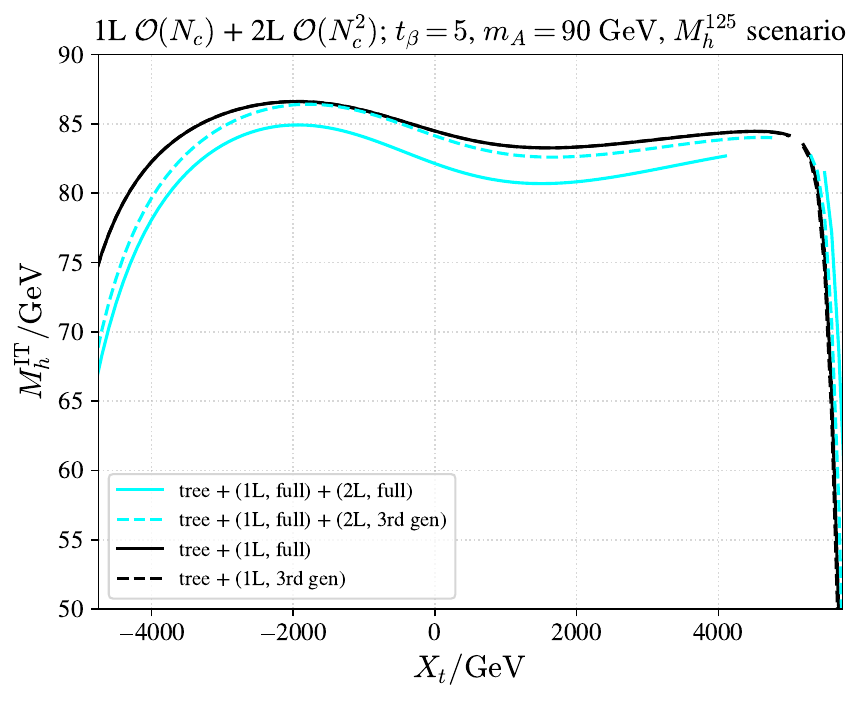"}
        \caption{}
        \label{fig:MH125_Mh1_overview}
    \end{subfigure}
    \hfill
    \begin{subfigure}{.6\textwidth}
        \centering
        \includegraphics[width = \textwidth]{"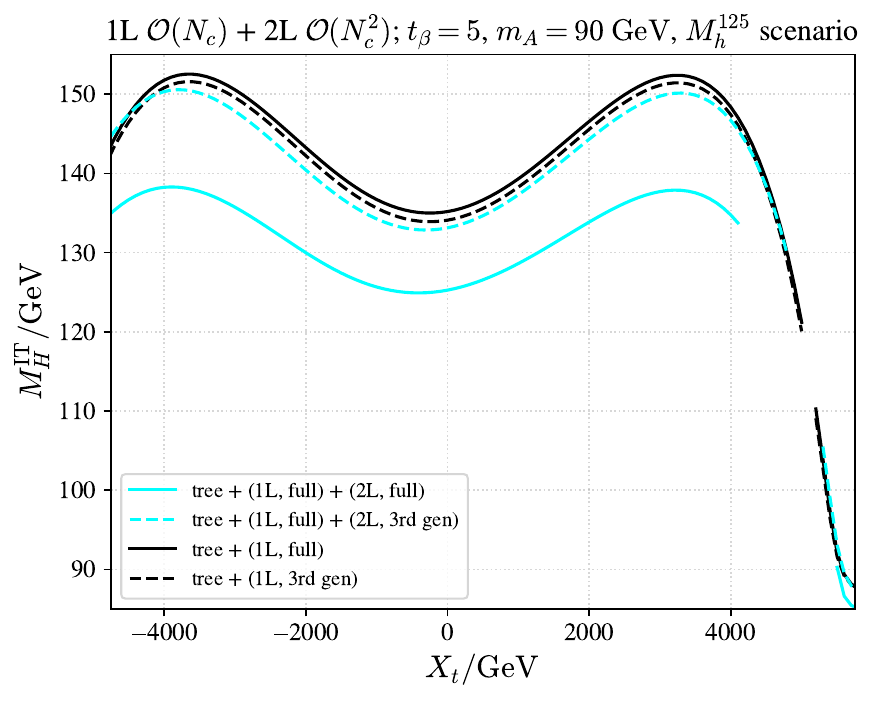"}
        \caption{}
        \label{fig:MH125_Mh2_overview}
    \end{subfigure}
    \caption{The dependence of the $\CP$-even Higgs boson masses, $M_h$ and $M_H$, on $A_q$ for the same scenario as in \fig{fig:MH125_FO_overview} but using a fixed-point iteration for determining the pole masses. We plot the masses against the stop-mixing parameter $X_t = A_t - \mu/\tb$. The cyan curves include the tree-level prediction as well as the one-loop contributions of $\order{\nc}$ and the two-loop contributions of $\order{\nc^2}$. The black curves show the prediction of the Higgs boson masses omitting the two-loop contributions. The solid curves have been generated by including all considered contributions. The dashed curves include only contributions from the third generation of quarks and squarks.}
    \label{fig:MH125_IT_overview}
\end{figure}

The results of the fixed-point iteration are shown in \fig{fig:MH125_IT_overview} for $M_h$ (upper plot) and $M_H$ (lower plot). The plots contain gaps in the region $X_t > 4 \tev$ because, for these parameter points, the fixed-point iteration did not converge within the desired relative precision ($10^{-5}$) after a designated number of steps ($\sim 1000$).

In both plots, the two-loop predictions are much closer to the one-loop results than they were in \fig{fig:MH125_FO_overview}. In contrast to the fixed-order method, for which the hierarchy between the tree-level mass eigenstates $h$ and $H$ got inverted at around $X_t \approx 5 \tev$, we now also have a clear separation between the masses. With the iterated procedure, the lighter mass $M_h$ is always lower than $90 \gev$ (\fig{fig:MH125_Mh1_overview}) while the heavier mass exceeds $120 \gev$ for $X_t < 4 \tev$ (\fig{fig:MH125_Mh2_overview}).

\begin{figure}
    \centering
    \includegraphics[width = .6\textwidth]{"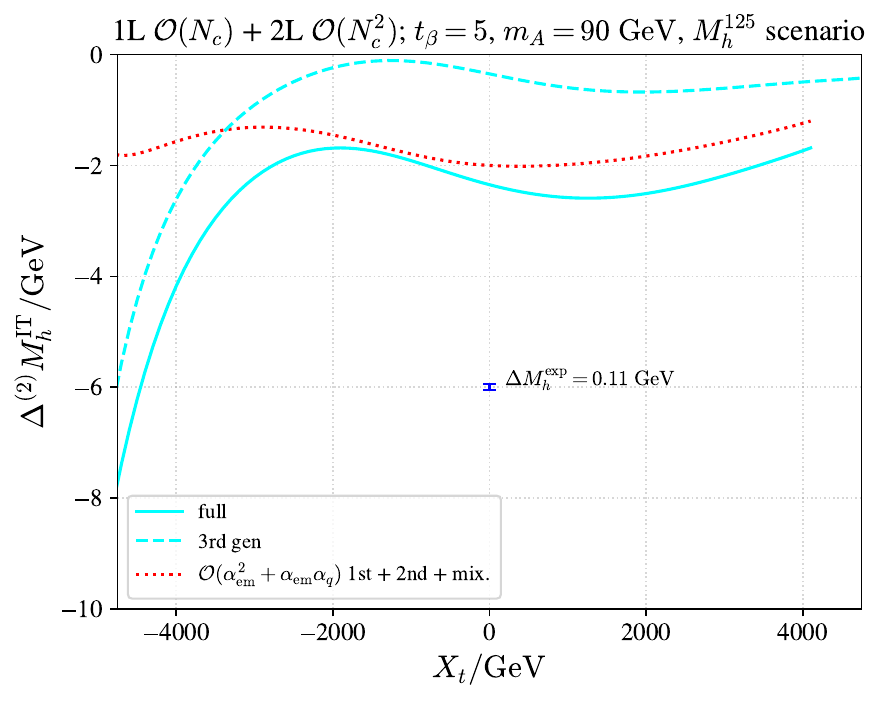"}
    \caption{Two-loop contributions to the light Higgs boson mass $M_h$ for the same scenario as in \fig{fig:MH125_FO_overview} but using a fixed-point iteration for determining the pole masses. The solid cyan curve includes all two-loop contributions at $\order{\nc^2}$, the dashed cyan curve incorporates only contributions from the third generation. The red curve shows the contributions that stem from the inclusion of the first and second generation of squarks as well as generation mixing. These contributions vanish in the gaugeless limit.}
    \label{fig:MH125_Mh1_ctrb}
\end{figure}

A remarkable feature of the iterated results is the large difference between the two-loop predictions which include either all (s)quarks (solid cyan) or only the third generation (dashed cyan). For the light mass $M_h$, for which the two-loop contributions are shown in \fig{fig:MH125_Mh1_ctrb}, the inclusion of the first and second generation as well as generation mixing lowers the Higgs mass prediction by around $2 \gev$ across the whole considered range of $X_t$. The effect of these contributions exceeds the experimental uncertainty by one order of magnitude and is also responsible for the bulk of the two-loop corrections for $X_t > -3 \tev$. We can also see that they are two to three times larger than the contributions which stem from the third generation alone (dotted red curve vs dashed cyan curve in \fig{fig:MH125_Mh1_ctrb}). The numerical impact of the inclusion of the first and second generation as well as generation mixing is even larger for the heavier mass $M_H$, exceeding $10\gev$, see \fig{fig:MH125_Mh2_overview}.

In this scenario, we have showcased that a strict fixed-order method can lead to unreliable predictions for pole masses if the states that mix with each other have sufficiently similar tree-level masses. A fixed-point iteration, which determines the exact location of the propagator pole, remedies the issues of the fixed-order approach at the cost of mixing different loop orders and contributions. In our case, the inclusion of gauge and gauge-Yukawa-mixing contributions, which we calculated for the first time in this work, leads to a large shift of the Higgs boson masses in such a scenario (which is related to the suppression of the top-Yukawa contributions by $\tan\beta$). We stress again, however, that a scenario with two light $\CP$-even Higgs bosons is phenomenologically not viable while similar mixing scenarios can appear between the heavy $\CP$-even $H$ and $\CP$-odd $A$ bosons if $\CP$-violating phases are non-zero.

\section{Conclusions and outlook}
\label{cha:conclusions}
The MSSM is a well-motivated model for physics beyond the SM. The probably most striking feature of the MSSM (which holds also for non-minimal supersymmetric models) is that it relates the masses of the various Higgs bosons to other parameters. Therefore, predictions for Higgs boson masses can be obtained within the model. Different approaches have been pursued in order to obtain such predictions with sufficient accuracy, one of which is a perturbative pole mass calculation in terms of self-energy Feynman diagrams. So far, all one-loop contributions, a variety of two-loop terms, and leading three-loop contributions have been calculated for the MSSM Higgs boson masses. In this paper, we focused on so-far undetermined two-loop terms of $\order{(\alem + \alq)^2 \nc^2}$, which are expected to constitute the dominant part of those two-loop electroweak corrections that had not been known up to now. From this class of contributions, only the pure Yukawa subpart of $\order{\alq^2 \nc^2}$ was known so far, albeit restricted to the limit of vanishing external momentum.

The inclusion of pure gauge and gauge-Yukawa-mixing contributions required us to generalise a relation between two-loop mass counterterms of Higgs, (would-be) Goldstone, and gauge bosons in order to obtain a finite result for the Higgs boson self-energies. Specifically, in the case of complex parameters giving rise to $\CP$ violation, the generalised relation for the counterterms derived in this work is needed for the neutral Higgs boson self-energies, while already in the $\CP$-conserving case this relation is required in order to obtain a finite result for the masses of the charged Higgs bosons. To the best of our knowledge, this new relation had not been known in the literature up to now; the additional terms do not contribute at the order of perturbation theory analysed in the existing literature and, therefore, were not taken into account. Additionally, we generalised the modified two-loop relation to an expression that holds in all orders of perturbation theory (\eqn{eqn:mHpm_mA0_rel}).

In our calculation, we employed a mixed $\OS$--$\DRbar$ renormalisation scheme at the two-loop level. We investigated in this context how the choice of renormalisation scheme is related to the possible appearance of so-called evanescent terms, arising in particular from the $\order{\del}$ parts of one-loop counterterms, in the final prediction for the Higgs boson masses. We have explicitly demonstrated that, in a fixed-order prediction with on-shell self-energies, both a full $\OS$ renormalisation and a full $\DRbar$ renormalisation lead to a total cancellation of evanescent $\order{\del}$ parts of the loop integrals. In case a mixed renormalisation scheme is used, in general all two-loop parameter counterterms need to be defined in a momentum-subtraction scheme for this cancellation to take place. We stress that the dependence of the prediction for a physical mass on $\order{\del}$ contributions of counterterms is a spurious one in the sense that these terms always drop out in relations between physical observables. This can be viewed as a theoretical limitation of such predictions, in particular since the $\order{\del}$ terms may give rise to numerical instabilities or gauge-dependent contributions. Furthermore, a scheme involving uncancelled $\order{\del}$ contributions of counterterms cannot be translated into a different scheme via the usual kind of reparameterisation. Our results clearly demonstrate that particular care is necessary from two-loop order onwards in the application of mixed renormalisation schemes, which are widely used in the literature (both for calculations in the SM and beyond).

As we opted for a mixed renormalisation scheme, we required an $\OS$ definition at the one- and two-loop level for the parameter that is given by the ratio of the vacuum expectation values of the two Higgs doublets, $\tb$.\footnote{In the literature, this parameter is more commonly defined as a $\DRbar$ quantity.} We have performed the $\OS$ renormalisation via the decay $A \to \tau^- \tau^+$ by requiring that the absolute square of the associated physical amplitude should not receive any higher order corrections. If this decay were to be observed, a numerical value for $\tb^\OS$ could be extracted from its measurement. We explicitly checked that our definition of the one-loop (\eqn{eqn:deltaOL_tanbeta}) and the two-loop (\eqn{eqn:deltaTL_tanbeta}) counterterm does not depend on the choice of the field renormalisations. As expected, this definition of $\tb$ leads to a total cancellation of the evanescent $\order{\del}$ terms of loop integrals in the Higgs boson pole mass prediction at the two-loop order.

In our numerical analysis we have investigated the effects of our newly calculated contributions on the predictions for the neutral MSSM Higgs boson masses in four different $\CP$-conserving scenarios. We have compared their size with the already known pure Yukawa contributions of $\order{\nc^2}$ and also with the experimental uncertainty of the mass measurement of the detected Higgs boson. While expectedly smaller than the pure Yukawa contributions (by typically an order of magnitude), the pure gauge and gauge-Yukawa-mixing terms turned out to be larger than or at least of a similar size as the experimental uncertainty. Their inclusion in MSSM Higgs mass predictions will therefore be important for improving the theoretical uncertainty. In the fourth scenario, we have also showcased the fact that a strict fixed-order treatment becomes insufficient for cases where there is large mixing between particles that are nearly mass-degenerate at lowest order. We showed that only the fixed-point iteration leads to a reliable prediction in this case, but at the cost of mixing different orders of perturbation theory. Depending on the investigated scenario, the appropriate method has to be chosen and the drawbacks of each approach have to be taken into account.

In the future, the newly calculated corrections are planned to be combined with the other known higher-order corrections and a resummation of large logarithms. This combined result is planned to be implemented into the code \texttt{FeynHiggs} \cite{Heinemeyer:1998yj,Heinemeyer:1998np,Hahn:2009zz,Degrassi:2002fi,Frank:2006yh,Hahn:2013ria,Bahl:2016brp,Bahl:2017aev,Bahl:2018qog}.\footnote{Analytic expressions are available on request.} Based on this implementation and the existing routines in \texttt{FeynHiggs} \cite{Bahl:2019hmm}, the remaining theoretical uncertainties will be estimated. In addition, the derived fixed-order result can be used to derive so far unknown electroweak two-loop threshold corrections for the EFT approach, which will further reduce the theoretical uncertainties.

We, moreover, emphasise that the renormalisation of the MSSM Higgs--gauge sector that we have carried out in detail at the two-loop level includes full electroweak effects and goes significantly beyond the renormalisation of electroweak two-loop contributions in the MSSM performed elsewhere.  The obtained results should hence be useful for any future prediction heading in a similar direction. In particular, the presented results for the contributions at $\order{\nc^2}$ can readily be extended to various Higgs production and decay processes at the two-loop level, as it was demonstrated in this work for the decay process $A \to \tau^- \tau^+$. We furthermore expect that the ingredients of the analyses in this work can directly be transferred to a more general $n$-loop order calculation of electroweak $\order{\nc^n}$ terms as well; these contributions will similarly decompose into products of one-loop integrals for the Higgs and vector boson self-energies. It remains to be seen whether the renormalisation of the quark-squark sector, which for the two-loop predictions of the Higgs boson masses carried out in this work was needed only at the one-loop order, can as easily be extended to the two-loop case. 

Since the renormalisation described in this work was performed allowing for $\CP$-violating phases of MSSM parameters, the study of $\CP$-violating scenarios requires only little additional work. The on-shell renormalisation of $\tb$ in terms of a charged Higgs boson decay like $H^+ \to \tau^+ \nu_\tau$ (instead of $A \to \tau^- \tau^+$) is expected to be straightforward. The relevant Slavnov-Taylor identities involving charged particles are included in \appx{app:ST_id} alongside the ones for the neutral particles.

Electroweak two-loop contributions of $\order{\nc^2}$ also appear in extensions of the MSSM, like e.g.\ the \textit{Next-to-Minimal Supersymmetric Standard Model} (NMSSM). Our results can therefore serve as a building block for similar calculations in extended SUSY models as well.

Besides the particular relevance for supersymmetric models, our work has more generally led to new insights about the renormalisation in spontaneously broken gauge theories at two-loop order and beyond. In particular, the features of unstable particles and particle mixing are present in a wide variety of physical models and their proper treatment is paramount in order to provide accurate theoretical predictions.

\section*{Acknowledgements}
\sloppy{We thank Johannes Braathen for useful discussions. H.~B.\ acknowledges support from the Alexander von Humboldt foundation. D.~M.\ and G.~W.\ acknowledge support by the Deutsche Forschungsgemeinschaft (DFG, German Research Foundation) under Germany's Excellence Strategy --- EXC 2121 ``Quantum Universe'' --- 390833306. 
This work has been partially funded by the Deutsche Forschungsgemeinschaft 
(DFG, German Research Foundation) --- 491245950. 
}

\appendix

\section{The \texorpdfstring{$\DRbar$}{DRbar} scheme at the two-loop level}
\label{app:reg_ren}
It has been noted that---starting at the two-loop level---the definition of modified minimal subtraction is no longer unique \cite{Sperling:2013fwy,Collins:2011zzd}. This can be seen as follows. For dimensional reasons, all one-loop integrals are proportional to $\Mudim^{2\del}$, owing to the way the regularisation scale is introduced. The factor of $(4\pi)^\del$ always appears in the same fashion as well, as it stems from a combination of the prefactor $C$ (see \eqn{eqn:loop_prefactor}) of the integrals and the angular integral in $D = 4 - 2\del$ dimensions. However, different combinations of Gamma functions occur depending on the considered integral. Therefore, different definitions of the modified regularisation scale $\MudimBar$ and, hence, the $\DRbar$ scheme exist. The following conventions, among others, are found in the literature:
\begin{subequations}
\allowdisplaybreaks
\begin{align}
    \MudimBar^{2\del} ={}& \left( 4 \pi e^{- \EulerGamma} \right)^\del \Mudim^{2\del} \text{\cite{Martin:2003qz}}, \\
    \MudimBar^{2\del} ={}& \frac{(4 \pi)^\del}{\Gamma(1 - \del)} \Mudim^{2\del} \text{\cite{Collins:2011zzd,Sperling:2013fwy}}, \\
    \MudimBar^{2\del} ={}& \frac{(4 \pi)^\del \Gamma^2(1 - \del) \Gamma(1 + \del)}{\Gamma(1 - 2\del)} \Mudim^{2\del} \text{\cite{Hahn:LT215}}.
\end{align}
\end{subequations}
All these conventions agree at $\order{\del}$. The second and third conventions agree at $\order{\del^2}$, but differ from the first one at that order. At $\order{\del^3}$, all conventions differ. While all conventions are able to get rid of any $\log(4\pi)$ or $\EulerGamma$ terms, other irrational constants, which appear at higher orders, cannot be removed simultaneously by any choice. It can be shown, however, that differences of $\order{\del^2}$ in the definition of $\MudimBar$ do not alter the value of a renormalised Green function after taking the limit $\del \to 0$ \cite{Collins:2011zzd}. Therefore, the exact choice of how to define $\MudimBar$ matters only for technical reasons.

We work with the first of the aforementioned conventions,
\begin{equation}
    \MudimBar^2 = 4 \pi e^{-\EulerGamma}  \Mudim^2.
\end{equation}

This replacement ``hides'' all appearances of $\log(4 \pi)$ and $\EulerGamma$ in the newly defined regularisation scale $\MudimBar$. 

In the $\DR$ renormalisation scheme, the counterterms contain only divergent parts without any irrational constants. At the one- and two-loop level, they can be cast into the form
\begin{subequations}
\begin{align}
    \deltaOL p^\DR ={}& \frac{A}{\del}, \\
    \deltaTL p^\DR ={}& \frac{B}{\del^2} + \frac{C}{\del},
\end{align}
\end{subequations}
respectively, with the coefficients $A$, $B$ and $C$. With our convention for $\MudimBar$, we write the one- and two-loop $\DRbar$ counterterms as
\begin{subequations}
\allowdisplaybreaks
\begin{align}
\begin{split}
    \deltaOL p^{\DRbar} \equiv{}& \left(  4 \pi e^{-\EulerGamma}\right)^\del \frac{\overline A}{\del} \\
    ={}& A \left( \frac{1}{\del} + \log(4 \pi) - \EulerGamma + \frac{\del}{2}[\log(4 \pi) - \EulerGamma]^2 + \order{\del^2} \right),
\end{split} \\
\begin{split}
    \deltaTL p^{\DRbar} \equiv{}& \left(  4 \pi e^{-\EulerGamma}\right)^{2 \del} \left( \frac{\overline B}{\del^2} + \frac{\overline C}{\del} \right) \\
    ={}& B \left( \frac{1}{\del^2} - 2[\log(4 \pi) - \EulerGamma]^2 + \order{\del} \right) \\
    &+ C \left( \frac{1}{\del} + 2 [\log(4 \pi) - \EulerGamma] + \order{\del} \right).
\end{split}
\end{align}
\end{subequations}
The coefficients $\overline A$, $\overline B$, and $\overline C$ were determined by comparing the divergent parts of $\DR$ and $\DRbar$ counterterms, which have to agree in order to obtain finite results in either renormalisation scheme. It should be noted that, in this definition of the $\DRbar$ scheme, the one-loop counterterm contains terms of $\order{\del}$. These terms are important for a cancellation of the irrational constants at the two-loop level.

The $\DRbar$ coefficients $\overline A$, $\overline B$, and $\overline C$ do not contain the irrational constants $\log(4 \pi)$ and $\EulerGamma$. This prescription agrees with the one found in \citere{Collins:2011zzd} and was derived independently. In \citere{Collins:2011zzd}, the idea of adding one factor $S_\del = (  4 \pi e^{-\EulerGamma})^\del$ for each loop in the counterterms is presented as well.\footnote{In \citere{Collins:2011zzd}, the different but equivalent convention $S_\del = (4\pi)^\del/\Gamma(1 - \del)$ is used.}

\section{The quark and squark sector}
\label{app:squark_ren}
In this section, we fix the notation for the quark and squark sector in the MSSM. We give the renormalisation transformations and the resulting expressions for the renormalised squark self-energy diagrams up to the one-loop order. We present three different renormalisation schemes for the case of massive quarks and we illustrate how the renormalisation has to be modified in the massless case.

Throughout the whole paper, we assume that mass terms do not mix quarks and squarks of different generations. This implies a unit CKM matrix and squark mass matrices which are diagonal in flavour space. The generalisation of our results to the case of non-zero mixing between the generations would easily be possible but is expected to not yield any new insights. Quartic interaction terms between squark flavours of different generations nevertheless lead to Higgs self-energy diagrams with generation mixing.

The quark sector requires no renormalisation at our considered order of perturbation theory, so we focus solely on the squark sector of the MSSM.

\subsection{Tree-level}

The bilinear squark Lagrangian reads
\begin{equation}
    \mathcal{L}_\text{squark}^\text{bil.} = \sum_{\tilde q} \left( \partial_\mu \tilde q_L^* \partial^\mu \tilde q_L + \partial_\mu \tilde q_R^* \partial^\mu \tilde q_R -
    \begin{pmatrix}
    \tilde q_L^* & \tilde q_R^*
    \end{pmatrix}
    \textbf{M}_{\tilde q}^2
    \begin{pmatrix}
    \tilde q_L \\ \tilde q_R
    \end{pmatrix} \right),
\end{equation}
where the sum runs over the squark flavors $\tilde t, \tilde b, \tilde c, \tilde s, \tilde u, \tilde d$. As we do not consider generation mixing, the squark mass matrices are flavour diagonal. We denote the elements of a squark mass matrix by
\begin{equation}
\label{eqn:squarkmassmatrix}
    \textbf{M}_{\tilde q}^2 =
    \begin{pmatrix}
    \left(\textbf{M}_{\tilde q}^2\right)_{LL} & \left(\textbf{M}_{\tilde q}^2\right)_{LR} \\
    \left(\textbf{M}_{\tilde q}^2\right)_{RL} & \left(\textbf{M}_{\tilde q}^2\right)_{RR}
    \end{pmatrix}.
\end{equation}
The squark fields carry non-vanishing quantum numbers and are thus complex scalar fields; the mass matrices are, in the most general case, not symmetric but hermitian so that their eigenvalues---the physical squark masses---are real. The mass matrices for up- and down-type squarks read
\begin{subequations}
\label{eqs:squark_mass_matrices_params}
\allowdisplaybreaks
\begin{align}
	\textbf{M}_{\tilde u_g}^2 ={}&
    \begin{pmatrix}
    M_{\tilde q_g}^2 + m_{u_g}^2 + M_Z^2 \cos(2\beta) (\frac{1}{2} - \frac{2}{3} \sw^2) & m_{u_g} X_{u_g}^* \\
    m_{u_g} X_{u_g} & M_{\tilde u_g}^2 + m_{u_g}^2 + \frac{2}{3} M_Z^2 \cos(2\beta) \sw^2
    \end{pmatrix}, \\
	\textbf{M}_{\tilde d_g}^2 ={}&
    \begin{pmatrix}
    M_{\tilde q_g}^2 + m_{d_g}^2 + M_Z^2 \cos(2\beta) (-\frac{1}{2} + \frac{1}{3} \sw^2) & m_{d_g} X_{d_g}^* \\
    m_{d_g} X_{d_g} & M_{\tilde d_g}^2 + m_{d_g}^2 - \frac{1}{3} M_Z^2 \cos(2\beta) \sw^2
    \end{pmatrix}.
\end{align}
\end{subequations}
The index $g$ labels the three generations of matter such that $m_{u_3} = m_t$ and $X_{d_2} = X_s$, for instance. We do not use this convention for the soft SUSY-breaking masses $M_{\tilde q_g}^2$, $M_{\tilde u_g}^2$, and $M_{\tilde d_g}^2$, i.e.\ there is no parameter $M_{\tilde t}^2$, so that we can distinguish between the respective left- ($M_{\tilde q_3}^2$) and right-handed ($M_{\tilde u_3}^2$) mass terms. Moreover, we introduced the common abbreviations
\begin{subequations}
\begin{align}
    X_{u_g} ={}& A_{u_g} - \muC \cot(\beta), \\ X_{d_g} ={}& A_{d_g} - \muC \tan(\beta).
\end{align}
\end{subequations}
The parameters $M_{\tilde q_g}^2, M_{\tilde u_g}^2, M_{\tilde d_g}^2, A_{u_g}, A_{d_g}$ break supersymmetry softly. In the most general scenario of SUSY breaking, they would be $3 \times 3$ matrices in generation space. In our calculation, as mentioned above, we neglect this mixing between generations and assume these matrices to be diagonal. 

To change from the gauge eigenbasis to the mass eigenbasis, we introduce the unitary transformation
\begin{equation}
\label{eqn:squark_rotation_matrix}
	\begin{pmatrix}
    \tilde q_1 \\ \tilde q_2
    \end{pmatrix} =
    \begin{pmatrix}
    c_{\tilde q} & - s_{\tilde q} e^{-\text{i} \phi_{\tilde q}} \\
    s_{\tilde q} e^{\text{i} \phi_{\tilde q}} & c_{\tilde q}
    \end{pmatrix} 
	\begin{pmatrix}
    \tilde q_L \\ \tilde q_R
    \end{pmatrix} \equiv U_{\tilde q}
	\begin{pmatrix}
    \tilde q_L \\ \tilde q_R
    \end{pmatrix}
\end{equation}
for each squark flavor $\tilde q$. Here we introduced the abbreviations $c_{\tilde q} = \cos(\theta_{\tilde q})$ and $s_{\tilde q} = \sin(\theta_{\tilde q})$. The bilinear squark Lagrangian in terms of the mass eigenbasis reads
\begin{equation}
    \mathcal{L}_\text{squark}^\text{bil.} = \sum_{\tilde q} \left( \partial_\mu \tilde q_1^* \partial^\mu \tilde q_1 + \partial_\mu \tilde q_2^* \partial^\mu \tilde q_2 -
    \begin{pmatrix}
    \tilde q_1^* & \tilde q_2^*
    \end{pmatrix}
    \textbf{D}_{\tilde q}^2
    \begin{pmatrix}
    \tilde q_1 \\ \tilde q_2
    \end{pmatrix} \right),
\end{equation}
where
\begin{equation}
	\textbf{D}_{\tilde q}^2 = U_{\tilde q} \textbf{M}_{\tilde q}^2 U_{\tilde q}^\dagger \equiv
    \begin{pmatrix}
    m_{\tilde q_1}^2 & m_{\tilde q_{12}}^2 \\
    m_{\tilde q_{21}}^2 & m_{\tilde q_2}^2
    \end{pmatrix},
    \quad m_{\tilde q_{21}}^2 = m_{\tilde q_{12}}^{2*}.
\end{equation}
The angles $\theta_{\tilde q} \in [0,\tfrac{\pi}{2}]$ and $\phi_{\tilde q} \in (-\pi,\pi]$ are then determined by the conditions
\begin{equation}
    m_{\tilde q_{12}}^2 = 0 \quad \wedge \quad m_{\tilde q_1}^2 \leq m_{\tilde q_2}^2.
\end{equation}
To give explicit expressions for $\theta_{\tilde q}$ and $\phi_{\tilde q}$, we have to distinguish between the degenerate and the non-degenerate case. The two squark mass eigenvalues are degenerate if and only if the matrix $\textbf{M}_{\tilde q}^2$ is proportional to the identity matrix, in which case no rotation is needed, and $U_{\tilde q}$ can simply be chosen as unity. This happens if both $m_q \Xq = 0$ and $\left(\textbf{M}_{\tilde q}^2\right)_{LL} = \left(\textbf{M}_{\tilde q}^2\right)_{RR}$ are fulfilled simultaneously. When working in the gaugeless limit, assuming a vanishing quark mass and a universal SUSY scale $M_\text{SUSY}^2 = M_{\tilde q_g}^2 = M_{\tilde u_g}^2 = M_{\tilde d_g}^2$, this is always the case.

In the case of non-degenerate masses, the mass ordering ensures $m_{\tilde q_1}^2 < m_{\tilde q_2}^2$, and we can write
\begin{subequations}
\label{eqs:quark_mix_angles}
\begin{align}
	\exp(\text{i}\phi_{\tilde q}) ={}& \frac{\Xq}{|\Xq|}, \\
	\cos(2\theta_{\tilde q}) ={}&\frac{\left(\textbf{M}_{\tilde q}^2\right)_{RR} - \left(\textbf{M}_{\tilde q}^2\right)_{LL}}{m_{\tilde q_2}^2-m_{\tilde q_1}^2}, \\
	\sin(2\theta_{\tilde q}) ={}& \frac{2m_q |\Xq|}{m_{\tilde q_2}^2-m_{\tilde q_1}^2}.
\end{align}
\end{subequations}
The angles can then uniquely be determined from
\begin{subequations}
\begin{alignat}{2}
	\phi_{\tilde q} ={}& \text{Arg}(\Xq), \quad &-\pi <{}& \phi_{\tilde q} \leq \pi, \\
	\theta_{\tilde q} ={}& \frac{1}{2} \arccos \frac{\left(\textbf{M}_{\tilde q}^2\right)_{RR} - \left(\textbf{M}_{\tilde q}^2\right)_{LL}}{m_{\tilde q_2}^2-m_{\tilde q_1}^2}, \quad &0 \leq{}& \theta_{\tilde q} \leq \tfrac{\pi}{2}.
\end{alignat}
\end{subequations}
Mathematically, $\phi_{\tilde q}$ is undefined if $\Xq$ vanishes, and we set it to 0 for simplicity in this case.

\subsection{Renormalisation transformations}
For a full two-loop prediction of the Higgs boson masses, the one-loop renormalisation of the squark sector is needed. As we are only interested in electroweak corrections of $\order{(\alem + \alq)^2 \nc^2}$, no lepton/slepton/quark renormalisation constants are needed.

We renormalise the squark mass matrices and fields via
\begin{subequations}
\allowdisplaybreaks
\begin{align}
    \textbf{D}_{\tilde q}^2 \to{}& \textbf{D}_{\tilde q}^2 +
    \begin{pmatrix}
    \delta m_{\tilde q_1}^2 & \delta m_{\tilde q_{12}}^2 \\
    \delta m_{\tilde q_{21}}^2 & \delta m_{\tilde q_2}^2
    \end{pmatrix}, \\
    \begin{pmatrix}
    \tilde q_1 \\ \tilde q_2
    \end{pmatrix} \to{}
    &\begin{pmatrix}
    \sqrt{1 + \delta Z_{\tilde q_{11}}} & \frac{1}{2} \delta Z_{\tilde q_{12}} \\
    \frac{1}{2} \delta Z_{\tilde q_{21}} &  \sqrt{1 + \delta Z_{\tilde q_{22}}}
    \end{pmatrix}
    \begin{pmatrix}
    \tilde q_1 \\ \tilde q_2
    \end{pmatrix}, \\
    \begin{pmatrix}
    \tilde q_1^* \\ \tilde q_2^*
    \end{pmatrix} \to{}&
    \begin{pmatrix}
    \sqrt{1 + \delta Z_{\tilde q_{11}}} & \frac{1}{2} \delta \bar Z_{\tilde q_{21}} \\
    \frac{1}{2} \delta \bar Z_{\tilde q_{12}} &  \sqrt{1 + \delta Z_{\tilde q_{22}}}
    \end{pmatrix}
    \begin{pmatrix}
    \tilde q_1^* \\ \tilde q_2^*
    \end{pmatrix}.
\end{align} 
\end{subequations}

It should be noted that we introduce separate off-diagonal field counterterms for the squark and anti-squark fields. This follows the convention of \citeres{Fritzsche:2011nr, Fritzsche:2013fta}, where it enables the inclusion of absorptive contributions into the field counterterms. If the absorptive parts are left out, the off-diagonal field counterterms are related by $\delta \bar Z_{\tilde q_{ij}} = \delta Z_{\tilde q_{ji}}^*$.

\subsection{Renormalisation at the one-loop level}
\label{ssec:squark_1L_ren}

\begin{figure}[t]
	\centering
	\includegraphics[width = .3\textwidth]{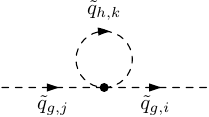}
	\caption{Unrenormalised one-loop squark self-energy $\Sigma^{(1)}_{\tilde q_{g,i} \tilde q_{g,j}}$. 
    The indices $g$ and $h$ are flavour labels, which we make explicit here since the squark in the loop, $\tilde q_h$, can have a different flavour than the external squark. The indices $i$, $j$, and $k$ label the two mass eigenstates. We note the convention for naming the self-energy, where the first label corresponds to the outgoing squark in the diagram.}
	\label{fig:squark_1L}
\end{figure}

In this section, we give an overview over the most important renormalisation schemes for the squark sector. We present expressions for the relevant one-loop renormalisation constants; they enter the prediction for the two-loop Higgs boson masses through the sub-loop part of two-loop Higgs self-energies.

With the renormalisation transformations given in the previous section, the renormalised one-loop squark self-energies read
\begin{subequations}
\begin{align}
    \hat \Sigma^{(1)}_{\tilde q_1 \tilde q_1}(p^2) ={}& \Sigma^{(1)}_{\tilde q_1 \tilde q_1}(p^2) + \deltaOL Z_{\tilde q_{11}} (p^2 - m_{\tilde q_1}^2) - \deltaOL m_{\tilde q_1}^2, \\
\begin{split}
    \hat \Sigma^{(1)}_{\tilde q_1 \tilde q_2}(p^2) ={}& \Sigma^{(1)}_{\tilde q_1 \tilde q_2}(p^2) + \frac{1}{2} \deltaOL Z_{\tilde q_{12}} (p^2 - m_{\tilde q_1}^2) \\
    &+ \frac{1}{2} \deltaOL \bar Z_{\tilde q_{12}} (p^2 - m_{\tilde q_2}^2) - \deltaOL m_{\tilde q_{12}}^2,
\end{split} \\
\begin{split}
    \hat \Sigma^{(1)}_{\tilde q_2 \tilde q_1}(p^2) ={}& \Sigma^{(1)}_{\tilde q_2 \tilde q_1}(p^2) + \frac{1}{2} \deltaOL Z_{\tilde q_{21}} (p^2 - m_{\tilde q_2}^2) \\
    &+ \frac{1}{2} \deltaOL \bar Z_{\tilde q_{21}} (p^2 - m_{\tilde q_1}^2) - \deltaOL m_{\tilde q_{21}}^2,
\end{split} \\
    \hat \Sigma^{(1)}_{\tilde q_2 \tilde q_2}(p^2) ={}& \Sigma^{(1)}_{\tilde q_2 \tilde q_2}(p^2) + \deltaOL Z_{\tilde q_{22}} (p^2 - m_{\tilde q_2}^2) - \deltaOL m_{\tilde q_2}^2.
\end{align} 
\end{subequations}

The only topology contributing to the $\Sigma^{(1)}_{\tilde q_i \tilde q_j}$ self-energy at $\order{\nc}$ is shown in \fig{fig:squark_1L}. These diagrams are independent of the external momentum, and no field renormalisation constants are needed to yield finite renormalised one-loop self-energies. Since in our calculation squarks appear as internal particles only, any finite part of their field renormalisation constants cancels in the sub-loop renormalisation of the two-loop Higgs (and vector boson) self-energies, and we will set them to zero for simplicity:
\begin{subequations}
\begin{align}
    \deltaOL Z_{\tilde q_{ij}} ={}& 0, \\
    \deltaOL \bar Z_{\tilde q_{ij}} ={}& 0.
\end{align}
\end{subequations}
While this choice is not necessary, it simplifies the algebraic expressions and, if an on-shell renormalisation scheme is chosen for the squark masses, it implies that the squark self-energies vanish for arbitrary external momenta.

By virtue of the momentum independence, the self-energies are free from absorptive contributions and they can only be complex because of the involved couplings. Consequently, the off-diagonal unrenormalised squark self-energies are related via
\begin{equation}
    \Sigma^{(1)*}_{\tilde q_1 \tilde q_2} \overset{\order{\nc}}{=} \Sigma^{(1)}_{\tilde q_2 \tilde q_1},
\end{equation}
which holds for the renormalised self-energies as well. Due to the absence of absorptive contributions, the diagonal self-energies are real.

\subsubsection{Renormalisation conditions and counterterms for the massive case}

In this paper, unless explicitly stated otherwise, we work with a massive third generation of quarks, while the first two generations are treated as massless. For the third generation squark sector, we consider several different renormalisation schemes; an on-shell scheme ($\OS$), a $\DRbar$ scheme, and a mixed scheme. In each of these schemes, we allow for either sbottom mass $m_{\tilde b_i}^2$ to be used as input parameter, which amounts to a total of six different renormalisation schemes for the third generation squark sector.

Regarding the choice of renormalisation conditions, it is useful to count the number of independent parameters first. A set of independent parameters in the stop-sbottom sector is for example given by $\{M_{\tilde q_3}^2, M_{\tilde u_3}^2, M_{\tilde d_3}^2, \At, \Ab\}$, of which the trilinear couplings can be complex. This requires us to impose seven (real) renormalisation conditions. Independent of the chosen renormalisation scheme for the squark sector, we require that
\begin{equation}
    \text{$\mu$, $\Ab$ are renormalised in the $\DRbar$ scheme.}
\end{equation}
We derive the $\DRbar$ expressions for $\mu$ and $A_b$ in \sct{ssec:DRbarMUEAq}.

In all schemes, the stop masses and one of the sbottom masses are used as independent input parameters. We label the independent sbottom with $n$ and the dependent sbottom with $f = 3 - n$. In the case of the $\order{(\alem + \alq)^2 \nc^2}$ contributions that are of interest to us, the quark sector is not renormalised. For the sake of completeness, we include the vanishing quark mass counterterms in the expressions below.

\paragraph{(i) $\OS$ scheme.}
We use on-shell definitions for the stop parameters and the $n$th sbottom mass:
\begin{subequations}
\begin{alignat}{2}
    \deltaOL m_{\tilde t_i}^2 ={}& \Re \Sigma^{(1)}_{\tilde t_i \tilde t_i} &&\overset{\order{\nc}}{=} \Sigma^{(1)}_{\tilde t_i \tilde t_i}, \quad i \in \{1,2\}, \\
    \deltaOL m_{\tilde b_n}^2 ={}& \Re \Sigma^{(1)}_{\tilde b_n \tilde b_n} &&\overset{\order{\nc}}{=} \Sigma^{(1)}_{\tilde b_n \tilde b_n}, \\
    \deltaOL m_{\tilde t_{12}}^2 ={}& \, \TRe\ \Sigma^{(1)}_{\tilde t_1 \tilde t_2} &&\overset{\order{\nc}}{=} \Sigma^{(1)}_{\tilde t_1 \tilde t_2}, \\
    \deltaOL m_{\tilde t_{21}}^2 ={}& \deltaOL m_{\tilde t_{12}}^{2*} &&\overset{\order{\nc}}{=} \Sigma^{(1)}_{\tilde t_2 \tilde t_1}.
\end{alignat}
\end{subequations}
The operator $\TRe$ takes the real part of the loop integrals but leaves complex couplings unaffected. In our calculation, the squark self-energies are momentum independent and so we do not specify any momentum at which the self-energies are to be evaluated. It should be noted that the renormalisation condition for the off-diagonal stop mass terms implies that the whole renormalised self-energy vanishes only in our calculation of $\nc$ contributions (because we chose $\delta Z_{\tilde q_{ij}} = \delta \bar Z_{\tilde q_{ij}} = 0$); in a setting where the squark self-energies are momentum-dependent, an unphysical $\MOM$ scheme is often used instead, see \citeres{Fritzsche:2013fta,Heinemeyer:2010mm,Heinemeyer:2007aq,Rzehak:2005zz,Bahl:2022kzs}.

In this scheme, the $\At$ counterterm is a dependent quantity:
\begin{equation}
\begin{split}
    \deltaOL \At ={}& \frac{1}{m_t}\left[ U_{\tilde t_{11}} U_{\tilde t_{12}}^* \left( \deltaOL m_{\tilde t_1}^2 - \deltaOL m_{\tilde t_2}^2 \right) + U_{\tilde t_{21}} U_{\tilde t_{12}}^* \deltaOL m_{\tilde t_{12}}^2 + U_{\tilde t_{11}} U_{\tilde t_{22}}^* \deltaOL m_{\tilde t_{21}}^2 \right] \\ 
    &- \frac{\Xt \, \deltaOL m_t}{m_t} +  \frac{\deltaOL \muC}{\tb} - \frac{\muC \deltaOL \tb}{\tb^2}.
\end{split}
\end{equation}

\paragraph{(ii) $\DRbar$ scheme.}
In this scheme, we use $\At$ to formulate a renormalisation condition instead of $m_{\tilde t_{12}}^2$. Now, all the input counterterms are defined in the $\DRbar$ scheme:
\begin{equation}
    \text{$m_{\tilde t_1}^2$, $m_{\tilde t_2}^2$, $m_{\tilde b_n}^2$, and $\At$ are renormalised in the $\DRbar$ scheme.}
\end{equation}
The $\DRbar$ counterterms for the masses are obtained by simply discarding the finite parts of the $\OS$ counterterms. The $\DRbar$ expression for $\deltaOL A_t$ is derived in \sct{ssec:DRbarMUEAq}. Finally, the $m_{\tilde t_{12}}^2$ counterterm is a dependent quantity now:
\begin{equation}
\begin{split}
    \deltaOL m_{\tilde t_{12}}^2 ={}& \frac{1}{|U_{\tilde t_{11}}|^2 - |U_{\tilde t_{12}}|^2} \Big[ U_{\tilde t_{11}} U_{\tilde t_{21}}^* \left( \deltaOL m_{\tilde t_1}^2 - \deltaOL m_{\tilde t_2}^2 \right) \\
    &+ U_{\tilde t_{11}} U_{\tilde t_{22}}^* \left( m_t \, \deltaOL \XtC + \XtC \, \deltaOL m_t \right) - U_{\tilde t_{12}} U_{\tilde t_{21}}^* \left( m_t \, \deltaOL \Xt + \Xt \, \deltaOL m_t \right) \Big].
\end{split}
\end{equation}

\paragraph{(iii) Mixed scheme.}
The input counterterms are the same as in the $\DRbar$ scheme, but now
\begin{equation}
    \text{$m_{\tilde t_1}^2$, $m_{\tilde t_2}^2$, and $m_{\tilde b_n}^2$ are renormalised on-shell. $\At$ is renormalised $\DRbar$.}
\end{equation}
$\deltaOL m_{\tilde t_{12}}^2$ is calculated by the same expression as in the $\DRbar$ scheme. In this scheme, just as in the $\DRbar$ scheme, $\mu$, $\tb$, and both $\At$ and $\Ab$ are $\DRbar$ quantities. Consequently, $\Xt$ and $\Xb$ are $\DRbar$ quantities as well.

\paragraph{Scheme-independent relations.}
In all three schemes, the counterterms for the sbottom masses $m_{\tilde b_f}^2$ and $m_{\tilde b_{12}}^2$ are dependent quantities and as such they have to be expressed in terms of the input counterterms. To find the expression for the remaining sbottom mass counterterm, we make use of a relation between the $LL$ entries (see \eqn{eqn:squarkmassmatrix}) of the stop and sbottom mass matrix. Both entries contain the SUSY-breaking parameter $M_{\tilde q_3}^2$, yielding
\begin{subequations}
\begin{align}
\begin{split}
    M_{\tilde q_3}^2 ={}&
    \left(\textbf{M}_{\tilde t}^2\right)_{LL} - m_t^2 - M_Z^2 \ctb \left( \frac{1}{2} - \frac{2}{3}\sw^2 \right) \\
    ={}& \left(\textbf{M}_{\tilde b}^2\right)_{LL} - m_b^2 + M_Z^2 \ctb \left( \frac{1}{2}-\frac{1}{3}\sw^2 \right)
\end{split} \\
    \Leftrightarrow \left(\textbf{M}_{\tilde b}^2\right)_{LL} ={}& \left(\textbf{M}_{\tilde t}^2\right)_{LL} - m_t^2 + m_b^2 - M_W^2 \ctb.
\end{align}
\end{subequations}

In order to simplify the notation, we introduce the auxiliary renormalisation constant
\begin{equation}
\begin{split}
    \deltaOL \! \left(\textbf{M}_{\tilde b}^2\right)_{LL} ={}& |U_{\tilde t_{11}}|^2 \deltaOL m_{\tilde t_1}^2 + |U_{\tilde t_{12}}|^2 \deltaOL m_{\tilde t_2}^2 - 2 \, \Re \left\{ U_{\tilde t_{22}} U_{\tilde t_{12}}^* \deltaOL m_{\tilde t_{12}}^2 \right\} \\
    & - 2 m_t \deltaOL m_t + 2 m_b \deltaOL m_b - \ctb \deltaOL M_W^2 + 4 M_W^2 \sib \cb^3 \deltaOL \tb.
\label{eqn:LL_sbottom_mass_CT}
\end{split}
\end{equation}

This constant allows us to write the dependent sbottom mass counterterm as
\begin{equation}
\begin{split}
    \deltaOL m_{\tilde b_f}^2 ={}& \frac{1}{|U_{\tilde b_{1f}}|^2} \Big[  |U_{\tilde b_{1n}}|^2 \deltaOL m_{\tilde b_n}^2 + (n-f) \Big( 2 \, \Re\left\{ U_{\tilde b_{11}} U_{\tilde b_{12}}^* \left( m_b \, \deltaOL \XbC + \XbC \, \deltaOL m_b \right) \right\} \\
    &+ \left( |U_{\tilde b_{11}}|^2 - |U_{\tilde b_{12}}|^2 \right) \deltaOL \! \left(\textbf{M}_{\tilde b}^2\right)_{LL} \Big) \Big].
\label{eqn:dep_sbottom_mass_CT}
\end{split}
\end{equation}

Lastly, the counterterm for $m_{\tilde b_{12}}^2$ reads
\begin{equation}
\begin{split}
    \deltaOL m_{\tilde b_{12}}^2 ={}& \frac{1}{|U_{\tilde b_{11}}|^2 - |U_{\tilde b_{12}}|^2} \Big[ U_{\tilde b_{11}} U_{\tilde b_{21}}^* \left( \deltaOL m_{\tilde b_1}^2 - \deltaOL m_{\tilde b_2}^2 \right) \\
    &+ U_{\tilde b_{11}} U_{\tilde b_{22}}^* \left( m_b \, \deltaOL \XbC + \XbC \, \deltaOL m_b \right) - U_{\tilde b_{12}} U_{\tilde b_{21}}^* \left( m_b \, \deltaOL \Xb + \Xb \, \deltaOL m_b \right) \Big].
\end{split}
\end{equation}

\subsubsection{Renormalisation in the massless case}

To extract all terms of order $\order{(\alem + \alq)^2 \nc^2}$ in the Higgs boson mass prediction at the two-loop level, the first and second generation of quarks and squarks need to be taken into account as well. These generations contribute even if their quarks are assumed to be massless. As there are also two-loop Higgs self-energies with both a third generation squark in one loop and a first/second generation squark in the other, those contributions cannot simply be obtained by taking the results for the third generation and applying the massless limit. Instead, the whole calculation has to be done anew. To this end, we assume both the first and second generation of quarks to be massless and again a diagonal CKM matrix.

As for the third generation of squarks, seven independent real parameters appear in the squark mass matrices of each of the first two generations. In the massless limit, however, the trilinear counterterms $\deltaOL A_q$ and correspondingly the off-diagonal mass counterterms $\deltaOL m_{\tilde q_{12}}^2$ do not contribute in the sub-loop renormalisation of the Higgs self-energies. This leaves us with three independent parameters in each generation. If we assume generation $g$ to be massless, these are $M_{\tilde q_g}^2$, $M_{\tilde u_g}^2$, and $M_{\tilde d_g}^2$. As before, we fix these parameters by imposing renormalisation conditions on the diagonal squark self-energies. When working with the massive third generation, we used both the stop and one of the sbottom masses as independent input parameters. The remaining sbottom mass was then fixed by virtue of the $SU(2)$ symmetry of the SUSY breaking parameter $M_{\tilde q_3}^2$. For the massless first two generations, this procedure needs to be adapted; the squark mass matrices are diagonal in the massless limit and so the corresponding rotation matrices $U_{\tilde q}$ are either purely diagonal or purely off-diagonal. One of the mass eigenstates thus corresponds to the left-handed gauge eigenstate $\tilde q_L$, the other one to the right-handed gauge eigenstate $\tilde q_R$. The $SU(2)$ symmetry fixes the mass of the left-handed down-type squark in terms of the left-handed up-type squark, and its mass counterterm cannot be chosen independently. This issue becomes clear when looking at \eqn{eqn:dep_sbottom_mass_CT}; this expression for the dependent mass counterterm $\deltaOL m_{\tilde b_f}^2$ is undefined if $U_{\tilde b_{1f}} = 0$. This can be circumvented by removing the freedom of choice as for which mass is treated independently. We demonstrate the procedure for the second generation and an on-shell renormalisation of the input mass (the first generation and $\DRbar$ renormalisation are treated in analogous fashion). The scalar charm quark mass counterterms are
\begin{equation}
    \deltaOL m_{\tilde c_i}^2 = \Re \Sigma^{(1)}_{\tilde c_i \tilde c_i} = \Sigma^{(1)}_{\tilde c_i \tilde c_i}, \quad i \in \{1,2\}.
\end{equation}
We cannot freely choose which of the two scalar strange quark masses is used as input. Instead, we fix our choice by the form of $U_{\tilde s}$ in order to avoid divergent and thus meaningless expressions:

\paragraph{$U_{\tilde s}$ is diagonal.}
This means that $U_{\tilde s_{12}} = U_{\tilde s_{21}} = 0$ and the second generation analogue of \eqn{eqn:dep_sbottom_mass_CT} is only meaningful if $f = 1, n = 2$ is chosen. We arrive at
\begin{subequations}
\begin{align}
    \deltaOL m_{\tilde s_1}^2 ={}& \deltaOL \! \left(\textbf{M}_{\tilde s}^2\right)_{LL},  \\
    \deltaOL m_{\tilde s_2}^2 ={}& \Sigma^{(1)}_{\tilde s_2 \tilde s_2}.
\end{align}
\end{subequations}

\paragraph{$U_{\tilde s}$ is purely off-diagonal.}
This means that $U_{\tilde s_{11}} = U_{\tilde s_{22}} = 0$ and the second generation analogue of \eqn{eqn:dep_sbottom_mass_CT} is only meaningful if $f = 2, n = 1$ is chosen. We arrive at
\begin{subequations}
\begin{align}
    \deltaOL m_{\tilde s_1}^2 ={}& \Sigma^{(1)}_{\tilde s_1 \tilde s_1},  \\
    \deltaOL m_{\tilde s_2}^2 ={}& \deltaOL \! \left(\textbf{M}_{\tilde s}^2\right)_{LL}.
\end{align}
\end{subequations}

We can combine both cases in the formulae
\begin{subequations}
\begin{align}
    \deltaOL m_{\tilde s_1}^2 ={}& |U_{\tilde s_{12}}|^2 \Sigma^{(1)}_{\tilde s_1 \tilde s_1} + |U_{\tilde s_{11}}|^2 \deltaOL \! \left(\textbf{M}_{\tilde s}^2\right)_{LL},  \\
    \deltaOL m_{\tilde s_2}^2 ={}& |U_{\tilde s_{11}}|^2 \Sigma^{(1)}_{\tilde s_2 \tilde s_2} + |U_{\tilde s_{12}}|^2 \deltaOL \! \left(\textbf{M}_{\tilde s}^2\right)_{LL}.
\end{align}
\end{subequations}
$\deltaOL \! \left(\textbf{M}_{\tilde s}^2\right)_{LL}$ is obtained from \eqn{eqn:LL_sbottom_mass_CT} by replacing the third generation labels by second generation labels.

\subsection{\texorpdfstring{$\DRbar$}{DRbar} renormalisation of \texorpdfstring{$\mu$}{mu} and \texorpdfstring{$A_q$}{Aq}}
\label{ssec:DRbarMUEAq}

For a full renormalisation of the squark sector, the counterterms $\deltaOL \mu$, $\deltaOL \At$, and $\deltaOL \Ab$ need to be fixed. The higgsino mass parameter $\mu$ is typically defined via the chargino-neutralino sector (see e.g.\ \citere{Fritzsche:2013fta}). As that sector is otherwise irrelevant to our calculation, we choose a $\DRbar$ renormalisation for $\mu$. As can be seen in \citere{Fritzsche:2013fta}, the $\OS$ expression for the $\mu$ counterterm in a CCN scheme involves elements of both chargino rotation matrices, which transform the gauge eigenstates into mass eigenstates. Taking the divergent part of the $\OS$ counterterm yields an expression which is still rather complicated as the rotation matrix elements do not easily cancel out algebraically.

As the higgsino mass parameter enters the $h\tilde t \tilde t^*$ vertex, an expression for its counterterm can also be obtained from the renormalisation of this vertex. This approach naturally leads to an expression for $\deltaOL \At$ as well and avoids the need to deal with the chargino rotation matrices. Therefore, we calculate the amplitudes $h \to \tilde t_1 \tilde t_2^*$ and $H \to \tilde t_1 \tilde t_2^*$ at the one-loop level and determine $\deltaOL \mu^\DRbar$ and $\deltaOL \At^\DRbar$ from requiring both amplitudes to be finite.

As both amplitudes $\hat \Gamma^{(1)}_{h \tilde t_1 \tilde t_2^*}$ and $\hat \Gamma^{(1)}_{H \tilde t_1 \tilde t_2^*}$ involve both counterterms $\deltaOL \mu$ and $\deltaOL \At$, we have to form appropriate combinations of both amplitudes:
\begin{subequations}
\begin{align}
    &\ca \hat \Gamma^{(1)}_{H \tilde t_1 \tilde t_2^*} - \sa \hat \Gamma^{(1)}_{h \tilde t_1 \tilde t_2^*}: \quad \text{$\deltaOL \At$ drops out}, \\
    &\ca \hat \Gamma^{(1)}_{h \tilde t_1 \tilde t_2^*} + \sa \hat \Gamma^{(1)}_{H \tilde t_1 \tilde t_2^*}: \quad \text{$\deltaOL \mu$ drops out}.
\end{align}
\end{subequations}
Now the first expression is used to determine $\deltaOL \mu$ and the second one to determine $\deltaOL \At$. The same procedure works for $\deltaOL \Ab$ as well:
\begin{subequations}
\begin{align}
    &\ca \hat \Gamma^{(1)}_{H \tilde b_1 \tilde b_2^*} - \sa \hat \Gamma^{(1)}_{h \tilde b_1 \tilde b_2^*}: \quad \text{$\deltaOL \mu$ drops out}, \\
    &\ca \hat \Gamma^{(1)}_{h \tilde b_1 \tilde b_2^*} + \sa \hat \Gamma^{(1)}_{H \tilde b_1 \tilde b_2^*}: \quad \text{$\deltaOL \Ab$ drops out}.
\end{align}
\end{subequations}
Taking into account also the other counterterms contributing to the Higgs--sfermion amplitudes, we obtain
\begin{subequations}
\allowdisplaybreaks
\begin{align}
    \frac{\deltaOL \mu^\DRbar}{\mu} ={}& \frac{\alem \nc}{16 \pi M_W^2 \sw^2} \Big( \frac{m_t^2}{\sib^2} + \frac{m_b^2}{\cb^2} \Big) \frac{1}{\del}, \\
    \frac{\deltaOL \At^\DRbar}{\At} ={}& \frac{\alem \nc}{8 \pi M_W^2 \sw^2} \frac{m_t^2}{\sib^2} \frac{1}{\del}, \\
    \frac{\deltaOL \Ab^\DRbar}{\Ab} ={}& \frac{\alem \nc}{8 \pi M_W^2 \sw^2} \frac{m_b^2}{\cb^2} \frac{1}{\del}.
\end{align}
\end{subequations}

\section{One- and two-loop field and parameter counterterms}
\label{app:CTs}
In this appendix, we list the one- and two-loop expressions for the renormalised tadpoles and self-energies in the Higgs--gauge sector that appear in our calculation. They are obtained by applying the renormalisation transformations given in \cha{cha:mssm_ren} to the Higgs--gauge Lagrangian.

We also give the one- and two-loop expressions for the counterterms of all Higgs boson mass parameters appearing in \eqn{eqn:Higgs_Lag_mass}. The counterterms will be expressed as combinations of the one- and two-loop counterterms of the input parameters. The renormalisation of the input parameters is explained in \scts{ssec:higgs_gauge_1L_ren} and \ref{ssec:higgs_gauge_2L_ren}. We also relate the renormalisation constants of the Higgs fields to the Higgs doublet counterterms.

\subsection{Renormalised one-loop tadpoles and self-energies}
\label{ssec:tad_se_ren_1L}

The one-loop one-point vertex functions are renormalised by the tadpole counterterms:
\begin{subequations}
\allowdisplaybreaks
\begin{align}
    \hat \Gamma^{(1)}_h ={}& \Gamma^{(1)}_h + \deltaOL T_h, \\
    \hat \Gamma^{(1)}_H ={}& \Gamma^{(1)}_H + \deltaOL T_H, \\
    \hat \Gamma^{(1)}_A ={}& \Gamma^{(1)}_A + \deltaOL T_A, \\
    \hat \Gamma^{(1)}_G ={}& \Gamma^{(1)}_G + \deltaOL T_G.
\end{align}
\end{subequations}

The neutral $\CP$-even self-energies are
\begin{subequations}
\allowdisplaybreaks
\begin{align}
    \hat \Sigma^{(1)}_{hh}(p^2) ={}& \Sigma^{(1)}_{hh}(p^2) + \deltaOL Z_{hh} (p^2 - m_h^2) - \deltaOL m_h^2, \\
    \hat \Sigma^{(1)}_{hH}(p^2) ={}& \Sigma^{(1)}_{hH}(p^2) + \deltaOL Z_{hH} \left( p^2 - \frac{m_h^2 + m_H^2}{2} \right) - \deltaOL m_{hH}^2, \\
    \hat \Sigma^{(1)}_{HH}(p^2) ={}& \Sigma^{(1)}_{HH}(p^2) + \deltaOL Z_{HH} (p^2 - m_H^2) - \deltaOL m_H^2,
\end{align}
\end{subequations}
while the $\CP$-odd self-energies read
\begin{subequations}
\allowdisplaybreaks
\begin{align}
    \hat \Sigma^{(1)}_{AA}(p^2) ={}& \Sigma^{(1)}_{AA}(p^2) + \deltaOL Z_{AA} (p^2 - m_A^2) - \deltaOL m_A^2, \\
    \hat \Sigma^{(1)}_{AG}(p^2) ={}& \Sigma^{(1)}_{AG}(p^2) + \deltaOL Z_{AG} \left( p^2 - \frac{m_A^2}{2} \right) - \deltaOL m_{AG}^2, \\
    \hat \Sigma^{(1)}_{GG}(p^2) ={}& \Sigma^{(1)}_{GG}(p^2) + \deltaOL Z_{GG} \, p^2 - \deltaOL m_G^2.
\end{align}
\end{subequations}

In the case of $\CP$ violation, the self-energies
\begin{subequations}
\allowdisplaybreaks
\begin{align}
    \hat \Sigma^{(1)}_{hA}(p^2) ={}& \Sigma^{(1)}_{hA}(p^2) -\deltaOL m_{hA}^2, \\
    \hat \Sigma^{(1)}_{hG}(p^2) ={}& \Sigma^{(1)}_{hG}(p^2) -\deltaOL m_{hG}^2, \\
    \hat \Sigma^{(1)}_{HA}(p^2) ={}& \Sigma^{(1)}_{HA}(p^2) -\deltaOL m_{HA}^2, \\
    \hat \Sigma^{(1)}_{HG}(p^2) ={}& \Sigma^{(1)}_{HG}(p^2) -\deltaOL m_{HG}^2
\end{align}
\end{subequations}
do not vanish. The neutral self-energies are symmetric such that for instance ${\hat \Sigma^{(1)}_{Hh} = \hat \Sigma^{(1)}_{hH}}$.

The charged Higgs self-energies are
\begin{subequations}
\allowdisplaybreaks
\begin{align}
    \hat \Sigma^{(1)}_{H^-H^+}(p^2) ={}& \Sigma^{(1)}_{H^-H^+}(p^2) + \deltaOL Z_{H^-H^+} \left( p^2 - m_{H^\pm}^2 \right) - \deltaOL m_{H^\pm}^2, \\
    \hat \Sigma^{(1)}_{H^-G^+}(p^2) ={}& \Sigma^{(1)}_{H^-G^+}(p^2) + \deltaOL Z_{H^-G^+} \left( p^2 - \frac{m_{H^\pm}^2}{2} \right) - \deltaOL m_{H^-G^+}^2, \\
    \hat \Sigma^{(1)}_{G^-H^+}(p^2) ={}& \Sigma^{(1)}_{G^-H^+}(p^2) + \deltaOL Z_{G^-H^+} \left( p^2 - \frac{m_{H^\pm}^2}{2} \right) - \deltaOL m_{G^-H^+}^2, \\
    \hat \Sigma^{(1)}_{G^-G^+}(p^2) ={}& \Sigma^{(1)}_{G^-G^+}(p^2) + \deltaOL Z_{G^-G^+} \, p^2 - \deltaOL m_{G^\pm}^2.
\end{align}
\end{subequations}
They are symmetric in the sense that
\begin{equation}
    \hat \Sigma^{(1)}_{G^-H^+} = \hat \Sigma^{(1)}_{H^+G^-}
\end{equation}
but in general
\begin{equation}
    \big( \hat \Sigma^{(1)}_{G^-H^+}(p^2) \big)^* \neq  \hat \Sigma^{(1)}_{H^-G^+}(p^2). 
\end{equation}
Instead, we have
\begin{equation}
    \big( \hat \Sigma^{(1)}_{G^-H^+}(p^2) \big)^* =  \TCo\ \hat \Sigma^{(1)}_{H^-G^+}(p^2),
\end{equation}
where $\TCo$ takes the complex conjugate of loop integrals only and leaves complex couplings unaffected. Loop integrals are complex quantities for sufficiently large external momenta and so the $\TCo$ must not be left out.

Vector boson self-energies are Lorentz tensors of rank two. We decompose them into a transverse and into a longitudinal component
\begin{equation}
    \Sigma^{\mu\nu}(p) = \left( -g^{\mu\nu} + \frac{p^\mu p^\nu}{p^2} \right) \Sigma^T(p^2) - \frac{p^\mu p^\nu}{p^2} \Sigma^L(p^2),
\end{equation}
using the same convention as in \citere{Weiglein:1993hd}.

The renormalised transverse parts of the gauge boson self-energies are
\begin{subequations}
\allowdisplaybreaks
\begin{align}
\label{eqn:gaga_1L_ren}
    \hat \Sigma^{T,(1)}_{\gamma \gamma}(p^2) ={}& \Sigma^{T,(1)}_{\gamma \gamma}(p^2) + \deltaOL Z_{\gamma \gamma} \, p^2, \\
\begin{split}
    \hat \Sigma^{T,(1)}_{\gamma Z}(p^2) ={}& \Sigma^{T,(1)}_{\gamma Z}(p^2) + \tfrac{1}{2} \deltaOL Z_{\gamma Z} \, p^2 + \tfrac{1}{2} \deltaOL Z_{Z \gamma} (p^2 - M_Z^2) \\
    ={}& \hat \Sigma^{T,(1)}_{Z \gamma}(p^2),
\end{split} \\
    \hat \Sigma^{T,(1)}_{ZZ}(p^2) ={}& \Sigma^{T,(1)}_{ZZ}(p^2) + \deltaOL Z_{ZZ} (p^2 - M_Z^2) - \deltaOL M_Z^2, \\
\begin{split}
    \hat \Sigma^{T,(1)}_{W^-W^+}(p^2) ={}& \Sigma^{T,(1)}_{W^-W^+}(p^2) + \deltaOL Z_{WW} (p^2 - M_W^2) - \deltaOL M_W^2 \\
    ={}& \hat \Sigma^{T,(1)}_{W^+W^-}(p^2).
\label{eqn:WW_ren_SE_1L}
\end{split}
\end{align}
\end{subequations}

The transverse parts of vector self-energies are relevant in our calculation already at the one-loop level, as they are used to determine the mass and field counterterms of the gauge bosons. The longitudinal vector boson self-energies enter a Higgs boson mass prediction at the three-loop order and higher, and will not be discussed here.

For a two-loop calculation, we also need self-energies which mix scalars and vectors. Their Lorentz decomposition reads
\begin{subequations}
\begin{align}
    \Sigma^\mu_{SV}(p) = p^\mu \Sigma^L_{SV}(p^2), \\
    \Sigma^\mu_{VS}(p) = p^\mu \Sigma^L_{VS}(p^2),
\end{align}
\end{subequations}
where $\Sigma^\mu_{SV}(p)$ denotes a self-energy with incoming vector $V^\dagger$ and outgoing scalar $S$, and $\Sigma^\mu_{VS}(p)$ denotes a self-energy with incoming scalar $S^\dagger$ and outgoing vector $V$. We have four neutral scalar--vector self-energies
\begin{subequations}
\allowdisplaybreaks
\begin{align}
\begin{split}
    \hat \Sigma^{L,(1)}_{A\gamma}(p^2) ={}& \Sigma^{L,(1)}_{A\gamma}(p^2) \\
    ={}&- \hat \Sigma^{L,(1)}_{\gamma A}(p^2) \overset{\order{\nc}}{=} 0,
\end{split} \\
\begin{split}
\label{eqn:AZ-mixing_1L}
    \hat \Sigma^{L,(1)}_{AZ}(p^2) ={}& \Sigma^{L,(1)}_{AZ}(p^2) - \imag M_Z \left( \tfrac{1}{2} \deltaOL Z_{AG} + \cb^2 \deltaOL \tb \right) \\
    ={}& - \hat \Sigma^{L,(1)}_{ZA}(p^2),
\end{split} \\
\begin{split}
    \hat \Sigma^{L,(1)}_{G \gamma}(p^2) ={}& \Sigma^{L,(1)}_{G \gamma}(p^2) - \frac{\imag M_Z}{2} \deltaOL Z_{Z \gamma} \\
    ={}& - \hat \Sigma^{L,(1)}_{\gamma G}(p^2) \overset{\order{\nc}}{=} 0,
\end{split} \\
\begin{split}
    \hat \Sigma^{L,(1)}_{GZ}(p^2) ={}& \Sigma^{L,(1)}_{GZ}(p^2) - \frac{\imag M_Z}{2} \left( \frac{\deltaOL M_Z^2}{M_Z^2} + \deltaOL Z_{ZZ} + \deltaOL Z_{GG} \right) \\
    ={}& - \hat \Sigma^{L,(1)}_{ZG}(p^2).
\end{split}
\end{align}
\end{subequations}

The charged self-energies are
\begin{subequations}
\allowdisplaybreaks
\begin{align}
\begin{split}
\label{eqn:HW-mixing_1L}
    \hat \Sigma^{L,(1)}_{H^-W^+}(p^2) ={}& \Sigma^{L,(1)}_{H^-W^+}(p^2) + M_W \left( \tfrac{1}{2} \deltaOL Z_{H^-G^+}  + \cb^2 \deltaOL \tb \right) \\
    ={}& - \hat \Sigma^{L,(1)}_{W^+H^-}(p^2),
\end{split} \\
\begin{split}
    \hat \Sigma^{L,(1)}_{W^-H^+}(p^2) ={}& \Sigma^{L,(1)}_{W^-H^+}(p^2) + M_W \left( \tfrac{1}{2} \deltaOL Z_{G^-H^+}  + \cb^2 \deltaOL \tb \right) \\
    ={}& - \hat \Sigma^{L,(1)}_{H^+W^-}(p^2).
\end{split} \\
\begin{split}
    \hat \Sigma^{L,(1)}_{G^-W^+}(p^2) ={}& \Sigma^{L,(1)}_{G^-W^+}(p^2) + \frac{M_W}{2} \left( \frac{\deltaOL M_W^2}{M_W^2} + \deltaOL Z_{WW} + \deltaOL Z_{G^-G^+} \right) \\
    ={}& - \hat \Sigma^{L,(1)}_{W^+G^-}(p^2),
\end{split} \\
\begin{split}
    \hat \Sigma^{L,(1)}_{W^-G^+}(p^2) ={}& \Sigma^{L,(1)}_{W^-G^+}(p^2) + \frac{M_W}{2} \left( \frac{\deltaOL M_W^2}{M_W^2} + \deltaOL Z_{WW} + \deltaOL Z_{G^-G^+} \right) \\
    ={}& - \hat \Sigma^{L,(1)}_{G^+W^-}(p^2).
\end{split}
\end{align}
\end{subequations}
Again, conjugated diagrams are related via
\begin{equation}
    \big( \hat \Sigma^{L,(1)}_{W^-H^+}(p^2) \big)^* = \TCo\ \hat \Sigma^{L,(1)}_{H^-W^+}(p^2).
\end{equation}

Not all of the self-energies presented in this section are actually needed for our calculation. We have numerically shown the finiteness of all given one-loop self-energies as a cross-check and provide the derived expressions for the sake of completeness and for future reference.

\subsection{One-loop counterterms}

The one-loop mass counterterms read
\begin{subequations}
\allowdisplaybreaks
\begin{align}
\begin{split}
    \deltaOL m_h^2 ={}& \deltaOL m_A^2 \camb^2 + \deltaOL M_Z^2 \sapb^2 \\
    &+ \deltaOL \tb \cb^2 \big( m_A^2 \stamb + M_Z^2 \stapb \big) \\
    &+ \frac{e \samb}{2 M_W \sw} \big( \deltaOL T_h (1 + \camb^2) + \deltaOL T_H \camb \samb \big),
\end{split} \\
\begin{split}
    \deltaOL m_{hH}^2 ={}& \deltaOL m_A^2 \camb \samb - \deltaOL M_Z^2 \capb \sapb \\
    &- \deltaOL \tb \cb^2 \big( m_A^2 \ctamb + M_Z^2 \ctapb \big) \\
    &- \frac{e}{2 M_W \sw} \big( \deltaOL T_h \camb^3 - \deltaOL T_H \samb^3 \big),
\end{split} \\
\begin{split}
    \deltaOL m_H^2 ={}& \deltaOL m_A^2 \samb^2 + \deltaOL M_Z^2 \capb^2 \\
    &- \deltaOL \tb \cb^2 \big( m_A^2 \stamb + M_Z^2 \stapb \big) \\
    &- \frac{e \camb}{2 M_W \sw} \big( \deltaOL T_h \camb \samb + \deltaOL T_H (1 + \samb^2) \big),
\end{split} \\
\begin{split}
    \deltaOL m_{AG}^2 ={}& -\deltaOL \tb \cb^2 m_A^2 \\
    &- \frac{e}{2 M_W \sw} \big( \deltaOL T_h \camb + \deltaOL T_H \samb  \big),
\end{split} \\
    \deltaOL m_G^2 ={}& \frac{e}{2 M_W \sw} \big( \deltaOL T_h \samb - \deltaOL T_H \camb  \big), \\
    \deltaOL m_{hA}^2 ={}& \frac{e \samb}{2 M_W \sw} \deltaOL T_A, \\
    \deltaOL m_{HA}^2 ={}& - \frac{e \camb}{2 M_W \sw} \deltaOL T_A, \\
    \deltaOL m_{hG}^2 ={}& \frac{e \camb}{2 M_W \sw} \deltaOL T_A = - \deltaOL m_{HA}^2, \\
    \deltaOL m_{HG}^2 ={}& \frac{e \samb}{2 M_W \sw} \deltaOL T_A = \deltaOL m_{hA}^2, \\
\begin{split}
    \deltaOL m_{H^-G^+}^2 ={}& -\deltaOL \tb \cb^2 m_{H^\pm}^2 \\
    &- \frac{e}{2 M_W \sw} \big( \deltaOL T_h \camb + \deltaOL T_H \samb - \imag \deltaOL T_A  \big),
\end{split} \\
\begin{split}
    \deltaOL m_{G^-H^+}^2 ={}& -\deltaOL \tb \cb^2 m_{H^\pm}^2 \\
    &- \frac{e}{2 M_W \sw} \big( \deltaOL T_h \camb + \deltaOL T_H \samb + \imag \deltaOL T_A  \big) \\
    ={}& \big( \deltaOL m_{H^-G^+}^2 \big)^*,
\end{split} \\
    \deltaOL m_{G^\pm}^2 ={}& \frac{e}{2 M_W \sw} \big( \deltaOL T_h \samb - \deltaOL T_H \camb  \big) = \deltaOL m_G^2.
\end{align}
\end{subequations}
The one-loop field counterterms are given by
\begin{subequations}
\allowdisplaybreaks
\begin{align}
    \deltaOL Z_{hh} ={}& \sa^2 \deltaOL \zho + \ca^2 \deltaOL \zht, \\
    \deltaOL Z_{hH} ={}& \sa \ca \big( \deltaOL \zht - \deltaOL \zho \big), \\
    \deltaOL Z_{HH} ={}& \ca^2 \deltaOL \zho + \sa^2 \deltaOL \zht, \\
    \deltaOL Z_{AA} ={}& \sib^2 \deltaOL \zho + \cb^2 \deltaOL \zht, \\
    \deltaOL Z_{AG} ={}& \sib \cb \big( \deltaOL \zht - \deltaOL \zho \big), \\
    \deltaOL Z_{GG} ={}& \cb^2 \deltaOL \zho + \sib^2 \deltaOL \zht, \\
    \deltaOL Z_{H^-H^+} ={}& \sib^2 \deltaOL \zho + \cb^2 \deltaOL \zht, \\
    \deltaOL Z_{H^-G^+} ={}& \sib \cb \big( \deltaOL \zht - \deltaOL \zho \big), \\
    \deltaOL Z_{G^-H^+} ={}& \sib \cb \big( \deltaOL \zht - \deltaOL \zho \big), \\
    \deltaOL Z_{G^-G^+} ={}& \cb^2 \deltaOL \zho + \sib^2 \deltaOL \zht.
\end{align}
\end{subequations}

\subsection{Renormalised two-loop tadpoles and self-energies}
\label{ssec:tad_se_ren_2L}

At the two-loop level, the renormalised one-point vertex functions read
\begin{subequations}
\allowdisplaybreaks
\begin{align}
    \hat \Gamma^{(2)}_h ={}& \Gamma^{(2)}_h + \deltaTL T_h + \tfrac{1}{2} \deltaOL Z_{hh} \deltaOL T_h + \tfrac{1}{2} \deltaOL Z_{hH} \deltaOL T_H, \\
    \hat \Gamma^{(2)}_H ={}& \Gamma^{(2)}_H + \deltaTL T_H + \tfrac{1}{2} \deltaOL Z_{HH} \deltaOL T_H + \tfrac{1}{2} \deltaOL Z_{hH} \deltaOL T_h, \\
    \hat \Gamma^{(2)}_A ={}& \Gamma^{(2)}_A + \deltaTL T_A + \tfrac{1}{2} \deltaOL Z_{AA} \deltaOL T_A + \tfrac{1}{2} \deltaOL Z_{AG} \deltaOL T_G, \\
    \hat \Gamma^{(2)}_G ={}& \Gamma^{(2)}_G + \deltaTL T_G + \tfrac{1}{2} \deltaOL Z_{GG} \deltaOL T_G + \tfrac{1}{2} \deltaOL Z_{AG} \deltaOL T_A.
\end{align}
\end{subequations}

The renormalised $\CP$-even two-loop self-energies read
\begin{subequations}
\allowdisplaybreaks
\begin{align}
\begin{split}
    \hat \Sigma^{(2)}_{hh}(p^2) ={}& \Sigma^{(2)}_{hh}(p^2) + \deltaTL Z_{hh} (p^2 - m_h^2) - \deltaTL m_h^2 \\
    &+ \tfrac{1}{4} \left( \deltaOL Z_{hH} \right)^2 (p^2 - m_H^2) - \deltaOL Z_{hh} \deltaOL m_h^2 - \deltaOL Z_{hH} \deltaOL m_{hH}^2,
\end{split} \\
\begin{split}
    \hat \Sigma^{(2)}_{hH}(p^2) ={}& \Sigma^{(2)}_{hH}(p^2) + \deltaTL Z_{hH} \left( p^2 - \frac{m_h^2 + m_H^2}{2} \right) - \deltaTL m_{hH}^2 \\
    &+ \tfrac{1}{4} \deltaOL Z_{hH} \deltaOL Z_{hh} (p^2 - m_h^2) + \tfrac{1}{4} \deltaOL Z_{hH} \deltaOL Z_{HH} (p^2 - m_H^2) \\
    &- \deltaOL Z_{hH} \frac{\deltaOL m_h^2 + \deltaOL m_H^2}{2} - \frac{\deltaOL Z_{hh} + \deltaOL Z_{HH}}{2} \deltaOL m_{hH}^2,
\end{split} \\
\begin{split}
    \hat \Sigma^{(2)}_{HH}(p^2) ={}& \Sigma^{(2)}_{HH}(p^2) + \deltaTL Z_{HH} (p^2 - m_H^2) - \deltaTL m_H^2 \\
    &+ \tfrac{1}{4} \left( \deltaOL Z_{hH} \right)^2 (p^2 - m_h^2) - \deltaOL Z_{HH} \deltaOL m_H^2 - \deltaOL Z_{hH} \deltaOL m_{hH}^2.
\end{split}
\end{align}
\end{subequations}

Similarly, the $\CP$-odd self-energies are
\begin{subequations}
\allowdisplaybreaks
\begin{align}
\begin{split}
\label{eqn:A0A0_2L_ren}
    \hat \Sigma^{(2)}_{AA}(p^2) ={}& \Sigma^{(2)}_{AA}(p^2) + \deltaTL Z_{AA} (p^2 - m_A^2) - \deltaTL m_A^2 \\
    &+ \tfrac{1}{4} \left( \deltaOL Z_{AG} \right)^2 p^2 - \deltaOL Z_{AA} \deltaOL m_A^2 - \deltaOL Z_{AG} \deltaOL m_{AG}^2,
\end{split} \\
\begin{split}
    \hat \Sigma^{(2)}_{AG}(p^2) ={}& \Sigma^{(2)}_{AG}(p^2) + \deltaTL Z_{AG} \left( p^2 - \frac{m_A^2}{2} \right) - \deltaTL m_{AG}^2 \\
    &+ \tfrac{1}{4} \deltaOL Z_{AG} \deltaOL Z_{AA} (p^2 - m_A^2) + \tfrac{1}{4} \deltaOL Z_{AG} \deltaOL Z_{GG} \, p^2 \\
    &- \deltaOL Z_{AG} \frac{\deltaOL m_A^2 + \deltaOL m_G^2}{2} - \frac{\deltaOL Z_{AA} + \deltaOL Z_{GG}}{2} \deltaOL m_{AG}^2,
\end{split} \\
\begin{split}
    \hat \Sigma^{(2)}_{GG}(p^2) ={}& \Sigma^{(2)}_{GG}(p^2) + \deltaTL Z_{GG} \, p^2 - \deltaTL m_G^2 \\
    &+ \tfrac{1}{4} \left( \deltaOL Z_{AG} \right)^2 (p^2 - m_A^2) - \deltaOL Z_{GG} \deltaOL m_G^2 - \deltaOL Z_{AG} \deltaOL m_{AG}^2.
\end{split}
\end{align}
\end{subequations}

Finally, the $\CP$-mixing self-energies read
\begin{subequations}
\allowdisplaybreaks
\begin{align}
\begin{split}
    \hat \Sigma^{(2)}_{hA}(p^2) ={}& \Sigma^{(2)}_{hA}(p^2) - \deltaTL m_{hA}^2 \\
    &- \frac{\deltaOL Z_{hh} + \deltaOL Z_{AA}}{2} \deltaOL m_{hA}^2 - \tfrac{1}{2} Z_{hH} m_{HA}^2 - \tfrac{1}{2} Z_{AG} m_{hG}^2,
\end{split} \\
\begin{split}
    \hat \Sigma^{(2)}_{hG}(p^2) ={}& \Sigma^{(2)}_{hG}(p^2) - \deltaTL m_{hG}^2 \\
    &- \frac{\deltaOL Z_{hh} + \deltaOL Z_{GG}}{2} \deltaOL m_{hG}^2 - \tfrac{1}{2} Z_{hH} m_{HG}^2 - \tfrac{1}{2} Z_{AG} m_{hA}^2,
\end{split} \\
\begin{split}
    \hat \Sigma^{(2)}_{HA}(p^2) ={}& \Sigma^{(2)}_{HA}(p^2) - \deltaTL m_{HA}^2 \\
    &- \frac{\deltaOL Z_{HH} + \deltaOL Z_{AA}}{2} \deltaOL m_{HA}^2 - \tfrac{1}{2} Z_{hH} m_{hA}^2 - \tfrac{1}{2} Z_{AG} m_{HG}^2,
\end{split} \\
\begin{split}
    \hat \Sigma^{(2)}_{HG}(p^2) ={}& \Sigma^{(2)}_{HG}(p^2) - \deltaTL m_{HG}^2 \\
    &- \frac{\deltaOL Z_{HH} + \deltaOL Z_{GG}}{2} \deltaOL m_{HG}^2 - \tfrac{1}{2} Z_{hH} m_{hG}^2 - \tfrac{1}{2} Z_{AG} m_{HA}^2.
\end{split}
\end{align}
\end{subequations}

The renormalised charged self-energies are
\begin{subequations}
\allowdisplaybreaks
\begin{align}
\begin{split}
    \hat \Sigma^{(2)}_{H^-H^+}(p^2) ={}& \Sigma^{(2)}_{H^-H^+}(p^2) + \deltaTL Z_{H^-H^+} (p^2 - m_{H^\pm}^2) - \deltaTL m_{H^\pm}^2 \\
    &+ \frac{\deltaOL Z_{H^-G^+} \deltaOL Z_{G^-H^+}}{4} p^2 - \deltaOL Z_{H^-H^+} \deltaOL m_{H^\pm}^2 \\
    &- \tfrac{1}{2} \deltaOL Z_{H^-G^+} \deltaOL m_{G^-H^+}^2 - \tfrac{1}{2} \deltaOL Z_{G^-H^+} \deltaOL m_{H^-G^+}^2,
\end{split} \\
\begin{split}
    \hat \Sigma^{(2)}_{H^-G^+}(p^2) ={}& \Sigma^{(2)}_{H^-G^+}(p^2) + \deltaTL Z_{H^-G^+} \left( p^2 - \frac{m_{H^\pm}^2}{2} \right) - \deltaTL m_{H^-G^+}^2 \\
    &+ \tfrac{1}{4} \deltaOL Z_{H^-G^+} \deltaOL Z_{H^-H^+} (p^2 - m_{H^\pm}^2) \\
    &+ \tfrac{1}{4} \deltaOL Z_{H^-G^+} \deltaOL Z_{G^-G^+} \, p^2 \\
    &- \deltaOL Z_{H^-G^+} \frac{\deltaOL m_{H^\pm}^2 + \deltaOL m_{G^\pm}^2}{2} \\
    &- \frac{\deltaOL Z_{H^-H^+} + \deltaOL Z_{G^-G^+}}{2} \deltaOL m_{H^-G^+}^2,
\end{split} \\
\begin{split}
    \hat \Sigma^{(2)}_{G^-H^+}(p^2) ={}& \Sigma^{(2)}_{G^-H^+}(p^2) + \deltaTL Z_{G^-H^+} \left( p^2 - \frac{m_{H^\pm}^2}{2} \right) - \deltaTL m_{G^-H^+}^2 \\
    &+ \tfrac{1}{4} \deltaOL Z_{G^-H^+} \deltaOL Z_{H^-H^+} (p^2 - m_{H^\pm}^2) \\
    &+ \tfrac{1}{4} \deltaOL Z_{G^-H^+} \deltaOL Z_{G^-G^+} \, p^2 \\
    &- \deltaOL Z_{G^-H^+} \frac{\deltaOL m_{H^\pm}^2 + \deltaOL m_{G^\pm}^2}{2} \\
    &- \frac{\deltaOL Z_{H^-H^+} + \deltaOL Z_{G^-G^+}}{2} \deltaOL m_{G^-H^+}^2,
\end{split} \\
\begin{split}
    \hat \Sigma^{(2)}_{G^-G^+}(p^2) ={}& \Sigma^{(2)}_{G^-G^+}(p^2) + \deltaTL Z_{G^-G^+} \, p^2 - \deltaTL m_{G^\pm}^2 \\
    &+ \frac{\deltaOL Z_{H^-G^+} \deltaOL Z_{G^-H^+}}{4} (p^2 - m_{H^\pm}^2) - \deltaOL Z_{G^-G^+} \deltaOL m_{G^\pm}^2 \\
    &- \tfrac{1}{2} \deltaOL Z_{H^-G^+} \deltaOL m_{G^-H^+}^2 - \tfrac{1}{2} \deltaOL Z_{G^-H^+} \deltaOL m_{H^-G^+}^2.
\end{split}
\end{align}
\end{subequations}

The renormalised transverse parts of the two-loop gauge boson self-energies are
\begin{subequations}
\allowdisplaybreaks
\begin{align}
\label{eqn:photon_se_ren}
    \hat \Sigma^{T,(2)}_{\gamma \gamma}(p^2) ={}& \Sigma^{T,(2)}_{\gamma \gamma}(p^2) + \deltaTL Z_{\gamma \gamma} \, p^2 + \tfrac{1}{4} \left( \deltaOL Z_{Z \gamma} \right)^2 (p^2 - M_Z^2),
    \\
\begin{split}
    \hat \Sigma^{T,(2)}_{\gamma Z}(p^2) ={}& \Sigma^{T,(2)}_{\gamma Z}(p^2) + \tfrac{1}{2} \left( \deltaTL Z_{\gamma Z} + \tfrac{1}{2} \deltaOL Z_{\gamma Z} \deltaOL Z_{\gamma \gamma} \right) p^2 \\
    &+ \tfrac{1}{2} \left( \deltaTL Z_{Z \gamma} + \tfrac{1}{2} \deltaOL Z_{Z \gamma} \deltaOL Z_{ZZ} \right) (p^2 - M_Z^2) \\
    &- \tfrac{1}{2} \deltaOL Z_{Z \gamma} \deltaOL M_Z^2 \\
    ={}& \hat \Sigma^{T,(2)}_{Z \gamma}(p^2),
\end{split} \\
\begin{split}
    \hat \Sigma^{T,(2)}_{ZZ}(p^2) ={}& \Sigma^{T,(2)}_{ZZ}(p^2) + \deltaTL Z_{ZZ} (p^2 - M_Z^2) - \deltaTL M_Z^2 \\
    &+ \tfrac{1}{4} \left( \deltaOL Z_{\gamma Z} \right)^2 p^2  - \deltaOL Z_{ZZ} \deltaOL M_Z^2,
\end{split} \\
\begin{split}
    \hat \Sigma^{T,(2)}_{W^-W^+}(p^2) ={}& \Sigma^{T,(2)}_{W^-W^+}(p^2) + \deltaTL Z_{WW} (p^2 - M_W^2) - \deltaTL M_W^2 \\
    &- \deltaOL Z_{WW} \deltaOL M_W^2 \\
    ={}& \hat \Sigma^{T,(2)}_{W^+W^-}(p^2).
\label{eqn:WW_ren_SE_2L}
\end{split}
\end{align}
\end{subequations}

At the two-loop level, the renormalised transverse part of the photon self-energy receives a non-vanishing contribution at $p^2 = 0$ from the mixing with the $Z$ boson. Another non-vanishing contribution stems from the sub-loop part of the unrenormalised self-energy. To demonstrate this, we write the unrenormalised self-energy as
\begin{equation}
\label{eqn:photon_se_unr}
    \Sigma^{T,(2)}_{\gamma \gamma}(p^2) = \widetilde \Sigma^{T,(2)}_{\gamma \gamma}(p^2) + \deltaOL Z_{\gamma \gamma} \Sigma^{T,(1)}_{\gamma \gamma}(p^2) + \deltaOL Z_{Z \gamma} \Sigma^{T,(1)}_{\gamma Z}(p^2),
\end{equation}
where $\widetilde \Sigma^{T,(2)}_{\gamma \gamma}(p^2)$ does not contain any field renormalisation constants. While the second term on the right-hand side vanishes at zero momentum due to a Slavnov-Taylor identity, the third term will usually give a non-vanishing contribution.\footnote{For the set of contributions considered in this paper, $\Sigma^{T,(1)}_{\gamma Z}(0) = 0$.} The third term of \eqn{eqn:photon_se_ren} and the third term of \eqn{eqn:photon_se_unr} will drop out once the effective two-loop self-energy, which we defined in \eqn{eqn:gaga_eff_se}, is considered.

The renormalised two-loop self-energy vanishes at zero momentum if an on-shell renormalisation is chosen for $\deltaOL Z_{Z \gamma}$. In our calculation, the on-shell condition leads to $\delta Z_{Z \gamma} = 0$.

The two-loop Higgs--vector mixing self-energies
\begin{subequations}
\begin{align}
\begin{split}
    \hat \Sigma^{L,(2)}_{AZ}(p^2) ={}& \Sigma^{L,(2)}_{AZ}(p^2) - \imag M_Z \Big(\cb^2 \deltaTL \tb - \cb^3 \sib ( \deltaOL \tb )^2 \\
    &+ \tfrac{1}{2} \deltaTL Z_{AG} + \tfrac{1}{2} \cb^2 \deltaOL \tb \deltaOL Z_{AA} \Big) \\
    &- \frac{\imag M_Z}{2} \left(\cb^2 \deltaOL \tb + \tfrac{1}{2} \deltaOL Z_{AG} \right) \left( \frac{\deltaOL M_Z^2}{M_Z^2} + \deltaOL Z_{ZZ} \right) \\
    ={}&- \hat \Sigma^{L,(2)}_{ZA}(p^2),
\end{split} \\
\begin{split}
    \hat \Sigma^{L,(2)}_{H^-W^+}(p^2) ={}& \Sigma^{L,(2)}_{H^-W^+}(p^2) + M_W \Big(\cb^2 \deltaTL \tb - \cb^3 \sib ( \deltaOL \tb )^2 \\
    &+ \tfrac{1}{2} \deltaTL Z_{H^-G^+} + \tfrac{1}{2} \cb^2 \deltaOL \tb \deltaOL Z_{H^-H^+} \Big) \\
    &+ \frac{M_W}{2} \left(\cb^2 \deltaOL \tb + \tfrac{1}{2} \deltaOL Z_{H^-G^+} \right) \left( \frac{\deltaOL M_W^2}{M_W^2} + \deltaOL Z_{WW} \right) \\
    ={}&- \hat \Sigma^{L,(2)}_{W^+H^-}(p^2)
\end{split}
\end{align}
\end{subequations}
are used in some schemes to determine the two-loop counterterm for $\tb$, see \sct{sec:tanbeta_ren}.

\subsection{Two-loop counterterms}

The two-loop mass counterterms are given by
\begin{subequations}
\allowdisplaybreaks
\begin{align}
\begin{split}
    \deltaTL m_h^2 ={}& \deltaTL m_A^2 \camb^2 + \deltaTL M_Z^2 \sapb^2 \\
    &+ \deltaTL \tb \cb^2 \big( m_A^2 \stamb + M_Z^2 \stapb \big) \\
    &+ \deltaOL \tb \cb^2 \big( \deltaOL m_A^2 \stamb + \deltaOL M_Z^2 \stapb \big) \\
    &+ \tfrac{1}{2} \big( \deltaOL \tb \big)^2 \cb^3 \Big[ m_A^2 \samb (3 s_{\al - 2\beta} -\sa) + 2 M_Z^2 c_{2\al + 3\beta} \Big] \\
    &+ \frac{e \samb}{2 M_W \sw} \Big[ \big( \deltaTL T_h + \deltaOL Z_\text{w} \deltaOL T_h \big)(1 + \camb^2) \\
    &\hphantom{{}+ \frac{e \samb}{2 M_W \sw} \Big[{}}+ \big( \deltaTL T_H + \deltaOL Z_\text{w} \deltaOL T_H \big) \camb \samb \\
    &\hphantom{{}+ \frac{e \samb}{2 M_W \sw} \Big[{}}+ \deltaOL \tb \cb^2 \big( \deltaOL T_h \camb + \deltaOL T_H \samb \big) \samb \Big],
\end{split} \\
\begin{split}
    \deltaTL m_{hH}^2 ={}& \deltaTL m_A^2 \camb \samb - \deltaTL M_Z^2 \capb \sapb \\
    &- \deltaTL \tb \cb^2 \big( m_A^2 \ctamb + M_Z^2 \ctapb \big) \\
    &- \deltaOL \tb \cb^2 \big( \deltaOL m_A^2 \ctamb + \deltaOL M_Z^2 \ctapb \big) \\
    &- \tfrac{1}{2} \big( \deltaOL \tb \big)^2 \cb^3 \Big[ m_A^2 \frac{\camb (3 s_{\al - 2\beta} -\sa) + \samb (3 c_{\al - 2\beta} -\ca)}{2} \\
    &\hphantom{{}- \tfrac{1}{4} \big( \deltaOL \tb \big)^2 \cb^3 \Big[{}}- 2 M_Z^2 s_{2\al + 3\beta} \Big] \\
    &- \frac{e}{2 M_W \sw} \Big[ \big( \deltaTL T_h + \deltaOL Z_\text{w} \deltaOL T_h \big) \camb^3 \\
    &\hphantom{{}- \frac{e}{2 M_W \sw} \Big[{}}- \big( \deltaTL T_H +\deltaOL Z_\text{w} \deltaOL T_H \big) \samb^3 \\
    &\hphantom{{}- \frac{e}{2 M_W \sw} \Big[{}}+ \deltaOL \tb \cb^2 \big( \deltaOL T_h \camb + \deltaOL T_H \samb \big) \camb \samb \Big],
\end{split} \\
\begin{split}
    \deltaTL m_H^2 ={}& \deltaTL m_A^2 \samb^2 + \deltaTL M_Z^2 \capb^2 \\
    &- \deltaTL \tb \cb^2 \big( m_A^2 \stamb + M_Z^2 \stapb \big) \\
    &- \deltaOL \tb \cb^2 \big( \deltaOL m_A^2 \stamb + \deltaOL M_Z^2 \stapb \big) \\
    &+ \tfrac{1}{2} \big( \deltaOL \tb \big)^2 \cb^3 \Big[ m_A^2 \camb (3 c_{\al - 2\beta} -\ca) - 2 M_Z^2 c_{2\al + 3\beta} \Big] \\
    &- \frac{e \camb}{2 M_W \sw} \Big[ \big( \deltaTL T_h + \deltaOL Z_\text{w} \deltaOL T_h \big) \camb \samb \\
    &\hphantom{{}- \frac{e \camb}{2 M_W \sw} \Big[{}}+ \big( \deltaTL T_H + \deltaOL Z_\text{w} \deltaOL T_H \big) (1 + \samb^2) \\
    &\hphantom{{}- \frac{e \camb}{2 M_W \sw} \Big[{}}- \deltaOL \tb \cb^2 \big( \deltaOL T_h \camb + \deltaOL T_H \samb \big) \camb \Big],
\end{split} \\
\begin{split}
    \deltaTL m_{AG}^2 ={}& -\deltaTL \tb \cb^2 m_A^2 -\deltaOL \tb \cb^2 \deltaOL m_A^2 + \big( \deltaOL \tb \big)^2 \cb^3 \sib m_A^2 \\
    &- \frac{e}{2 M_W \sw} \Big[ \big( \deltaTL T_h + \deltaOL Z_\text{w} \deltaOL T_h \big) \camb \\
    &\hphantom{{}- \frac{e}{2 M_W \sw} \Big[{}}+ \big( \deltaTL T_H + \deltaOL Z_\text{w} \deltaOL T_H \big) \samb \Big],
\end{split} \\
\begin{split}
    \deltaTL m_G^2 ={}& \big( \deltaOL \tb \big)^2 \cb^4 m_A^2 \\
    &+ \frac{e}{2 M_W \sw} \Big[ \big( \deltaTL T_h + \deltaOL Z_\text{w} \deltaOL T_h \big) \samb \\
    &\hphantom{{}+ \frac{e}{2 M_W \sw} \Big[{}}- \big( \deltaTL T_H + \deltaOL Z_\text{w} \deltaOL T_H \big) \camb \\
    &\hphantom{{}+ \frac{e}{2 M_W \sw} \Big[{}}+ \deltaOL \tb \cb^2 \big( \deltaOL T_h \camb + \deltaOL T_H \samb \big) \Big],
\end{split} \\
    \deltaTL m_{hA}^2 ={}& \frac{e \samb}{2 M_W \sw} \big( \deltaTL T_A + \deltaOL Z_\text{w} \deltaOL T_A \big), \\
    \deltaTL m_{HA}^2 ={}& - \frac{e \camb}{2 M_W \sw} \big( \deltaTL T_A + \deltaOL Z_\text{w} \deltaOL T_A \big), \\
    \deltaTL m_{hG}^2 ={}& \frac{e \camb}{2 M_W \sw} \big( \deltaTL T_A + \deltaOL Z_\text{w} \deltaOL T_A \big) = - \deltaTL m_{HA}^2, \\
    \deltaTL m_{HG}^2 ={}& \frac{e \samb}{2 M_W \sw} \big( \deltaTL T_A + \deltaOL Z_\text{w} \deltaOL T_A \big) = \deltaTL m_{hA}^2, \\
\begin{split}
    \deltaTL m_{H^-G^+}^2 ={}& -\deltaTL \tb \cb^2 m_{H^\pm}^2 -\deltaOL \tb \cb^2 \deltaOL m_{H^\pm}^2 + \big( \deltaOL \tb \big)^2 \cb^3 \sib m_{H^\pm}^2 \\
    &- \frac{e}{2 M_W \sw} \Big[ \big( \deltaTL T_h + \deltaOL Z_\text{w} \deltaOL T_h \big) \camb \\
    &\hphantom{{}- \frac{e}{2 M_W \sw} \Big[{}}+ \big( \deltaTL T_H + \deltaOL Z_\text{w} \deltaOL T_H \big) \samb \\
    &\hphantom{{}- \frac{e}{2 M_W \sw} \Big[{}}- \imag \big( \deltaTL T_A + \deltaOL Z_\text{w} \deltaOL T_A \big) \Big],
\end{split} \\
\begin{split}
    \deltaTL m_{G^-H^+}^2 ={}& -\deltaTL \tb \cb^2 m_{H^\pm}^2 -\deltaOL \tb \cb^2 \deltaOL m_{H^\pm}^2 + \big( \deltaOL \tb \big)^2 \cb^3 \sib m_{H^\pm}^2 \\
    &- \frac{e}{2 M_W \sw} \Big[ \big( \deltaTL T_h + \deltaOL Z_\text{w} \deltaOL T_h \big) \camb \\
    &\hphantom{{}- \frac{e}{2 M_W \sw} \Big[{}}+ \big( \deltaTL T_H + \deltaOL Z_\text{w} \deltaOL T_H \big) \samb \\
    &\hphantom{{}- \frac{e}{2 M_W \sw} \Big[{}}+ \imag \big( \deltaTL T_A + \deltaOL Z_\text{w} \deltaOL T_A \big) \Big] \\
    ={}& \big( \deltaTL m_{H^-G^+}^2 \big)^*,
\end{split} \\
\begin{split}
    \deltaTL m_{G^\pm}^2 ={}& \big( \deltaOL \tb \big)^2 \cb^4 m_{H^\pm}^2 \\
    &+ \frac{e}{2 M_W \sw} \Big[ \big( \deltaTL T_h + \deltaOL Z_\text{w} \deltaOL T_h \big) \samb \\
    &\hphantom{{}+ \frac{e}{2 M_W \sw} \Big[{}}- \big( \deltaTL T_H + \deltaOL Z_\text{w} \deltaOL T_H \big) \camb \\
    &\hphantom{{}+ \frac{e}{2 M_W \sw} \Big[{}}+ \deltaOL \tb \cb^2 \big( \deltaOL T_h \camb + \deltaOL T_H \samb \big) \Big].
\end{split}
\end{align}
\end{subequations}
The two-loop field counterterms read
\begin{subequations}
\begin{align}
\begin{split}
    \deltaTL Z_{hh} ={}& \sa^2 \left( \deltaTL \zho - \frac{1}{4} \left( \deltaOL \zho \right)^2 \right) + \ca^2 \left( \deltaTL \zht - \frac{1}{4} \left( \deltaOL \zht \right)^2 \right) \\
    &+ \frac{1}{4} \left( \deltaOL Z_{hh} \right)^2,
\end{split} \\
    \deltaTL Z_{hH} ={}& \sa \ca \left[ \deltaTL \zht - \frac{1}{4} \left( \deltaOL \zht \right)^2 - \deltaTL \zho + \frac{1}{4} \left( \deltaOL \zho \right)^2 \right], \\
\begin{split}
    \deltaTL Z_{HH} ={}& \ca^2 \left( \deltaTL \zho - \frac{1}{4} \left( \deltaOL \zho \right)^2 \right) + \sa^2 \left( \deltaTL \zht - \frac{1}{4} \left( \deltaOL \zht \right)^2 \right) \\
    &+ \frac{1}{4} \left( \deltaOL Z_{HH} \right)^2,
\end{split} \\
\begin{split}
    \deltaTL Z_{AA} ={}& \sib^2 \left( \deltaTL \zho - \frac{1}{4} \left( \deltaOL \zho \right)^2 \right) + \cb^2 \left( \deltaTL \zht - \frac{1}{4} \left( \deltaOL \zht \right)^2 \right) \\
    &+ \frac{1}{4} \left( \deltaOL Z_{AA} \right)^2,
\end{split} \\
    \deltaTL Z_{AG} ={}& \sib \cb \left[ \deltaTL \zht - \frac{1}{4} \left( \deltaOL \zht \right)^2 - \deltaTL \zho + \frac{1}{4} \left( \deltaOL \zho \right)^2 \right], \\
\begin{split}
    \deltaTL Z_{GG} ={}& \cb^2 \left( \deltaTL \zho - \frac{1}{4} \left( \deltaOL \zho \right)^2 \right) + \sib^2 \left( \deltaTL \zht - \frac{1}{4} \left( \deltaOL \zht \right)^2 \right) \\
    &+ \frac{1}{4} \left( \deltaOL Z_{GG} \right)^2,
\end{split} \\
\begin{split}
    \deltaTL Z_{H^-H^+} ={}& \sib^2 \left( \deltaTL \zho - \frac{1}{4} \left( \deltaOL \zho \right)^2 \right) + \cb^2 \left( \deltaTL \zht - \frac{1}{4} \left( \deltaOL \zht \right)^2 \right) \\
    &+ \frac{1}{4} \left( \deltaOL Z_{H^-H^+} \right)^2,
\end{split} \\
    \deltaTL Z_{H^-G^+} ={}& \sib \cb \left[ \deltaTL \zht - \frac{1}{4} \left( \deltaOL \zht \right)^2 - \deltaTL \zho + \frac{1}{4} \left( \deltaOL \zho \right)^2 \right], \\
    \deltaTL Z_{G^-H^+} ={}& \sib \cb \left[ \deltaTL \zht - \frac{1}{4} \left( \deltaOL \zht \right)^2 - \deltaTL \zho + \frac{1}{4} \left( \deltaOL \zho \right)^2 \right], \\
\begin{split}
    \deltaTL Z_{G^-G^+} ={}& \cb^2 \left( \deltaTL \zho - \frac{1}{4} \left( \deltaOL \zho \right)^2 \right) + \sib^2 \left( \deltaTL \zht - \frac{1}{4} \left( \deltaOL \zht \right)^2 \right) \\
    &+ \frac{1}{4} \left( \deltaOL Z_{GG} \right)^2.
\end{split}
\end{align}
\end{subequations}

\section{Slavnov-Taylor identities for scalar-vector mixing}
\label{app:ST_id}
Slavnov-Taylor (ST) identities are the generalisation of the abelian Ward-Takahashi (WT) identities to non-abelian gauge theories. While WT identities follow from gauge symmetry, ST identities are a consequence of the Becchi-Rouet-Stora-Tyutin (BRST) symmetry, which is an extension of gauge symmetry after gauge fixing. For the present discussion, we will not derive ST identities from BRST invariance but simply check relations between the relevant self-energies algebraically or numerically.

In \citeres{Dabelstein:1995js,Logan:2002jh}, MSSM Slavnov-Taylor identities for self-energies in the $AGZ$ and the $H^\pm G^\pm W^\pm$ system are given. As was pointed out in \citere{Baro:2008bg}, the identities given in \citeres{Dabelstein:1995js,Logan:2002jh} hold only in a linear gauge and on-shell. \citeres{Baro:2008bg,Williams:2011bu} give ST identities also for off-shell momenta. 

In our analysis, we only consider self-energy contributions of $\order{\nc}$ and in a general $R_\xi$ gauge. Therefore, no diagrams with electroweak particles in the loops appear, and the self-energies are gauge-parameter independent.

Our starting point are the equations
\begin{subequations}
\allowdisplaybreaks
\begin{align}
    \Sigma^{(1),\text{notad}}_{AG}(p^2) - \imag \frac{p^2}{M_Z} \Sigma^{L,(1),\text{notad}}_{AZ}(p^2) \overset{\order{\nc}}{=}{}& \frac{e}{2 \sw M_W} \big( \Gamma^{(1)}_h \camb + \Gamma^{(1)}_H \samb \big), \\
    \Sigma^{(1),\text{notad}}_{H^-G^+}(p^2) - \frac{p^2}{M_W} \Sigma^{L,(1),\text{notad}}_{H^-W^+}(p^2) \overset{\order{\nc}}{=}{}& \frac{e}{2 \sw M_W} \big( \Gamma^{(1)}_h \camb + \Gamma^{(1)}_H \samb - \imag \Gamma^{(1)}_A \big), \\
    \Sigma^{(1),\text{notad}}_{G^-H^+}(p^2) - \frac{p^2}{M_W} \Sigma^{L,(1),\text{notad}}_{W^-H^+}(p^2) \overset{\order{\nc}}{=}{}& \frac{e}{2 \sw M_W} \big( \Gamma^{(1)}_h \camb + \Gamma^{(1)}_H \samb + \imag \Gamma^{(1)}_A \big),
\end{align}
\end{subequations}
which we have explicitly verified. The superscript `$\text{notad}$' denotes that we do not include tadpole contributions in the respective self-energy. Instead, the tadpole contributions appear explicitly in the form of unrenormalised one-point functions on the right-hand side of the equations.

Using on-shell definitions for the one-loop tadpole counterterms, $\deltaOL T_i = - \Gamma^{(1)}_i$, we can write
\begin{subequations}
\allowdisplaybreaks
\begin{align}
\label{eqn:AGZ_ST_identity}
    \Sigma^{(1),\text{notad}}_{AG}(p^2) - \imag \frac{p^2}{M_Z} \Sigma^{L,(1),\text{notad}}_{AZ}(p^2) \overset{\order{\nc}}{=}{}& \deltaOL m_{AG}^2 + \deltaOL \tb \cb^2 m_A^2, \\
    \Sigma^{(1),\text{notad}}_{H^-G^+}(p^2) - \frac{p^2}{M_W} \Sigma^{L,(1),\text{notad}}_{H^-W^+}(p^2) \overset{\order{\nc}}{=}{}& \deltaOL m_{H^-G^+}^2 + \deltaOL \tb \cb^2 m_{H^\pm}^2, \\
    \Sigma^{(1),\text{notad}}_{G^-H^+}(p^2) - \frac{p^2}{M_W} \Sigma^{L,(1),\text{notad}}_{W^-H^+}(p^2) \overset{\order{\nc}}{=}{}& \deltaOL m_{G^-H^+}^2 + \deltaOL \tb \cb^2 m_{H^\pm}^2,
\end{align}
\end{subequations}
where he have used the expressions for the one-loop Higgs mass counterterms which were introduced in \appx{app:CTs}.

For the renormalised self-energies (the definitions are given in \sct{ssec:tad_se_ren_1L}), we arrive at the following off-shell Slavnov-Taylor identities
\begin{subequations}
\label{eqn:ST_id_off-shell}
\allowdisplaybreaks
\begin{align}
    \hat \Sigma^{(1)}_{AG}(p^2) - \imag \frac{p^2}{M_Z} \hat \Sigma^{L,(1)}_{AZ}(p^2) \overset{\order{\nc}}{=}{}& (p^2 - m_A^2)\big(\tfrac{1}{2} \deltaOL Z_{AG} - \deltaOL \tb \cb^2 \big), \\
    \hat \Sigma^{(1)}_{H^-G^+}(p^2) - \frac{p^2}{M_W} \hat \Sigma^{L,(1)}_{H^-W^+}(p^2) \overset{\order{\nc}}{=}{}& (p^2 - m_{H^\pm}^2)\big(\tfrac{1}{2} \deltaOL Z_{H^-G^+} - \deltaOL \tb \cb^2 \big), \\
    \hat \Sigma^{(1)}_{G^-H^+}(p^2) - \frac{p^2}{M_W} \hat \Sigma^{L,(1)}_{W^-H^+}(p^2) \overset{\order{\nc}}{=}{}& (p^2 - m_{H^\pm}^2)\big(\tfrac{1}{2} \deltaOL Z_{G^-H^+} - \deltaOL \tb \cb^2 \big),
\end{align}
\end{subequations}
which are in agreement with the ones given in \citere{Baro:2008bg}.

The right-hand side of \eqs{eqn:ST_id_off-shell} vanishes in the $\DRbar$ scheme for any value of $p^2$, see \sct{sec:tanbeta_ren}. This is in agreement with \citere{Williams:2011bu}, which employs the $\DRbar$ version. As we allow for an on-shell renormalisation of $\tb$ while keeping the field counterterms defined in the minimal $\DRbar$ scheme, the right-hand side will not vanish in general. The on-shell ST identities
\begin{subequations}
\label{eqn:ST_id_on-shell}
\allowdisplaybreaks
\begin{align}
    \hat \Sigma^{(1)}_{AG}(m_A^2) - \imag \frac{m_A^2}{M_Z} \hat \Sigma^{L,(1)}_{AZ}(m_A^2) ={}& 0, \\
    \hat \Sigma^{(1)}_{H^-G^+}(m_{H^\pm}^2) - \frac{m_{H^\pm}^2}{M_W} \hat \Sigma^{L,(1)}_{H^-W^+}(m_{H^\pm}^2) ={}& 0, \\
    \hat \Sigma^{(1)}_{G^-H^+}(m_{H^\pm}^2) - \frac{m_{H^\pm}^2}{M_W} \hat \Sigma^{L,(1)}_{W^-H^+}(m_{H^\pm}^2) ={}& 0
\end{align}
\end{subequations}
hold independently of the renormalisation chosen for $\tb$ and the Higgs field counterterms. In a linear gauge, they are also valid if terms of $\order{\nc^0}$ are taken into account \cite{Baro:2008bg,Williams:2011bu}.

\section{One-loop integrals}
\label{app:loop_int}
For the definitions of the one-loop integrals, we follow the conventions used in \citere{Denner:1991kt}. Here, we provide additional details on the $B_0'$ integral, which is used less often than the $A_0$ and $B_0$ integrals. It is defined as
\begin{equation}
B_0'(p^2, m_1^2, m_2^2) \equiv \frac{\partial}{\partial m_1^2} B_0(p^2, m_1^2, m_2^2).
\end{equation}
To derive an expression for the $B_0'$ integral, we insert a factor $1 = D^{-1} \frac{\partial q^\mu}{\partial q^\mu}$ into the definition of the $B_0$ integral, and we integrate by parts:
\begin{equation}
    B_0(p^2, m_1^2, m_2^2) = - \frac{C}{D} \int \dd[D]{q} q^\mu \frac{\partial}{\partial q^\mu} \frac{1}{[q^2 - m_1^2][(q + p)^2 - m_2^2]},
\end{equation}
where we introduced the abbreviations
\begin{subequations}
\begin{align}
\label{eqn:loop_prefactor}
    C = \frac{16 \pi^2}{\imag} \frac{\Mudim^{2\del}}{(2\pi)^D}, \\
    D = 4 - 2\del.
\end{align}
\end{subequations}

Solving the resulting set of equations for $B_0'(p^2,m_1^2,m_2^2)$, we obtain
\begin{equation}
\begin{aligned}
    B_0'(p^2,m_1^2,m_2^2) ={}& \frac{-1}{\lambda(p^2, m_1^2, m_2^2)} \bigg[ (p^2 - m_1^2 + m_2^2)(1 - 2 \del) B_0(p^2,m_1^2,m_2^2) \\
    &- 2(1-\del) A_0(m_2^2) - \frac{p^2-m_1^2-m_2^2}{m_1^2}(1-\del) A_0(m_1^2) \bigg],
    \label{eqn:B0P_red}
\end{aligned}
\end{equation}
where $\lambda$ is the K\"{a}ll\'{e}n function, $\lambda(a,b,c) = a^2+b^2+c^2-2ab-2ac-2bc$. From \eqn{eqn:B0P_red}, a few special cases are readily derived:
\begin{subequations}
\allowdisplaybreaks
\begin{align}
    B_0'(p^2,m^2,m^2) ={}& \frac{(1-2\del) B_0(p^2,m^2,m^2)-B_0(0,m^2,m^2)}{4m^2-p^2}, \\
    B_0'(p^2,m^2,0) ={}& \frac{(1-2\del) B_0(p^2,m^2,0) - B_0(0,m^2,m^2)}{m^2-p^2}, \\
    B_0'(0,m^2,m^2) ={}& \frac{-\del}{2m^2} B_0(0,m^2,m^2) .
\end{align}
\end{subequations}
As we explained above, for identical mass arguments we take the derivative with respect to the first mass argument before setting the masses equal.

Setting $m_1^2 = 0$ and $m_2^2 = m^2$ in \eqn{eqn:B0P_red}, we obtain the relation
\begin{multline}
    B_0(0,0,0) - (p^2-m^2)B_0'(p^2,0,m^2) = 
    \frac{(p^2+m^2)(1-2\del)B_0(p^2,m^2,0)-2(1 - \del) A_0(m^2)}{p^2-m^2}.
\end{multline}
$B_0(0,0,0)$ and $B_0'(p^2,0,m^2)$ are both IR-divergent, while the integrals on the right-hand side are IR-finite.

\section{Generation of plots}
\label{app:PlotGen}
In this appendix, we explain how the different plots shown in \cha{cha:two-loop} were obtained. For any given point in parameter space, we always calculate the Higgs boson mass square at the one- ($1L$) and two-loop order ($2L$) including different generations of quarks and squarks:
\begin{itemize}
    \item{Only the third generation quarks ($t$ and $b$) and squarks are included ($3g$).}
    \item{Quarks and squarks of all generations are included ($ag$).}
\end{itemize}

We perform these calculations in several different limits:
\begin{itemize}
    \item{Full prediction at $\order{\nc}$/$\order{\nc^2}$ ($full$).}
    \item{Gaugeless limit, $\alem = 0$ ($gl$). We numerically take this limit by replacing $M_Z \to \widehat M_Z$ (where $\widehat M_Z$ is a dummy variable), $M_W \to M_W (\widehat M_Z/M_Z)$, and $\alem \to \alem (\widehat M_Z/M_Z)^2$, where $\widehat M_Z \ll M_Z$. In scenarios 1 and 2 we use $\widehat M_Z = M_S/1000$, in scenario 3 $\widehat M_Z = \max\{|A_q|/1000, 1\gev\}$}, and in scenario 4 we set $\widehat M_Z = M_S/500$.
    \item{Limit of vanishing bottom mass, $m_b = 0$ ($bl$). We employ this numerically by replacing $m_b \to m_b/10 = 0.418 \gev$.}
    \item{$\alem = 0$ and $m_b = 0$ ($gl+bl$). For this limit, we simply combine the aforementioned prescriptions for the gaugeless limit and the limit of vanishing bottom mass.}
\end{itemize}

We use the symbol $(M_{h_i}^2)^{l,g,c}$ for each of the different predictions for the squared Higgs boson masses, where
\begin{equation}
\begin{aligned}
    h_i \in{}& \{ h, H \}, \\
    l \in{}& \{ 1L, 2L \}, \\
    g \in{}& \{ 3g, ag \}, \\
    c \in{}& \{ full, gl, bl, gl+bl \}.
\end{aligned}
\end{equation}

As the first and second generation are assumed to be massless, there is no difference between working with the third or all generations in the gaugeless limit: 
\begin{subequations}
\allowdisplaybreaks
\begin{align}
    (M_{h_i}^2)^{l,3g,gl} ={}& (M_{h_i}^2)^{l,ag,gl}, \\
    (M_{h_i}^2)^{l,3g,gl+bl} ={}& (M_{h_i}^2)^{l,ag,gl+bl}.
\end{align}
\end{subequations}
We have used these identities to validate our implementation of the gaugeless limit. Nevertheless, we occasionally use gaugeless results with either one or three generations. This leaves us with six predictions for the Higgs boson mass at one-loop order\footnote{3g/full, 3g/gl, 3g/bl, 3g/gl+bl, ag/full, and ag/bl.} and the same number of predictions for the two-loop masses per parameter point. We combine them to obtain all contributions we are interested in.

In the plots for the Higgs boson masses in \cha{cha:two-loop} (these are \figs{fig:MH125_FO_overview}, \ref{fig:MH125_Mh1_overview}, and\ref{fig:MH125_Mh2_overview}), up to five curves are shown:
\begin{subequations}
\label{eqs:Mh2plots}
\allowdisplaybreaks
\begin{align}
    \text{cyan, solid: } M_{h_i} ={}& \sqrt{(M_{h_i}^2)^{2L,ag,full}}, \\
    \text{cyan, dashed: } M_{h_i} ={}& \sqrt{(M_{h_i}^2)^{2L,3g,full} - (M_{h_i}^2)^{1L,3g,full} + (M_{h_i}^2)^{1L,ag,full}}, \\
    \text{green, solid: } M_{h_i} ={}& \sqrt{(M_{h_i}^2)^{2L,ag,gl} - (M_{h_i}^2)^{1L,3g,gl} + (M_{h_i}^2)^{1L,ag,full}}, \\
    \text{green, dashed: } M_{h_i} ={}& \sqrt{(M_{h_i}^2)^{2L,3g,gl+bl} - (M_{h_i}^2)^{1L,3g,gl+bl} + (M_{h_i}^2)^{1L,ag,full}}, \\
    \text{black, solid: } M_{h_i} ={}& \sqrt{(M_{h_i}^2)^{1L,ag,full}}.
\end{align}
\end{subequations}

When we make a prediction for the two-loop mass in any given limit and for any number of fermion generations, the same properties are also applied to the tree-level and one-loop contribution. As we are interested in estimating the size of the newly calculated two-loop corrections, we subtract the appropriate one-loop prediction from the two-loop value and add the full $\order{N_c}$ one-loop result, which includes all generations of quarks and squarks.

In the plots for the two-loop contributions to the Higgs boson mass (these are \figs{fig:Aq0_big}, \ref{fig:Aqn2_big}, \ref{fig:Aqn2_sml}, \ref{fig:Aqvar_big}, \ref{fig:Aqvar_sml}, and \ref{fig:MH125_Mh1_ctrb}), twelve combinations are possible:
\begin{subequations}
\label{eqs:DeltaMhplots}
\allowdisplaybreaks
\begin{align}
    \text{cyan, solid: } \Delta^{(2)}M_{h_i} ={}& \sqrt{(M_{h_i}^2)^{2L,ag,full}} - \sqrt{(M_{h_i}^2)^{1L,ag,full}}, \\
    \begin{split}
    \text{cyan, dashed: } \Delta^{(2)}M_{h_i} ={}& \sqrt{(M_{h_i}^2)^{2L,3g,full} - (M_{h_i}^2)^{1L,3g,full} + (M_{h_i}^2)^{1L,ag,full}} \\
    &- \sqrt{(M_{h_i}^2)^{1L,ag,full}},
    \end{split} \\
    \begin{split}
    \text{magenta, solid: } \Delta^{(2)}M_{h_i} ={}& \sqrt{(M_{h_i}^2)^{2L,ag,bl} - (M_{h_i}^2)^{1L,ag,bl} + (M_{h_i}^2)^{1L,ag,full}} \\
    &- \sqrt{(M_{h_i}^2)^{1L,ag,full}},
    \end{split} \\
    \begin{split}
    \text{magenta, dashed: } \Delta^{(2)}M_{h_i} ={}& \sqrt{(M_{h_i}^2)^{2L,3g,bl} - (M_{h_i}^2)^{1L,3g,bl} + (M_{h_i}^2)^{1L,ag,full}} \\
    &- \sqrt{(M_{h_i}^2)^{1L,ag,full}},
    \end{split} \\
    \begin{split}
    \text{green, solid: } \Delta^{(2)}M_{h_i} ={}& \sqrt{(M_{h_i}^2)^{2L,ag,gl} - (M_{h_i}^2)^{1L,ag,gl} + (M_{h_i}^2)^{1L,ag,full}} \\
    &- \sqrt{(M_{h_i}^2)^{1L,ag,full}},
    \end{split} \\
    \begin{split}
    \text{green, dashed: } \Delta^{(2)}M_{h_i} ={}& \sqrt{(M_{h_i}^2)^{2L,3g,gl+bl} - (M_{h_i}^2)^{1L,3g,gl+bl} + (M_{h_i}^2)^{1L,ag,full}} \\
    &- \sqrt{(M_{h_i}^2)^{1L,ag,full}},
    \end{split} \\
    \begin{split}
    \text{green, dotted: } \Delta^{(2)}M_{h_i} ={}& \sqrt{(M_{h_i}^2)^{2L,ag,gl} - (M_{h_i}^2)^{1L,ag,gl} + (M_{h_i}^2)^{1L,ag,full}} \\ 
    &- \sqrt{(M_{h_i}^2)^{2L,3g,gl+bl} - (M_{h_i}^2)^{1L,3g,gl+bl} + (M_{h_i}^2)^{1L,ag,full}},
    \end{split} \\
    \begin{split}
    \text{red, solid: } \Delta^{(2)}M_{h_i} ={}& \sqrt{(M_{h_i}^2)^{2L,ag,full}} \\
    &- \sqrt{(M_{h_i}^2)^{2L,ag,gl} - (M_{h_i}^2)^{1L,ag,gl} + (M_{h_i}^2)^{1L,ag,full}},
    \end{split} \\
    \begin{split}
    \text{red, dashed: } \Delta^{(2)}M_{h_i} ={}& \sqrt{(M_{h_i}^2)^{2L,3g,full} - (M_{h_i}^2)^{1L,3g,full} + (M_{h_i}^2)^{1L,ag,full}} \\
    &- \sqrt{(M_{h_i}^2)^{2L,3g,gl} - (M_{h_i}^2)^{1L,3g,gl} + (M_{h_i}^2)^{1L,ag,full}},
    \end{split} \\
    \begin{split}
    \text{red, dotted: } \Delta^{(2)}M_{h_i} ={}& \sqrt{(M_{h_i}^2)^{2L,ag,full}} \\
    &- \sqrt{(M_{h_i}^2)^{2L,3g,full} - (M_{h_i}^2)^{1L,3g,full} + (M_{h_i}^2)^{1L,ag,full}},
    \end{split} \\
    \begin{split}
    \text{orange, solid: } \Delta^{(2)}M_{h_i} ={}& \sqrt{(M_{h_i}^2)^{2L,ag,full}} \\
    &- \sqrt{(M_{h_i}^2)^{2L,ag,bl} - (M_{h_i}^2)^{1L,ag,bl} + (M_{h_i}^2)^{1L,ag,full}},
    \end{split} \\
    \begin{split}
    \text{orange, dashed: } \Delta^{(2)}M_{h_i} ={}& \sqrt{(M_{h_i}^2)^{2L,3g,full}} \\
    &- \sqrt{(M_{h_i}^2)^{2L,3g,bl} - (M_{h_i}^2)^{1L,3g,bl} + (M_{h_i}^2)^{1L,ag,full}}.
    \end{split}
\end{align}
\end{subequations}

\clearpage

\addcontentsline{toc}{section}{Bibliography}
\bibliographystyle{jhep.bst}
\bibliography{bibliography}{}

\providecommand{\href}[2]{#2}\begingroup\raggedright\begin{thebibliography}{100}

\bibitem{ATLAS:2012yve}
{\scshape ATLAS} collaboration, \emph{{Observation of a new particle in the search for the Standard Model Higgs boson with the ATLAS detector at the LHC}}, \href{https://doi.org/10.1016/j.physletb.2012.08.020}{\emph{Phys. Lett. B} {\bfseries 716} (2012) 1} [\href{https://arxiv.org/abs/1207.7214}{{\ttfamily 1207.7214}}].

\bibitem{CMS:2012qbp}
{\scshape CMS} collaboration, \emph{{Observation of a New Boson at a Mass of 125 GeV with the CMS Experiment at the LHC}}, \href{https://doi.org/10.1016/j.physletb.2012.08.021}{\emph{Phys. Lett. B} {\bfseries 716} (2012) 30} [\href{https://arxiv.org/abs/1207.7235}{{\ttfamily 1207.7235}}].

\bibitem{ATLAS:2015yey}
{\scshape ATLAS, CMS} collaboration, \emph{{Combined Measurement of the Higgs Boson Mass in $pp$ Collisions at $\sqrt{s}=7$ and 8 TeV with the ATLAS and CMS Experiments}}, \href{https://doi.org/10.1103/PhysRevLett.114.191803}{\emph{Phys. Rev. Lett.} {\bfseries 114} (2015) 191803} [\href{https://arxiv.org/abs/1503.07589}{{\ttfamily 1503.07589}}].

\bibitem{ATLAS:2016neq}
{\scshape ATLAS, CMS} collaboration, \emph{{Measurements of the Higgs boson production and decay rates and constraints on its couplings from a combined ATLAS and CMS analysis of the LHC pp collision data at $ \sqrt{s}=7 $ and 8 TeV}}, \href{https://doi.org/10.1007/JHEP08(2016)045}{\emph{JHEP} {\bfseries 08} (2016) 045} [\href{https://arxiv.org/abs/1606.02266}{{\ttfamily 1606.02266}}].

\bibitem{ATLAS:2022vkf}
{\scshape ATLAS} collaboration, \emph{{A detailed map of Higgs boson interactions by the ATLAS experiment ten years after the discovery}}, \href{https://doi.org/10.1038/s41586-022-04893-w}{\emph{Nature} {\bfseries 607} (2022) 52} [\href{https://arxiv.org/abs/2207.00092}{{\ttfamily 2207.00092}}].

\bibitem{CMS:2022dwd}
{\scshape CMS} collaboration, \emph{{A portrait of the Higgs boson by the CMS experiment ten years after the discovery}}, \href{https://doi.org/10.1038/s41586-022-04892-x}{\emph{Nature} {\bfseries 607} (2022) 60} [\href{https://arxiv.org/abs/2207.00043}{{\ttfamily 2207.00043}}].

\bibitem{ATLAS:2023oaq}
{\scshape ATLAS} collaboration, \emph{{Combined measurement of the Higgs boson mass from the $H\to\gamma\gamma$ and $H\to ZZ^{*} \to 4\ell$ decay channels with the ATLAS detector using $\sqrt{s}$ = 7, 8 and 13 TeV $pp$ collision data}},  \href{https://arxiv.org/abs/2308.04775}{{\ttfamily 2308.04775}}.

\bibitem{ATLAS:2023owm}
{\scshape ATLAS} collaboration, \emph{{Measurement of the Higgs boson mass with $H \rightarrow \gamma\gamma$ decays in 140 fb$^{-1}$ of $\sqrt{s}=13$ TeV $pp$ collisions with the ATLAS detector}},  \href{https://arxiv.org/abs/2308.07216}{{\ttfamily 2308.07216}}.

\bibitem{CMSMhnew}
{\scshape CMS} collaboration, \emph{{Measurement of the Higgs boson mass and width using the four leptons final state}},  CMS PAS HIG-21-019 (2023).

\bibitem{Nilles:1983ge}
H.P.~Nilles, \emph{{Supersymmetry, Supergravity and Particle Physics}}, \href{https://doi.org/10.1016/0370-1573(84)90008-5}{\emph{Phys. Rept.} {\bfseries 110} (1984) 1}.

\bibitem{Haber:1984rc}
H.E.~Haber and G.L.~Kane, \emph{{The Search for Supersymmetry: Probing Physics Beyond the Standard Model}}, \href{https://doi.org/10.1016/0370-1573(85)90051-1}{\emph{Phys. Rept.} {\bfseries 117} (1985) 75}.

\bibitem{Inoue:1982ej}
K.~Inoue, A.~Kakuto, H.~Komatsu and S.~Takeshita, \emph{{Low-Energy Parameters and Particle Masses in a Supersymmetric Grand Unified Model}}, \href{https://doi.org/10.1143/PTP.67.1889}{\emph{Prog. Theor. Phys.} {\bfseries 67} (1982) 1889}.

\bibitem{Li:1984tc}
S.P.~Li and M.~Sher, \emph{{Upper Limit to the Lightest Higgs Mass in Supersymmetric Models}}, \href{https://doi.org/10.1016/0370-2693(84)90767-6}{\emph{Phys. Lett. B} {\bfseries 140} (1984) 339}.

\bibitem{Gunion:1989dp}
J.F.~Gunion and A.~Turski, \emph{{Corrections to Higgs Boson Mass Sum Rules from the Sfermion Sector of a Supersymmetric Model}}, \href{https://doi.org/10.1103/PhysRevD.40.2333}{\emph{Phys. Rev. D} {\bfseries 40} (1989) 2333}.

\bibitem{Berger:1989hg}
M.S.~Berger, \emph{{Radiative Corrections to Higgs Boson Mass Sum Rules in the Minimal Supersymmetric Extension to the Standard Model}}, \href{https://doi.org/10.1103/PhysRevD.41.225}{\emph{Phys. Rev. D} {\bfseries 41} (1990) 225}.

\bibitem{Okada:1990vk}
Y.~Okada, M.~Yamaguchi and T.~Yanagida, \emph{{Upper bound of the lightest Higgs boson mass in the minimal supersymmetric standard model}}, \href{https://doi.org/10.1143/ptp/85.1.1}{\emph{Prog. Theor. Phys.} {\bfseries 85} (1991) 1}.

\bibitem{Ellis:1990nz}
J.R.~Ellis, G.~Ridolfi and F.~Zwirner, \emph{{Radiative corrections to the masses of supersymmetric Higgs bosons}}, \href{https://doi.org/10.1016/0370-2693(91)90863-L}{\emph{Phys. Lett. B} {\bfseries 257} (1991) 83}.

\bibitem{Haber:1990aw}
H.E.~Haber and R.~Hempfling, \emph{{Can the mass of the lightest Higgs boson of the minimal supersymmetric model be larger than m(Z)?}}, \href{https://doi.org/10.1103/PhysRevLett.66.1815}{\emph{Phys. Rev. Lett.} {\bfseries 66} (1991) 1815}.

\bibitem{Barbieri:1991tk}
R.~Barbieri and M.~Frigeni, \emph{{The Supersymmetric Higgs searches at LEP after radiative corrections}}, \href{https://doi.org/10.1016/0370-2693(91)91106-6}{\emph{Phys. Lett. B} {\bfseries 258} (1991) 395}.

\bibitem{Ellis:1991zd}
J.R.~Ellis, G.~Ridolfi and F.~Zwirner, \emph{{On radiative corrections to supersymmetric Higgs boson masses and their implications for LEP searches}}, \href{https://doi.org/10.1016/0370-2693(91)90626-2}{\emph{Phys. Lett. B} {\bfseries 262} (1991) 477}.

\bibitem{Brignole:1991pq}
A.~Brignole, J.R.~Ellis, G.~Ridolfi and F.~Zwirner, \emph{{The supersymmetric charged Higgs boson mass and LEP phenomenology}}, \href{https://doi.org/10.1016/0370-2693(91)91287-6}{\emph{Phys. Lett. B} {\bfseries 271} (1991) 123}.

\bibitem{Chankowski:1991md}
P.H.~Chankowski, S.~Pokorski and J.~Rosiek, \emph{{Charged and neutral supersymmetric Higgs boson masses: Complete one loop analysis}}, \href{https://doi.org/10.1016/0370-2693(92)90522-6}{\emph{Phys. Lett. B} {\bfseries 274} (1992) 191}.

\bibitem{Brignole:1991wp}
A.~Brignole, \emph{{Radiative corrections to the supersymmetric charged Higgs boson mass}}, \href{https://doi.org/10.1016/0370-2693(92)90752-P}{\emph{Phys. Lett. B} {\bfseries 277} (1992) 313}.

\bibitem{Brignole:1992uf}
A.~Brignole, \emph{{Radiative corrections to the supersymmetric neutral Higgs boson masses}}, \href{https://doi.org/10.1016/0370-2693(92)91142-V}{\emph{Phys. Lett. B} {\bfseries 281} (1992) 284}.

\bibitem{Chankowski:1992er}
P.H.~Chankowski, S.~Pokorski and J.~Rosiek, \emph{{Complete on-shell renormalization scheme for the minimal supersymmetric Higgs sector}}, \href{https://doi.org/10.1016/0550-3213(94)90141-4}{\emph{Nucl. Phys. B} {\bfseries 423} (1994) 437} [\href{https://arxiv.org/abs/hep-ph/9303309}{{\ttfamily hep-ph/9303309}}].

\bibitem{Dabelstein:1994hb}
A.~Dabelstein, \emph{{The One loop renormalization of the MSSM Higgs sector and its application to the neutral scalar Higgs masses}}, \href{https://doi.org/10.1007/BF01624592}{\emph{Z. Phys. C} {\bfseries 67} (1995) 495} [\href{https://arxiv.org/abs/hep-ph/9409375}{{\ttfamily hep-ph/9409375}}].

\bibitem{Pierce:1996zz}
D.M.~Pierce, J.A.~Bagger, K.T.~Matchev and R.-j.~Zhang, \emph{{Precision corrections in the minimal supersymmetric standard model}}, \href{https://doi.org/10.1016/S0550-3213(96)00683-9}{\emph{Nucl. Phys. B} {\bfseries 491} (1997) 3} [\href{https://arxiv.org/abs/hep-ph/9606211}{{\ttfamily hep-ph/9606211}}].

\bibitem{Hempfling:1993qq}
R.~Hempfling and A.H.~Hoang, \emph{{Two loop radiative corrections to the upper limit of the lightest Higgs boson mass in the minimal supersymmetric model}}, \href{https://doi.org/10.1016/0370-2693(94)90948-2}{\emph{Phys. Lett. B} {\bfseries 331} (1994) 99} [\href{https://arxiv.org/abs/hep-ph/9401219}{{\ttfamily hep-ph/9401219}}].

\bibitem{Heinemeyer:1998jw}
S.~Heinemeyer, W.~Hollik and G.~Weiglein, \emph{{QCD corrections to the masses of the neutral CP - even Higgs bosons in the MSSM}}, \href{https://doi.org/10.1103/PhysRevD.58.091701}{\emph{Phys. Rev. D} {\bfseries 58} (1998) 091701} [\href{https://arxiv.org/abs/hep-ph/9803277}{{\ttfamily hep-ph/9803277}}].

\bibitem{Heinemeyer:1998kz}
S.~Heinemeyer, W.~Hollik and G.~Weiglein, \emph{{Precise prediction for the mass of the lightest Higgs boson in the MSSM}}, \href{https://doi.org/10.1016/S0370-2693(98)01116-2}{\emph{Phys. Lett. B} {\bfseries 440} (1998) 296} [\href{https://arxiv.org/abs/hep-ph/9807423}{{\ttfamily hep-ph/9807423}}].

\bibitem{Zhang:1998bm}
R.-J.~Zhang, \emph{{Two loop effective potential calculation of the lightest CP even Higgs boson mass in the MSSM}}, \href{https://doi.org/10.1016/S0370-2693(98)01575-5}{\emph{Phys. Lett. B} {\bfseries 447} (1999) 89} [\href{https://arxiv.org/abs/hep-ph/9808299}{{\ttfamily hep-ph/9808299}}].

\bibitem{Heinemeyer:1998np}
S.~Heinemeyer, W.~Hollik and G.~Weiglein, \emph{{The Masses of the neutral CP - even Higgs bosons in the MSSM: Accurate analysis at the two loop level}}, \href{https://doi.org/10.1007/s100529900006}{\emph{Eur. Phys. J. C} {\bfseries 9} (1999) 343} [\href{https://arxiv.org/abs/hep-ph/9812472}{{\ttfamily hep-ph/9812472}}].

\bibitem{Espinosa:1999zm}
J.R.~Espinosa and R.-J.~Zhang, \emph{{MSSM lightest CP even Higgs boson mass to O(alpha(s) alpha(t)): The Effective potential approach}}, \href{https://doi.org/10.1088/1126-6708/2000/03/026}{\emph{JHEP} {\bfseries 03} (2000) 026} [\href{https://arxiv.org/abs/hep-ph/9912236}{{\ttfamily hep-ph/9912236}}].

\bibitem{Carena:2000dp}
M.~Carena, H.E.~Haber, S.~Heinemeyer, W.~Hollik, C.E.M.~Wagner and G.~Weiglein, \emph{{Reconciling the two loop diagrammatic and effective field theory computations of the mass of the lightest CP - even Higgs boson in the MSSM}}, \href{https://doi.org/10.1016/S0550-3213(00)00212-1}{\emph{Nucl. Phys. B} {\bfseries 580} (2000) 29} [\href{https://arxiv.org/abs/hep-ph/0001002}{{\ttfamily hep-ph/0001002}}].

\bibitem{Espinosa:2000df}
J.R.~Espinosa and R.-J.~Zhang, \emph{{Complete two loop dominant corrections to the mass of the lightest CP even Higgs boson in the minimal supersymmetric standard model}}, \href{https://doi.org/10.1016/S0550-3213(00)00421-1}{\emph{Nucl. Phys. B} {\bfseries 586} (2000) 3} [\href{https://arxiv.org/abs/hep-ph/0003246}{{\ttfamily hep-ph/0003246}}].

\bibitem{Degrassi:2001yf}
G.~Degrassi, P.~Slavich and F.~Zwirner, \emph{{On the neutral Higgs boson masses in the MSSM for arbitrary stop mixing}}, \href{https://doi.org/10.1016/S0550-3213(01)00343-1}{\emph{Nucl. Phys. B} {\bfseries 611} (2001) 403} [\href{https://arxiv.org/abs/hep-ph/0105096}{{\ttfamily hep-ph/0105096}}].

\bibitem{Brignole:2001jy}
A.~Brignole, G.~Degrassi, P.~Slavich and F.~Zwirner, \emph{{On the O(alpha(t)**2) two loop corrections to the neutral Higgs boson masses in the MSSM}}, \href{https://doi.org/10.1016/S0550-3213(02)00184-0}{\emph{Nucl. Phys. B} {\bfseries 631} (2002) 195} [\href{https://arxiv.org/abs/hep-ph/0112177}{{\ttfamily hep-ph/0112177}}].

\bibitem{Brignole:2002bz}
A.~Brignole, G.~Degrassi, P.~Slavich and F.~Zwirner, \emph{{On the two loop sbottom corrections to the neutral Higgs boson masses in the MSSM}}, \href{https://doi.org/10.1016/S0550-3213(02)00748-4}{\emph{Nucl. Phys. B} {\bfseries 643} (2002) 79} [\href{https://arxiv.org/abs/hep-ph/0206101}{{\ttfamily hep-ph/0206101}}].

\bibitem{Dedes:2002dy}
A.~Dedes and P.~Slavich, \emph{{Two loop corrections to radiative electroweak symmetry breaking in the MSSM}}, \href{https://doi.org/10.1016/S0550-3213(03)00173-1}{\emph{Nucl. Phys. B} {\bfseries 657} (2003) 333} [\href{https://arxiv.org/abs/hep-ph/0212132}{{\ttfamily hep-ph/0212132}}].

\bibitem{Dedes:2003km}
A.~Dedes, G.~Degrassi and P.~Slavich, \emph{{On the two loop Yukawa corrections to the MSSM Higgs boson masses at large tan beta}}, \href{https://doi.org/10.1016/j.nuclphysb.2003.08.033}{\emph{Nucl. Phys. B} {\bfseries 672} (2003) 144} [\href{https://arxiv.org/abs/hep-ph/0305127}{{\ttfamily hep-ph/0305127}}].

\bibitem{Allanach:2004rh}
B.C.~Allanach, A.~Djouadi, J.L.~Kneur, W.~Porod and P.~Slavich, \emph{{Precise determination of the neutral Higgs boson masses in the MSSM}}, \href{https://doi.org/10.1088/1126-6708/2004/09/044}{\emph{JHEP} {\bfseries 09} (2004) 044} [\href{https://arxiv.org/abs/hep-ph/0406166}{{\ttfamily hep-ph/0406166}}].

\bibitem{Heinemeyer:2004xw}
S.~Heinemeyer, W.~Hollik, H.~Rzehak and G.~Weiglein, \emph{{High-precision predictions for the MSSM Higgs sector at O(alpha(b) alpha(s))}}, \href{https://doi.org/10.1140/epjc/s2005-02112-6}{\emph{Eur. Phys. J. C} {\bfseries 39} (2005) 465} [\href{https://arxiv.org/abs/hep-ph/0411114}{{\ttfamily hep-ph/0411114}}].

\bibitem{Frank:2013hba}
M.~Frank, L.~Galeta, T.~Hahn, S.~Heinemeyer, W.~Hollik, H.~Rzehak et~al., \emph{{Charged Higgs Boson Mass of the MSSM in the Feynman Diagrammatic Approach}}, \href{https://doi.org/10.1103/PhysRevD.88.055013}{\emph{Phys. Rev. D} {\bfseries 88} (2013) 055013} [\href{https://arxiv.org/abs/1306.1156}{{\ttfamily 1306.1156}}].

\bibitem{Hollik:2015ema}
W.~Hollik and S.~Pa{\ss}ehr, \emph{{Two-loop top-Yukawa-coupling corrections to the charged Higgs-boson mass in the MSSM}}, \href{https://doi.org/10.1140/epjc/s10052-015-3558-7}{\emph{Eur. Phys. J. C} {\bfseries 75} (2015) 336} [\href{https://arxiv.org/abs/1502.02394}{{\ttfamily 1502.02394}}].

\bibitem{Martin:2002wn}
S.P.~Martin, \emph{{Complete Two Loop Effective Potential Approximation to the Lightest Higgs Scalar Boson Mass in Supersymmetry}}, \href{https://doi.org/10.1103/PhysRevD.67.095012}{\emph{Phys. Rev. D} {\bfseries 67} (2003) 095012} [\href{https://arxiv.org/abs/hep-ph/0211366}{{\ttfamily hep-ph/0211366}}].

\bibitem{Martin:2002iu}
S.P.~Martin, \emph{{Two Loop Effective Potential for the Minimal Supersymmetric Standard Model}}, \href{https://doi.org/10.1103/PhysRevD.66.096001}{\emph{Phys. Rev. D} {\bfseries 66} (2002) 096001} [\href{https://arxiv.org/abs/hep-ph/0206136}{{\ttfamily hep-ph/0206136}}].

\bibitem{Martin:2004kr}
S.P.~Martin, \emph{{Strong and Yukawa two-loop contributions to Higgs scalar boson self-energies and pole masses in supersymmetry}}, \href{https://doi.org/10.1103/PhysRevD.71.016012}{\emph{Phys. Rev. D} {\bfseries 71} (2005) 016012} [\href{https://arxiv.org/abs/hep-ph/0405022}{{\ttfamily hep-ph/0405022}}].

\bibitem{Martin:2003qz}
S.P.~Martin, \emph{{Evaluation of two loop selfenergy basis integrals using differential equations}}, \href{https://doi.org/10.1103/PhysRevD.68.075002}{\emph{Phys. Rev. D} {\bfseries 68} (2003) 075002} [\href{https://arxiv.org/abs/hep-ph/0307101}{{\ttfamily hep-ph/0307101}}].

\bibitem{Martin:2005qm}
S.P.~Martin and D.G.~Robertson, \emph{{TSIL: A Program for the calculation of two-loop self-energy integrals}}, \href{https://doi.org/10.1016/j.cpc.2005.08.005}{\emph{Comput. Phys. Commun.} {\bfseries 174} (2006) 133} [\href{https://arxiv.org/abs/hep-ph/0501132}{{\ttfamily hep-ph/0501132}}].

\bibitem{Heinemeyer:2010mm}
S.~Heinemeyer, H.~Rzehak and C.~Schappacher, \emph{{Proposals for Bottom Quark/Squark Renormalization in the Complex MSSM}}, \href{https://doi.org/10.1103/PhysRevD.82.075010}{\emph{Phys. Rev. D} {\bfseries 82} (2010) 075010} [\href{https://arxiv.org/abs/1007.0689}{{\ttfamily 1007.0689}}].

\bibitem{Fritzsche:2011nr}
T.~Fritzsche, S.~Heinemeyer, H.~Rzehak and C.~Schappacher, \emph{{Heavy Scalar Top Quark Decays in the Complex MSSM: A Full One-Loop Analysis}}, \href{https://doi.org/10.1103/PhysRevD.86.035014}{\emph{Phys. Rev. D} {\bfseries 86} (2012) 035014} [\href{https://arxiv.org/abs/1111.7289}{{\ttfamily 1111.7289}}].

\bibitem{Fritzsche:2013fta}
T.~Fritzsche, T.~Hahn, S.~Heinemeyer, F.~von~der Pahlen, H.~Rzehak and C.~Schappacher, \emph{{The Implementation of the Renormalized Complex MSSM in FeynArts and FormCalc}}, \href{https://doi.org/10.1016/j.cpc.2014.02.005}{\emph{Comput. Phys. Commun.} {\bfseries 185} (2014) 1529} [\href{https://arxiv.org/abs/1309.1692}{{\ttfamily 1309.1692}}].

\bibitem{Borowka:2014wla}
S.~Borowka, T.~Hahn, S.~Heinemeyer, G.~Heinrich and W.~Hollik, \emph{{Momentum-dependent two-loop QCD corrections to the neutral Higgs-boson masses in the MSSM}}, \href{https://doi.org/10.1140/epjc/s10052-014-2994-0}{\emph{Eur. Phys. J. C} {\bfseries 74} (2014) 2994} [\href{https://arxiv.org/abs/1404.7074}{{\ttfamily 1404.7074}}].

\bibitem{Degrassi:2014pfa}
G.~Degrassi, S.~Di~Vita and P.~Slavich, \emph{{Two-loop QCD corrections to the MSSM Higgs masses beyond the effective-potential approximation}}, \href{https://doi.org/10.1140/epjc/s10052-015-3280-5}{\emph{Eur. Phys. J. C} {\bfseries 75} (2015) 61} [\href{https://arxiv.org/abs/1410.3432}{{\ttfamily 1410.3432}}].

\bibitem{Carter:2010hi}
J.~Carter and G.~Heinrich, \emph{{SecDec: A general program for sector decomposition}}, \href{https://doi.org/10.1016/j.cpc.2011.03.026}{\emph{Comput. Phys. Commun.} {\bfseries 182} (2011) 1566} [\href{https://arxiv.org/abs/1011.5493}{{\ttfamily 1011.5493}}].

\bibitem{Borowka:2012yc}
S.~Borowka, J.~Carter and G.~Heinrich, \emph{{Numerical Evaluation of Multi-Loop Integrals for Arbitrary Kinematics with SecDec 2.0}}, \href{https://doi.org/10.1016/j.cpc.2012.09.020}{\emph{Comput. Phys. Commun.} {\bfseries 184} (2013) 396} [\href{https://arxiv.org/abs/1204.4152}{{\ttfamily 1204.4152}}].

\bibitem{Borowka:2015ura}
S.~Borowka, T.~Hahn, S.~Heinemeyer, G.~Heinrich and W.~Hollik, \emph{{Renormalization scheme dependence of the two-loop QCD corrections to the neutral Higgs-boson masses in the MSSM}}, \href{https://doi.org/10.1140/epjc/s10052-015-3648-6}{\emph{Eur. Phys. J. C} {\bfseries 75} (2015) 424} [\href{https://arxiv.org/abs/1505.03133}{{\ttfamily 1505.03133}}].

\bibitem{Borowka:2018anu}
S.~Borowka, S.~Pa{\ss}ehr and G.~Weiglein, \emph{{Complete two-loop QCD contributions to the lightest Higgs-boson mass in the MSSM with complex parameters}}, \href{https://doi.org/10.1140/epjc/s10052-018-6055-y}{\emph{Eur. Phys. J. C} {\bfseries 78} (2018) 576} [\href{https://arxiv.org/abs/1802.09886}{{\ttfamily 1802.09886}}].

\bibitem{Pilaftsis:1998dd}
A.~Pilaftsis, \emph{{Higgs scalar - pseudoscalar mixing in the minimal supersymmetric standard model}}, \href{https://doi.org/10.1016/S0370-2693(98)00771-0}{\emph{Phys. Lett. B} {\bfseries 435} (1998) 88} [\href{https://arxiv.org/abs/hep-ph/9805373}{{\ttfamily hep-ph/9805373}}].

\bibitem{Demir:1999hj}
D.A.~Demir, \emph{{Effects of the supersymmetric phases on the neutral Higgs sector}}, \href{https://doi.org/10.1103/PhysRevD.60.055006}{\emph{Phys. Rev. D} {\bfseries 60} (1999) 055006} [\href{https://arxiv.org/abs/hep-ph/9901389}{{\ttfamily hep-ph/9901389}}].

\bibitem{Pilaftsis:1999qt}
A.~Pilaftsis and C.E.M.~Wagner, \emph{{Higgs bosons in the minimal supersymmetric standard model with explicit CP violation}}, \href{https://doi.org/10.1016/S0550-3213(99)00261-8}{\emph{Nucl. Phys. B} {\bfseries 553} (1999) 3} [\href{https://arxiv.org/abs/hep-ph/9902371}{{\ttfamily hep-ph/9902371}}].

\bibitem{Choi:2000wz}
S.Y.~Choi, M.~Drees and J.S.~Lee, \emph{{Loop corrections to the neutral Higgs boson sector of the MSSM with explicit CP violation}}, \href{https://doi.org/10.1016/S0370-2693(00)00421-4}{\emph{Phys. Lett. B} {\bfseries 481} (2000) 57} [\href{https://arxiv.org/abs/hep-ph/0002287}{{\ttfamily hep-ph/0002287}}].

\bibitem{Carena:2000yi}
M.~Carena, J.R.~Ellis, A.~Pilaftsis and C.E.M.~Wagner, \emph{{Renormalization group improved effective potential for the MSSM Higgs sector with explicit CP violation}}, \href{https://doi.org/10.1016/S0550-3213(00)00358-8}{\emph{Nucl. Phys. B} {\bfseries 586} (2000) 92} [\href{https://arxiv.org/abs/hep-ph/0003180}{{\ttfamily hep-ph/0003180}}].

\bibitem{Ibrahim:2000qj}
T.~Ibrahim and P.~Nath, \emph{{Corrections to the Higgs boson masses and mixings from chargino, W and charged Higgs exchange loops and large CP phases}}, \href{https://doi.org/10.1103/PhysRevD.63.035009}{\emph{Phys. Rev. D} {\bfseries 63} (2001) 035009} [\href{https://arxiv.org/abs/hep-ph/0008237}{{\ttfamily hep-ph/0008237}}].

\bibitem{Heinemeyer:2001qd}
S.~Heinemeyer, \emph{{The Higgs boson sector of the complex MSSM in the Feynman diagrammatic approach}}, \href{https://doi.org/10.1007/s100520100819}{\emph{Eur. Phys. J. C} {\bfseries 22} (2001) 521} [\href{https://arxiv.org/abs/hep-ph/0108059}{{\ttfamily hep-ph/0108059}}].

\bibitem{Carena:2001fw}
M.~Carena, J.R.~Ellis, A.~Pilaftsis and C.E.M.~Wagner, \emph{{Higgs Boson Pole Masses in the MSSM with Explicit CP Violation}}, \href{https://doi.org/10.1016/S0550-3213(02)00014-7}{\emph{Nucl. Phys. B} {\bfseries 625} (2002) 345} [\href{https://arxiv.org/abs/hep-ph/0111245}{{\ttfamily hep-ph/0111245}}].

\bibitem{Ibrahim:2002zk}
T.~Ibrahim and P.~Nath, \emph{{Neutralino exchange corrections to the Higgs boson mixings with explicit CP violation}}, \href{https://doi.org/10.1103/PhysRevD.66.015005}{\emph{Phys. Rev. D} {\bfseries 66} (2002) 015005} [\href{https://arxiv.org/abs/hep-ph/0204092}{{\ttfamily hep-ph/0204092}}].

\bibitem{Ellis:2004fs}
J.R.~Ellis, J.S.~Lee and A.~Pilaftsis, \emph{{CERN LHC signatures of resonant CP violation in a minimal supersymmetric Higgs sector}}, \href{https://doi.org/10.1103/PhysRevD.70.075010}{\emph{Phys. Rev. D} {\bfseries 70} (2004) 075010} [\href{https://arxiv.org/abs/hep-ph/0404167}{{\ttfamily hep-ph/0404167}}].

\bibitem{Frank:2006yh}
M.~Frank, T.~Hahn, S.~Heinemeyer, W.~Hollik, H.~Rzehak and G.~Weiglein, \emph{{The Higgs Boson Masses and Mixings of the Complex MSSM in the Feynman-Diagrammatic Approach}}, \href{https://doi.org/10.1088/1126-6708/2007/02/047}{\emph{JHEP} {\bfseries 02} (2007) 047} [\href{https://arxiv.org/abs/hep-ph/0611326}{{\ttfamily hep-ph/0611326}}].

\bibitem{Heinemeyer:2007aq}
S.~Heinemeyer, W.~Hollik, H.~Rzehak and G.~Weiglein, \emph{{The Higgs sector of the complex MSSM at two-loop order: QCD contributions}}, \href{https://doi.org/10.1016/j.physletb.2007.07.030}{\emph{Phys. Lett. B} {\bfseries 652} (2007) 300} [\href{https://arxiv.org/abs/0705.0746}{{\ttfamily 0705.0746}}].

\bibitem{Heinemeyer:2004by}
S.~Heinemeyer, W.~Hollik, F.~Merz and S.~Penaranda, \emph{{Electroweak precision observables in the MSSM with nonminimal flavor violation}}, \href{https://doi.org/10.1140/epjc/s2004-02006-1}{\emph{Eur. Phys. J. C} {\bfseries 37} (2004) 481} [\href{https://arxiv.org/abs/hep-ph/0403228}{{\ttfamily hep-ph/0403228}}].

\bibitem{Cao:2006xb}
J.~Cao, G.~Eilam, K.-i.~Hikasa and J.M.~Yang, \emph{{Experimental constraints on stop-scharm flavor mixing and implications in top-quark FCNC processes}}, \href{https://doi.org/10.1103/PhysRevD.74.031701}{\emph{Phys. Rev. D} {\bfseries 74} (2006) 031701} [\href{https://arxiv.org/abs/hep-ph/0604163}{{\ttfamily hep-ph/0604163}}].

\bibitem{Brignole:2015kva}
A.~Brignole, \emph{{The supersymmetric Higgs boson with flavoured A-terms}}, \href{https://doi.org/10.1016/j.nuclphysb.2015.07.025}{\emph{Nucl. Phys. B} {\bfseries 898} (2015) 644} [\href{https://arxiv.org/abs/1504.03273}{{\ttfamily 1504.03273}}].

\bibitem{Arana-Catania:2011rnb}
M.~Arana-Catania, S.~Heinemeyer, M.J.~Herrero and S.~Penaranda, \emph{{Higgs Boson masses and B-Physics Constraints in Non-Minimal Flavor Violating SUSY scenarios}}, \href{https://doi.org/10.1007/JHEP05(2012)015}{\emph{JHEP} {\bfseries 05} (2012) 015} [\href{https://arxiv.org/abs/1109.6232}{{\ttfamily 1109.6232}}].

\bibitem{Gomez:2014uha}
M.E.~G\'omez, T.~Hahn, S.~Heinemeyer and M.~Rehman, \emph{{Higgs masses and Electroweak Precision Observables in the Lepton-Flavor-Violating MSSM}}, \href{https://doi.org/10.1103/PhysRevD.90.074016}{\emph{Phys. Rev. D} {\bfseries 90} (2014) 074016} [\href{https://arxiv.org/abs/1408.0663}{{\ttfamily 1408.0663}}].

\bibitem{Hollik:2014wea}
W.~Hollik and S.~Pa\ss{}ehr, \emph{{Two-loop top-Yukawa-coupling corrections to the Higgs boson masses in the complex MSSM}}, \href{https://doi.org/10.1016/j.physletb.2014.04.026}{\emph{Phys. Lett. B} {\bfseries 733} (2014) 144} [\href{https://arxiv.org/abs/1401.8275}{{\ttfamily 1401.8275}}].

\bibitem{Hollik:2014bua}
W.~Hollik and S.~Pa\ss{}ehr, \emph{{Higgs boson masses and mixings in the complex MSSM with two-loop top-Yukawa-coupling corrections}}, \href{https://doi.org/10.1007/JHEP10(2014)171}{\emph{JHEP} {\bfseries 10} (2014) 171} [\href{https://arxiv.org/abs/1409.1687}{{\ttfamily 1409.1687}}].

\bibitem{Hahn:2015gaa}
T.~Hahn and S.~Pa\ss{}ehr, \emph{{Implementation of the $\mathcal{O}{\left(\alpha_t^2\right)}$ MSSM Higgs-mass corrections in $\texttt{FeynHiggs}$}}, \href{https://doi.org/10.1016/j.cpc.2017.01.026}{\emph{Comput. Phys. Commun.} {\bfseries 214} (2017) 91} [\href{https://arxiv.org/abs/1508.00562}{{\ttfamily 1508.00562}}].

\bibitem{Passehr:2017ufr}
S.~Pa{\ss}ehr and G.~Weiglein, \emph{{Two-loop top and bottom Yukawa corrections to the Higgs-boson masses in the complex MSSM}}, \href{https://doi.org/10.1140/epjc/s10052-018-5665-8}{\emph{Eur. Phys. J. C} {\bfseries 78} (2018) 222} [\href{https://arxiv.org/abs/1705.07909}{{\ttfamily 1705.07909}}].

\bibitem{Goodsell:2019zfs}
M.D.~Goodsell and S.~Pa\ss{}ehr, \emph{{All two-loop scalar self-energies and tadpoles in general renormalisable field theories}}, \href{https://doi.org/10.1140/epjc/s10052-020-7657-8}{\emph{Eur. Phys. J. C} {\bfseries 80} (2020) 417} [\href{https://arxiv.org/abs/1910.02094}{{\ttfamily 1910.02094}}].

\bibitem{Domingo:2021kud}
F.~Domingo and S.~Pa\ss{}ehr, \emph{{Fighting off field dependence in MSSM Higgs-mass corrections of order $\alpha _t\,\alpha _s$ and $\alpha _t^2$}}, \href{https://doi.org/10.1140/epjc/s10052-021-09450-9}{\emph{Eur. Phys. J. C} {\bfseries 81} (2021) 661} [\href{https://arxiv.org/abs/2105.01139}{{\ttfamily 2105.01139}}].

\bibitem{R:2021bml}
E.A.~Reyes~R. and R.~Fazio, \emph{{High-Precision Calculations of the Higgs Boson Mass}}, \href{https://doi.org/10.3390/particles5010006}{\emph{Particles} {\bfseries 5} (2022) 53} [\href{https://arxiv.org/abs/2112.15295}{{\ttfamily 2112.15295}}].

\bibitem{Barbieri:1990ja}
R.~Barbieri, M.~Frigeni and F.~Caravaglios, \emph{{The Supersymmetric Higgs for heavy superpartners}}, \href{https://doi.org/10.1016/0370-2693(91)91226-L}{\emph{Phys. Lett. B} {\bfseries 258} (1991) 167}.

\bibitem{Espinosa:1991fc}
J.R.~Espinosa and M.~Quiros, \emph{{Two loop radiative corrections to the mass of the lightest Higgs boson in supersymmetric standard models}}, \href{https://doi.org/10.1016/0370-2693(91)91056-2}{\emph{Phys. Lett. B} {\bfseries 266} (1991) 389}.

\bibitem{Casas:1994us}
J.A.~Casas, J.R.~Espinosa, M.~Quiros and A.~Riotto, \emph{{The Lightest Higgs boson mass in the minimal supersymmetric standard model}}, \href{https://doi.org/10.1016/0550-3213(94)00508-C}{\emph{Nucl. Phys. B} {\bfseries 436} (1995) 3} [\href{https://arxiv.org/abs/hep-ph/9407389}{{\ttfamily hep-ph/9407389}}].

\bibitem{Haber:1993an}
H.E.~Haber and R.~Hempfling, \emph{{The Renormalization group improved Higgs sector of the minimal supersymmetric model}}, \href{https://doi.org/10.1103/PhysRevD.48.4280}{\emph{Phys. Rev. D} {\bfseries 48} (1993) 4280} [\href{https://arxiv.org/abs/hep-ph/9307201}{{\ttfamily hep-ph/9307201}}].

\bibitem{Carena:1995bx}
M.~Carena, J.R.~Espinosa, M.~Quiros and C.E.M.~Wagner, \emph{{Analytical expressions for radiatively corrected Higgs masses and couplings in the MSSM}}, \href{https://doi.org/10.1016/0370-2693(95)00694-G}{\emph{Phys. Lett. B} {\bfseries 355} (1995) 209} [\href{https://arxiv.org/abs/hep-ph/9504316}{{\ttfamily hep-ph/9504316}}].

\bibitem{Carena:1995wu}
M.~Carena, M.~Quiros and C.E.M.~Wagner, \emph{{Effective potential methods and the Higgs mass spectrum in the MSSM}}, \href{https://doi.org/10.1016/0550-3213(95)00665-6}{\emph{Nucl. Phys. B} {\bfseries 461} (1996) 407} [\href{https://arxiv.org/abs/hep-ph/9508343}{{\ttfamily hep-ph/9508343}}].

\bibitem{Haber:1996fp}
H.E.~Haber, R.~Hempfling and A.H.~Hoang, \emph{{Approximating the radiatively corrected Higgs mass in the minimal supersymmetric model}}, \href{https://doi.org/10.1007/s002880050498}{\emph{Z. Phys. C} {\bfseries 75} (1997) 539} [\href{https://arxiv.org/abs/hep-ph/9609331}{{\ttfamily hep-ph/9609331}}].

\bibitem{Degrassi:2002fi}
G.~Degrassi, S.~Heinemeyer, W.~Hollik, P.~Slavich and G.~Weiglein, \emph{{Towards high precision predictions for the MSSM Higgs sector}}, \href{https://doi.org/10.1140/epjc/s2003-01152-2}{\emph{Eur. Phys. J. C} {\bfseries 28} (2003) 133} [\href{https://arxiv.org/abs/hep-ph/0212020}{{\ttfamily hep-ph/0212020}}].

\bibitem{Martin:2007pg}
S.P.~Martin, \emph{{Three-loop corrections to the lightest Higgs scalar boson mass in supersymmetry}}, \href{https://doi.org/10.1103/PhysRevD.75.055005}{\emph{Phys. Rev. D} {\bfseries 75} (2007) 055005} [\href{https://arxiv.org/abs/hep-ph/0701051}{{\ttfamily hep-ph/0701051}}].

\bibitem{Hahn:2013ria}
T.~Hahn, S.~Heinemeyer, W.~Hollik, H.~Rzehak and G.~Weiglein, \emph{{High-Precision Predictions for the Light CP -Even Higgs Boson Mass of the Minimal Supersymmetric Standard Model}}, \href{https://doi.org/10.1103/PhysRevLett.112.141801}{\emph{Phys. Rev. Lett.} {\bfseries 112} (2014) 141801} [\href{https://arxiv.org/abs/1312.4937}{{\ttfamily 1312.4937}}].

\bibitem{Draper:2013oza}
P.~Draper, G.~Lee and C.E.M.~Wagner, \emph{{Precise estimates of the Higgs mass in heavy supersymmetry}}, \href{https://doi.org/10.1103/PhysRevD.89.055023}{\emph{Phys. Rev. D} {\bfseries 89} (2014) 055023} [\href{https://arxiv.org/abs/1312.5743}{{\ttfamily 1312.5743}}].

\bibitem{Arkani-Hamed:2004ymt}
N.~Arkani-Hamed and S.~Dimopoulos, \emph{{Supersymmetric unification without low energy supersymmetry and signatures for fine-tuning at the LHC}}, \href{https://doi.org/10.1088/1126-6708/2005/06/073}{\emph{JHEP} {\bfseries 06} (2005) 073} [\href{https://arxiv.org/abs/hep-th/0405159}{{\ttfamily hep-th/0405159}}].

\bibitem{Giudice:2004tc}
G.F.~Giudice and A.~Romanino, \emph{{Split supersymmetry}}, \href{https://doi.org/10.1016/j.nuclphysb.2004.08.001}{\emph{Nucl. Phys. B} {\bfseries 699} (2004) 65} [\href{https://arxiv.org/abs/hep-ph/0406088}{{\ttfamily hep-ph/0406088}}].

\bibitem{Carena:2008rt}
M.~Carena, G.~Nardini, M.~Quiros and C.E.M.~Wagner, \emph{{The Effective Theory of the Light Stop Scenario}}, \href{https://doi.org/10.1088/1126-6708/2008/10/062}{\emph{JHEP} {\bfseries 10} (2008) 062} [\href{https://arxiv.org/abs/0806.4297}{{\ttfamily 0806.4297}}].

\bibitem{Binger:2004nn}
M.~Binger, \emph{{Higgs boson mass in split supersymmetry at two-loops}}, \href{https://doi.org/10.1103/PhysRevD.73.095001}{\emph{Phys. Rev. D} {\bfseries 73} (2006) 095001} [\href{https://arxiv.org/abs/hep-ph/0408240}{{\ttfamily hep-ph/0408240}}].

\bibitem{Bernal:2007uv}
N.~Bernal, A.~Djouadi and P.~Slavich, \emph{{The MSSM with heavy scalars}}, \href{https://doi.org/10.1088/1126-6708/2007/07/016}{\emph{JHEP} {\bfseries 07} (2007) 016} [\href{https://arxiv.org/abs/0705.1496}{{\ttfamily 0705.1496}}].

\bibitem{Giardino:2011aa}
P.P.~Giardino and P.~Lodone, \emph{{Threshold Corrections to Hard Supersymmetric Relations}}, \href{https://doi.org/10.1142/S0217732314501016}{\emph{Mod. Phys. Lett. A} {\bfseries 29} (2014) 1450101} [\href{https://arxiv.org/abs/1112.2635}{{\ttfamily 1112.2635}}].

\bibitem{Giudice:2011cg}
G.F.~Giudice and A.~Strumia, \emph{{Probing High-Scale and Split Supersymmetry with Higgs Mass Measurements}}, \href{https://doi.org/10.1016/j.nuclphysb.2012.01.001}{\emph{Nucl. Phys. B} {\bfseries 858} (2012) 63} [\href{https://arxiv.org/abs/1108.6077}{{\ttfamily 1108.6077}}].

\bibitem{Bagnaschi:2014rsa}
E.~Bagnaschi, G.F.~Giudice, P.~Slavich and A.~Strumia, \emph{{Higgs Mass and Unnatural Supersymmetry}}, \href{https://doi.org/10.1007/JHEP09(2014)092}{\emph{JHEP} {\bfseries 09} (2014) 092} [\href{https://arxiv.org/abs/1407.4081}{{\ttfamily 1407.4081}}].

\bibitem{Tamarit:2012ie}
C.~Tamarit, \emph{{Decoupling heavy sparticles in hierarchical SUSY scenarios: Two-loop Renormalization Group equations}},  \href{https://arxiv.org/abs/1204.2292}{{\ttfamily 1204.2292}}.

\bibitem{Benakli:2013msa}
K.~Benakli, L.~Darm\'e, M.D.~Goodsell and P.~Slavich, \emph{{A Fake Split Supersymmetry Model for the 126 GeV Higgs}}, \href{https://doi.org/10.1007/JHEP05(2014)113}{\emph{JHEP} {\bfseries 05} (2014) 113} [\href{https://arxiv.org/abs/1312.5220}{{\ttfamily 1312.5220}}].

\bibitem{Fox:2005yp}
P.J.~Fox, D.E.~Kaplan, E.~Katz, E.~Poppitz, V.~Sanz, M.~Schmaltz et~al., \emph{{Supersplit supersymmetry}},  \href{https://arxiv.org/abs/hep-th/0503249}{{\ttfamily hep-th/0503249}}.

\bibitem{Hall:2009nd}
L.J.~Hall and Y.~Nomura, \emph{{A Finely-Predicted Higgs Boson Mass from A Finely-Tuned Weak Scale}}, \href{https://doi.org/10.1007/JHEP03(2010)076}{\emph{JHEP} {\bfseries 03} (2010) 076} [\href{https://arxiv.org/abs/0910.2235}{{\ttfamily 0910.2235}}].

\bibitem{Cabrera:2011bi}
M.E.~Cabrera, J.A.~Casas and A.~Delgado, \emph{{Upper Bounds on Superpartner Masses from Upper Bounds on the Higgs Boson Mass}}, \href{https://doi.org/10.1103/PhysRevLett.108.021802}{\emph{Phys. Rev. Lett.} {\bfseries 108} (2012) 021802} [\href{https://arxiv.org/abs/1108.3867}{{\ttfamily 1108.3867}}].

\bibitem{Degrassi:2012ry}
G.~Degrassi, S.~Di~Vita, J.~Elias-Miro, J.R.~Espinosa, G.F.~Giudice, G.~Isidori et~al., \emph{{Higgs mass and vacuum stability in the Standard Model at NNLO}}, \href{https://doi.org/10.1007/JHEP08(2012)098}{\emph{JHEP} {\bfseries 08} (2012) 098} [\href{https://arxiv.org/abs/1205.6497}{{\ttfamily 1205.6497}}].

\bibitem{PardoVega:2015eno}
J.~Pardo~Vega and G.~Villadoro, \emph{{SusyHD: Higgs mass Determination in Supersymmetry}}, \href{https://doi.org/10.1007/JHEP07(2015)159}{\emph{JHEP} {\bfseries 07} (2015) 159} [\href{https://arxiv.org/abs/1504.05200}{{\ttfamily 1504.05200}}].

\bibitem{Bagnaschi:2017xid}
E.~Bagnaschi, J.~Pardo~Vega and P.~Slavich, \emph{{Improved determination of the Higgs mass in the MSSM with heavy superpartners}}, \href{https://doi.org/10.1140/epjc/s10052-017-4885-7}{\emph{Eur. Phys. J. C} {\bfseries 77} (2017) 334} [\href{https://arxiv.org/abs/1703.08166}{{\ttfamily 1703.08166}}].

\bibitem{Harlander:2018yhj}
R.V.~Harlander, J.~Klappert, A.D.~Ochoa~Franco and A.~Voigt, \emph{{The light CP-even MSSM Higgs mass resummed to fourth logarithmic order}}, \href{https://doi.org/10.1140/epjc/s10052-018-6351-6}{\emph{Eur. Phys. J. C} {\bfseries 78} (2018) 874} [\href{https://arxiv.org/abs/1807.03509}{{\ttfamily 1807.03509}}].

\bibitem{Bagnaschi:2019esc}
E.~Bagnaschi, G.~Degrassi, S.~Pa\ss{}ehr and P.~Slavich, \emph{{Full two-loop QCD corrections to the Higgs mass in the MSSM with heavy superpartners}}, \href{https://doi.org/10.1140/epjc/s10052-019-7417-9}{\emph{Eur. Phys. J. C} {\bfseries 79} (2019) 910} [\href{https://arxiv.org/abs/1908.01670}{{\ttfamily 1908.01670}}].

\bibitem{Bahl:2019wzx}
H.~Bahl, I.~Sobolev and G.~Weiglein, \emph{{Precise prediction for the mass of the light MSSM Higgs boson for the case of a heavy gluino}}, \href{https://doi.org/10.1016/j.physletb.2020.135644}{\emph{Phys. Lett. B} {\bfseries 808} (2020) 135644} [\href{https://arxiv.org/abs/1912.10002}{{\ttfamily 1912.10002}}].

\bibitem{Bahl:2020tuq}
H.~Bahl, I.~Sobolev and G.~Weiglein, \emph{{The light MSSM Higgs boson mass for large $\tan \beta $ and complex input parameters}}, \href{https://doi.org/10.1140/epjc/s10052-020-08637-w}{\emph{Eur. Phys. J. C} {\bfseries 80} (2020) 1063} [\href{https://arxiv.org/abs/2009.07572}{{\ttfamily 2009.07572}}].

\bibitem{Carena:2015uoe}
M.~Carena, J.~Ellis, J.S.~Lee, A.~Pilaftsis and C.E.M.~Wagner, \emph{{CP Violation in Heavy MSSM Higgs Scenarios}}, \href{https://doi.org/10.1007/JHEP02(2016)123}{\emph{JHEP} {\bfseries 02} (2016) 123} [\href{https://arxiv.org/abs/1512.00437}{{\ttfamily 1512.00437}}].

\bibitem{Murphy:2019qpm}
N.~Murphy and H.~Rzehak, \emph{{Higgs-boson masses and mixings in the MSSM with CP~violation and heavy SUSY particles}}, \href{https://doi.org/10.1140/epjc/s10052-022-11007-3}{\emph{Eur. Phys. J. C} {\bfseries 82} (2022) 1093} [\href{https://arxiv.org/abs/1909.00726}{{\ttfamily 1909.00726}}].

\bibitem{Gorbahn:2009pp}
M.~Gorbahn, S.~Jager, U.~Nierste and S.~Trine, \emph{{The supersymmetric Higgs sector and $B-\bar{B}$ mixing for large tan $\beta$}}, \href{https://doi.org/10.1103/PhysRevD.84.034030}{\emph{Phys. Rev. D} {\bfseries 84} (2011) 034030} [\href{https://arxiv.org/abs/0901.2065}{{\ttfamily 0901.2065}}].

\bibitem{Bahl:2018jom}
H.~Bahl and W.~Hollik, \emph{{Precise prediction of the MSSM Higgs boson masses for low M$_{A}$}}, \href{https://doi.org/10.1007/JHEP07(2018)182}{\emph{JHEP} {\bfseries 07} (2018) 182} [\href{https://arxiv.org/abs/1805.00867}{{\ttfamily 1805.00867}}].

\bibitem{Lee:2015uza}
G.~Lee and C.E.M.~Wagner, \emph{{Higgs bosons in heavy supersymmetry with an intermediate m$_A$}}, \href{https://doi.org/10.1103/PhysRevD.92.075032}{\emph{Phys. Rev. D} {\bfseries 92} (2015) 075032} [\href{https://arxiv.org/abs/1508.00576}{{\ttfamily 1508.00576}}].

\bibitem{Bahl:2020jaq}
H.~Bahl and I.~Sobolev, \emph{{Two-loop matching of renormalizable operators: general considerations and applications}}, \href{https://doi.org/10.1007/JHEP03(2021)286}{\emph{JHEP} {\bfseries 03} (2021) 286} [\href{https://arxiv.org/abs/2010.01989}{{\ttfamily 2010.01989}}].

\bibitem{BhupalDev:2014bir}
P.S.~Bhupal~Dev and A.~Pilaftsis, \emph{{Maximally Symmetric Two Higgs Doublet Model with Natural Standard Model Alignment}}, \href{https://doi.org/10.1007/JHEP12(2014)024}{\emph{JHEP} {\bfseries 12} (2014) 024} [\href{https://arxiv.org/abs/1408.3405}{{\ttfamily 1408.3405}}].

\bibitem{Bednyakov:2018cmx}
A.V.~Bednyakov, \emph{{On three-loop RGE for the Higgs sector of 2HDM}}, \href{https://doi.org/10.1007/JHEP11(2018)154}{\emph{JHEP} {\bfseries 11} (2018) 154} [\href{https://arxiv.org/abs/1809.04527}{{\ttfamily 1809.04527}}].

\bibitem{Schienbein:2018fsw}
I.~Schienbein, F.~Staub, T.~Steudtner and K.~Svirina, \emph{{Revisiting RGEs for general gauge theories}}, \href{https://doi.org/10.1016/j.nuclphysb.2018.12.001}{\emph{Nucl. Phys. B} {\bfseries 939} (2019) 1} [\href{https://arxiv.org/abs/1809.06797}{{\ttfamily 1809.06797}}].

\bibitem{Oredsson:2018yho}
J.~Oredsson and J.~Rathsman, \emph{{$\mathbb Z_2$ breaking effects in 2-loop RG evolution of 2HDM}}, \href{https://doi.org/10.1007/JHEP02(2019)152}{\emph{JHEP} {\bfseries 02} (2019) 152} [\href{https://arxiv.org/abs/1810.02588}{{\ttfamily 1810.02588}}].

\bibitem{Herren:2017uxn}
F.~Herren, L.~Mihaila and M.~Steinhauser, \emph{{Gauge and Yukawa coupling beta functions of two-Higgs-doublet models to three-loop order}}, \href{https://doi.org/10.1103/PhysRevD.97.015016}{\emph{Phys. Rev. D} {\bfseries 97} (2018) 015016} [\href{https://arxiv.org/abs/1712.06614}{{\ttfamily 1712.06614}}].

\bibitem{Bagnaschi:2015pwa}
E.~Bagnaschi, F.~Br\"ummer, W.~Buchm\"uller, A.~Voigt and G.~Weiglein, \emph{{Vacuum stability and supersymmetry at high scales with two Higgs doublets}}, \href{https://doi.org/10.1007/JHEP03(2016)158}{\emph{JHEP} {\bfseries 03} (2016) 158} [\href{https://arxiv.org/abs/1512.07761}{{\ttfamily 1512.07761}}].

\bibitem{Bagnaschi:2015hka}
E.~Bagnaschi et~al., \emph{{Benchmark scenarios for low $\tan \beta$ in the MSSM}},  LHCHXSWG-2015-002 (2015).

\bibitem{Bahl:2019ago}
H.~Bahl, S.~Liebler and T.~Stefaniak, \emph{{MSSM Higgs benchmark scenarios for Run 2 and beyond: the low $\tan \beta $ region}}, \href{https://doi.org/10.1140/epjc/s10052-019-6770-z}{\emph{Eur. Phys. J. C} {\bfseries 79} (2019) 279} [\href{https://arxiv.org/abs/1901.05933}{{\ttfamily 1901.05933}}].

\bibitem{Cheung:2014hya}
K.~Cheung, R.~Huo, J.S.~Lee and Y.-L.~Sming~Tsai, \emph{{Dark Matter in Split SUSY with Intermediate Higgses}}, \href{https://doi.org/10.1007/JHEP04(2015)151}{\emph{JHEP} {\bfseries 04} (2015) 151} [\href{https://arxiv.org/abs/1411.7329}{{\ttfamily 1411.7329}}].

\bibitem{Kwasnitza:2021idg}
T.~Kwasnitza and D.~St\"ockinger, \emph{{Resummation of terms enhanced by trilinear squark-Higgs couplings in the MSSM}}, \href{https://doi.org/10.1007/JHEP08(2021)070}{\emph{JHEP} {\bfseries 08} (2021) 070} [\href{https://arxiv.org/abs/2103.08616}{{\ttfamily 2103.08616}}].

\bibitem{Bahl:2016brp}
H.~Bahl and W.~Hollik, \emph{{Precise prediction for the light MSSM Higgs boson mass combining effective field theory and fixed-order calculations}}, \href{https://doi.org/10.1140/epjc/s10052-016-4354-8}{\emph{Eur. Phys. J. C} {\bfseries 76} (2016) 499} [\href{https://arxiv.org/abs/1608.01880}{{\ttfamily 1608.01880}}].

\bibitem{Bahl:2017aev}
H.~Bahl, S.~Heinemeyer, W.~Hollik and G.~Weiglein, \emph{{Reconciling EFT and hybrid calculations of the light MSSM Higgs-boson mass}}, \href{https://doi.org/10.1140/epjc/s10052-018-5544-3}{\emph{Eur. Phys. J. C} {\bfseries 78} (2018) 57} [\href{https://arxiv.org/abs/1706.00346}{{\ttfamily 1706.00346}}].

\bibitem{Bahl:2018ykj}
H.~Bahl, \emph{{Pole mass determination in presence of heavy particles}}, \href{https://doi.org/10.1007/JHEP02(2019)121}{\emph{JHEP} {\bfseries 02} (2019) 121} [\href{https://arxiv.org/abs/1812.06452}{{\ttfamily 1812.06452}}].

\bibitem{Bahl:2019hmm}
H.~Bahl, S.~Heinemeyer, W.~Hollik and G.~Weiglein, \emph{{Theoretical uncertainties in the MSSM Higgs boson mass calculation}}, \href{https://doi.org/10.1140/epjc/s10052-020-8079-3}{\emph{Eur. Phys. J. C} {\bfseries 80} (2020) 497} [\href{https://arxiv.org/abs/1912.04199}{{\ttfamily 1912.04199}}].

\bibitem{Bahl:2020mjy}
H.~Bahl, N.~Murphy and H.~Rzehak, \emph{{Hybrid calculation of the MSSM Higgs boson masses using the complex THDM as EFT}}, \href{https://doi.org/10.1140/epjc/s10052-021-08939-7}{\emph{Eur. Phys. J. C} {\bfseries 81} (2021) 128} [\href{https://arxiv.org/abs/2010.04711}{{\ttfamily 2010.04711}}].

\bibitem{Bagnaschi:2018ofa}
E.~Bagnaschi et~al., \emph{{MSSM Higgs Boson Searches at the LHC: Benchmark Scenarios for Run 2 and Beyond}}, \href{https://doi.org/10.1140/epjc/s10052-019-7114-8}{\emph{Eur. Phys. J. C} {\bfseries 79} (2019) 617} [\href{https://arxiv.org/abs/1808.07542}{{\ttfamily 1808.07542}}].

\bibitem{Sobolev:2020cjh}
I.~Sobolev, \emph{{Precise predictions for Higgs physics in supersymmetric models}}, Ph.D. thesis, Hamburg U., 2020.
\newblock 10.3204/PUBDB-2020-02962.

\bibitem{Athron:2016fuq}
P.~Athron, J.-h.~Park, T.~Steudtner, D.~St\"ockinger and A.~Voigt, \emph{{Precise Higgs mass calculations in (non-)minimal supersymmetry at both high and low scales}}, \href{https://doi.org/10.1007/JHEP01(2017)079}{\emph{JHEP} {\bfseries 01} (2017) 079} [\href{https://arxiv.org/abs/1609.00371}{{\ttfamily 1609.00371}}].

\bibitem{Staub:2017jnp}
F.~Staub and W.~Porod, \emph{{Improved predictions for intermediate and heavy Supersymmetry in the MSSM and beyond}}, \href{https://doi.org/10.1140/epjc/s10052-017-4893-7}{\emph{Eur. Phys. J. C} {\bfseries 77} (2017) 338} [\href{https://arxiv.org/abs/1703.03267}{{\ttfamily 1703.03267}}].

\bibitem{Athron:2017fvs}
P.~Athron, M.~Bach, D.~Harries, T.~Kwasnitza, J.-h.~Park, D.~St{\"o}ckinger et~al., \emph{{FlexibleSUSY 2.0: Extensions to investigate the phenomenology of SUSY and non-SUSY models}}, \href{https://doi.org/10.1016/j.cpc.2018.04.016}{\emph{Comput. Phys. Commun.} {\bfseries 230} (2018) 145} [\href{https://arxiv.org/abs/1710.03760}{{\ttfamily 1710.03760}}].

\bibitem{Kwasnitza:2020wli}
T.~Kwasnitza, D.~St\"ockinger and A.~Voigt, \emph{{Improved MSSM Higgs mass calculation using the 3-loop FlexibleEFTHiggs approach including $x_{t}$-resummation}}, \href{https://doi.org/10.1007/JHEP07(2020)197}{\emph{JHEP} {\bfseries 07} (2020) 197} [\href{https://arxiv.org/abs/2003.04639}{{\ttfamily 2003.04639}}].

\bibitem{Harlander:2019dge}
R.V.~Harlander, J.~Klappert and A.~Voigt, \emph{{The light CP-even MSSM Higgs mass including N$^\mathbf {3}$LO+N$^\mathbf {3}$LL QCD corrections}}, \href{https://doi.org/10.1140/epjc/s10052-020-7747-7}{\emph{Eur. Phys. J. C} {\bfseries 80} (2020) 186} [\href{https://arxiv.org/abs/1910.03595}{{\ttfamily 1910.03595}}].

\bibitem{Slavich:2020zjv}
P.~Slavich et~al., \emph{{Higgs-mass predictions in the MSSM and beyond}}, \href{https://doi.org/10.1140/epjc/s10052-021-09198-2}{\emph{Eur. Phys. J. C} {\bfseries 81} (2021) 450} [\href{https://arxiv.org/abs/2012.15629}{{\ttfamily 2012.15629}}].

\bibitem{Weiglein:1998jz}
G.~Weiglein, \emph{{Results for precision observables in the electroweak standard model at two loop order and beyond}}, {\emph{Acta Phys. Polon. B} {\bfseries 29} (1998) 2735} [\href{https://arxiv.org/abs/hep-ph/9807222}{{\ttfamily hep-ph/9807222}}].

\bibitem{AchimDipl}
A.~Stremplat, ``diploma thesis.'' Univ. of Karlsruhe, 1998.

\bibitem{Chen:2020xzx}
L.~Chen and A.~Freitas, \emph{{Leading fermionic three-loop corrections to electroweak precision observables}}, \href{https://doi.org/10.1007/JHEP07(2020)210}{\emph{JHEP} {\bfseries 07} (2020) 210} [\href{https://arxiv.org/abs/2002.05845}{{\ttfamily 2002.05845}}].

\bibitem{Chen:2020xot}
L.~Chen and A.~Freitas, \emph{{Mixed EW-QCD leading fermionic three-loop corrections at $\mathcal{O}(\alpha_s\alpha^2)$ to electroweak precision observables}}, \href{https://doi.org/10.1007/JHEP03(2021)215}{\emph{JHEP} {\bfseries 03} (2021) 215} [\href{https://arxiv.org/abs/2012.08605}{{\ttfamily 2012.08605}}].

\bibitem{Meuser:2023}
D.~Meuser, \emph{Complete $\mathcal{O}(N_c^2)$ two-loop contributions to the Higgs boson masses in the MSSM and aspects of two-loop renormalisation}, Ph.D. thesis, University of Hamburg, 2023.

\bibitem{Peccei:1977hh}
R.D.~Peccei and H.R.~Quinn, \emph{{CP Conservation in the Presence of Instantons}}, \href{https://doi.org/10.1103/PhysRevLett.38.1440}{\emph{Phys. Rev. Lett.} {\bfseries 38} (1977) 1440}.

\bibitem{Peccei:1977ur}
R.D.~Peccei and H.R.~Quinn, \emph{{Constraints Imposed by CP Conservation in the Presence of Instantons}}, \href{https://doi.org/10.1103/PhysRevD.16.1791}{\emph{Phys. Rev. D} {\bfseries 16} (1977) 1791}.

\bibitem{Dimopoulos:1995kn}
S.~Dimopoulos and S.D.~Thomas, \emph{{Dynamical relaxation of the supersymmetric CP violating phases}}, \href{https://doi.org/10.1016/0550-3213(96)00065-X}{\emph{Nucl. Phys. B} {\bfseries 465} (1996) 23} [\href{https://arxiv.org/abs/hep-ph/9510220}{{\ttfamily hep-ph/9510220}}].

\bibitem{Drees:2004jm}
M.~Drees, R.M.~Godbole and P.~Roy, \emph{{Theory and Phenomenology of Sparticles: An account of four-dimensional N=1 supersymmetry in High Energy Physics}}, World Scientific Publishing Co.\ Pte.\ Ltd.\ (2004).

\bibitem{Peskin:1995ev}
M.E.~Peskin and D.V.~Schroeder, \emph{{An Introduction to Quantum Field Theory}}, Addison-Wesley, Reading, USA (1995).

\bibitem{Denner:1991kt}
A.~Denner, \emph{{Techniques for calculation of electroweak radiative corrections at the one loop level and results for W physics at LEP-200}}, \href{https://doi.org/10.1002/prop.2190410402}{\emph{Fortsch. Phys.} {\bfseries 41} (1993) 307} [\href{https://arxiv.org/abs/0709.1075}{{\ttfamily 0709.1075}}].

\bibitem{Bharucha:2012nx}
A.~Bharucha, A.~Fowler, G.~Moortgat-Pick and G.~Weiglein, \emph{{Consistent on shell renormalisation of electroweakinos in the complex MSSM: LHC and LC predictions}}, \href{https://doi.org/10.1007/JHEP05(2013)053}{\emph{JHEP} {\bfseries 05} (2013) 053} [\href{https://arxiv.org/abs/1211.3134}{{\ttfamily 1211.3134}}].

\bibitem{Degrassi:2014sxa}
G.~Degrassi, P.~Gambino and P.P.~Giardino, \emph{{The $m_{\scriptscriptstyle W}-m_{\scriptscriptstyle Z}$ interdependence in the Standard Model: a new scrutiny}}, \href{https://doi.org/10.1007/JHEP05(2015)154}{\emph{JHEP} {\bfseries 05} (2015) 154} [\href{https://arxiv.org/abs/1411.7040}{{\ttfamily 1411.7040}}].

\bibitem{Bauberger:1997zz}
S.~Bauberger, \emph{{Two-loop contributions to muon decay}}, Ph.D. thesis, W{\"u}rzburg U., 1997.

\bibitem{Freitas:2002ja}
A.~Freitas, W.~Hollik, W.~Walter and G.~Weiglein, \emph{{Electroweak two loop corrections to the $M_W-M_Z$ mass correlation in the standard model}}, \href{https://doi.org/10.1016/S0550-3213(02)00243-2}{\emph{Nucl. Phys. B} {\bfseries 632} (2002) 189} [\href{https://arxiv.org/abs/hep-ph/0202131}{{\ttfamily hep-ph/0202131}}].

\bibitem{Awramik:2002vu}
M.~Awramik, M.~Czakon, A.~Onishchenko and O.~Veretin, \emph{{Bosonic corrections to Delta r at the two loop level}}, \href{https://doi.org/10.1103/PhysRevD.68.053004}{\emph{Phys. Rev. D} {\bfseries 68} (2003) 053004} [\href{https://arxiv.org/abs/hep-ph/0209084}{{\ttfamily hep-ph/0209084}}].

\bibitem{Dittmaier:2021loa}
S.~Dittmaier, \emph{{Electric charge renormalization to all orders}}, \href{https://doi.org/10.1103/PhysRevD.103.053006}{\emph{Phys. Rev. D} {\bfseries 103} (2021) 053006} [\href{https://arxiv.org/abs/2101.05154}{{\ttfamily 2101.05154}}].

\bibitem{Fuchs:2016swt}
E.~Fuchs and G.~Weiglein, \emph{{Breit-Wigner approximation for propagators of mixed unstable states}}, \href{https://doi.org/10.1007/JHEP09(2017)079}{\emph{JHEP} {\bfseries 09} (2017) 079} [\href{https://arxiv.org/abs/1610.06193}{{\ttfamily 1610.06193}}].

\bibitem{Hessenberger:2018xzo}
S.~Hessenberger, \emph{{Two-loop corrections to electroweak precision observables in Two-Higgs-Doublet-Models}}, Ph.D. thesis, Munich, Tech. U., 2018.

\bibitem{Freitas:2002pe}
A.~Freitas and D.~St{\"o}ckinger, \emph{{Gauge dependence and renormalization of tan beta}},  in \emph{{10th International Conference on Supersymmetry and Unification of Fundamental Interactions (SUSY02)}}, pp.~657--661, 10, 2002 [\href{https://arxiv.org/abs/hep-ph/0210372}{{\ttfamily hep-ph/0210372}}].

\bibitem{Freitas:2002um}
A.~Freitas and D.~St{\"o}ckinger, \emph{{Gauge dependence and renormalization of tan beta in the MSSM}}, \href{https://doi.org/10.1103/PhysRevD.66.095014}{\emph{Phys. Rev. D} {\bfseries 66} (2002) 095014} [\href{https://arxiv.org/abs/hep-ph/0205281}{{\ttfamily hep-ph/0205281}}].

\bibitem{Baro:2008bg}
N.~Baro, F.~Boudjema and A.~Semenov, \emph{{Automatised full one-loop renormalisation of the MSSM. I. The Higgs sector, the issue of tan(beta) and gauge invariance}}, \href{https://doi.org/10.1103/PhysRevD.78.115003}{\emph{Phys. Rev. D} {\bfseries 78} (2008) 115003} [\href{https://arxiv.org/abs/0807.4668}{{\ttfamily 0807.4668}}].

\bibitem{Passehr:2014xwu}
S.~Pa{\ss}ehr, \emph{{Two-Loop Corrections to the Higgs-Boson Masses in the Minimal Supersymmetric Standard Model with CP-Violation}}, Ph.D. thesis, Munich, Tech. U., 2014.

\bibitem{Bahl:2022kzs}
H.~Bahl, J.~Braathen and G.~Weiglein, \emph{{Theoretical concepts and measurement prospects for BSM trilinear couplings: a case study for scalar top quarks}}, \href{https://doi.org/10.1140/epjc/s10052-023-11839-7}{\emph{Eur. Phys. J. C} {\bfseries 83} (2023) 685} [\href{https://arxiv.org/abs/2212.11213}{{\ttfamily 2212.11213}}].

\bibitem{Bahl:2021rts}
H.~Bahl, J.~Braathen and G.~Weiglein, \emph{{External leg corrections as an origin of large logarithms}}, \href{https://doi.org/10.1007/JHEP02(2022)159}{\emph{JHEP} {\bfseries 02} (2022) 159} [\href{https://arxiv.org/abs/2112.11419}{{\ttfamily 2112.11419}}].

\bibitem{Kublbeck:1990xc}
J.~K{\"u}blbeck, M.~B{\"o}hm and A.~Denner, \emph{{Feyn Arts: Computer Algebraic Generation of Feynman Graphs and Amplitudes}}, \href{https://doi.org/10.1016/0010-4655(90)90001-H}{\emph{Comput. Phys. Commun.} {\bfseries 60} (1990) 165}.

\bibitem{Eck:1992ms}
H.~Eck and J.~Kublbeck, \emph{{Computeralgebraic generation of Feynman graphs and amplitudes}},  in \emph{{2nd International Workshop on Software Engineering, Artificial Intelligence and Expert Systems for High-energy and Nuclear Physics}}, 1992.

\bibitem{Hahn:2000kx}
T.~Hahn, \emph{{Generating Feynman diagrams and amplitudes with FeynArts 3}}, \href{https://doi.org/10.1016/S0010-4655(01)00290-9}{\emph{Comput. Phys. Commun.} {\bfseries 140} (2001) 418} [\href{https://arxiv.org/abs/hep-ph/0012260}{{\ttfamily hep-ph/0012260}}].

\bibitem{Hahn:1998yk}
T.~Hahn and M.~Perez-Victoria, \emph{{Automatized one loop calculations in four-dimensions and D-dimensions}}, \href{https://doi.org/10.1016/S0010-4655(98)00173-8}{\emph{Comput. Phys. Commun.} {\bfseries 118} (1999) 153} [\href{https://arxiv.org/abs/hep-ph/9807565}{{\ttfamily hep-ph/9807565}}].

\bibitem{Hahn:2016ebn}
T.~Hahn, S.~Pa\ss{}ehr and C.~Schappacher, \emph{{FormCalc 9 and Extensions}}, \href{https://doi.org/10.1088/1742-6596/762/1/012065}{\emph{PoS} {\bfseries LL2016} (2016) 068} [\href{https://arxiv.org/abs/1604.04611}{{\ttfamily 1604.04611}}].

\bibitem{Weiglein:1993hd}
G.~Weiglein, R.~Scharf and M.~B{\"o}hm, \emph{{Reduction of general two loop selfenergies to standard scalar integrals}}, \href{https://doi.org/10.1016/0550-3213(94)90325-5}{\emph{Nucl. Phys. B} {\bfseries 416} (1994) 606} [\href{https://arxiv.org/abs/hep-ph/9310358}{{\ttfamily hep-ph/9310358}}].

\bibitem{Weiglein:1995qs}
G.~Weiglein, R.~Mertig, R.~Scharf and M.~B{\"o}hm, \emph{{Computer algebraic calculation of two loop selfenergies in the electroweak standard model}},  in \emph{{2nd International Workshop on Software Engineering, Artificial Intelligence and Expert Systems for High-energy and Nuclear Physics}}, 5, 1995.

\bibitem{Schwartz:2014sze}
M.D.~Schwartz, \emph{{Quantum Field Theory and the Standard Model}}, Cambridge University Press (3, 2014).

\bibitem{Haber:1997if}
H.E.~Haber, \emph{{The Status of the minimal supersymmetric standard model and beyond}}, \href{https://doi.org/10.1016/S0920-5632(97)00688-9}{\emph{Nucl. Phys. B Proc. Suppl.} {\bfseries 62} (1998) 469} [\href{https://arxiv.org/abs/hep-ph/9709450}{{\ttfamily hep-ph/9709450}}].

\bibitem{ParticleDataGroup:2020ssz}
{\scshape Particle Data Group} collaboration, \emph{{Review of Particle Physics}}, \href{https://doi.org/10.1093/ptep/ptaa104}{\emph{PTEP} {\bfseries 2020} (2020) 083C01}.

\bibitem{Steinhauser:1998rq}
M.~Steinhauser, \emph{{Leptonic contribution to the effective electromagnetic coupling constant up to three loops}}, \href{https://doi.org/10.1016/S0370-2693(98)00503-6}{\emph{Phys. Lett. B} {\bfseries 429} (1998) 158} [\href{https://arxiv.org/abs/hep-ph/9803313}{{\ttfamily hep-ph/9803313}}].

\bibitem{Carena:1999xa}
M.~Carena, S.~Heinemeyer, C.E.M.~Wagner and G.~Weiglein, \emph{{Suggestions for improved benchmark scenarios for Higgs boson searches at LEP-2}},  in \emph{{Workshop on New Theoretical Developments for Higgs Physics at LEP-2}}, 12, 1999 [\href{https://arxiv.org/abs/hep-ph/9912223}{{\ttfamily hep-ph/9912223}}].

\bibitem{Carena:2002qg}
M.~Carena, S.~Heinemeyer, C.E.M.~Wagner and G.~Weiglein, \emph{{Suggestions for Benchmark Scenarios for MSSM Higgs Boson Searches at Hadron Colliders}}, \href{https://doi.org/10.1140/epjc/s2002-01084-3}{\emph{Eur. Phys. J. C} {\bfseries 26} (2003) 601} [\href{https://arxiv.org/abs/hep-ph/0202167}{{\ttfamily hep-ph/0202167}}].

\bibitem{Heinemeyer:1998yj}
S.~Heinemeyer, W.~Hollik and G.~Weiglein, \emph{{FeynHiggs: A Program for the calculation of the masses of the neutral CP even Higgs bosons in the MSSM}}, \href{https://doi.org/10.1016/S0010-4655(99)00364-1}{\emph{Comput. Phys. Commun.} {\bfseries 124} (2000) 76} [\href{https://arxiv.org/abs/hep-ph/9812320}{{\ttfamily hep-ph/9812320}}].

\bibitem{Hahn:2009zz}
T.~Hahn, S.~Heinemeyer, W.~Hollik, H.~Rzehak and G.~Weiglein, \emph{{FeynHiggs: A program for the calculation of MSSM Higgs-boson observables - Version 2.6.5}}, {\emph{Comput. Phys. Commun.} {\bfseries 180} (2009) 1426}.

\bibitem{Bahl:2018qog}
H.~Bahl, T.~Hahn, S.~Heinemeyer, W.~Hollik, S.~Pa\ss{}ehr, H.~Rzehak et~al., \emph{{Precision calculations in the MSSM Higgs-boson sector with FeynHiggs 2.14}}, \href{https://doi.org/10.1016/j.cpc.2019.107099}{\emph{Comput. Phys. Commun.} {\bfseries 249} (2020) 107099} [\href{https://arxiv.org/abs/1811.09073}{{\ttfamily 1811.09073}}].

\bibitem{Sperling:2013fwy}
M.~Sperling, \emph{{Renormierung von Vakuumerwartungswerten in spontan gebrochenen Eichtheorien}},  diploma thesis, Dresden, Tech. U., 2013.

\bibitem{Collins:2011zzd}
J.~Collins, \emph{{Foundations of perturbative QCD}}, vol.~32, Cambridge University Press (11, 2013).

\bibitem{Hahn:LT215}
T.~Hahn, ``Looptools 2.15 user’s guide.'' \url{https://feynarts.de/looptools/LT215Guide.pdf}, 2018.

\bibitem{Rzehak:2005zz}
H.A.~Rzehak, \emph{{Zwei-Schleifen-Beitr{\"a}ge im supersymmetrischen Higgs-Sektor}},  dissertation, Munich Tech. U., 6, 2005.

\bibitem{Dabelstein:1995js}
A.~Dabelstein, \emph{{Fermionic decays of neutral MSSM Higgs bosons at the one loop level}}, \href{https://doi.org/10.1016/0550-3213(95)00523-2}{\emph{Nucl. Phys. B} {\bfseries 456} (1995) 25} [\href{https://arxiv.org/abs/hep-ph/9503443}{{\ttfamily hep-ph/9503443}}].

\bibitem{Logan:2002jh}
H.E.~Logan and S.-f.~Su, \emph{{Associated Production of $H^{\pm}$ and $W^{\mp}$ in High-Energy $e^+ e^-$ Collisions in the Minimal Supersymmetric Standard Model}}, \href{https://doi.org/10.1103/PhysRevD.66.035001}{\emph{Phys. Rev. D} {\bfseries 66} (2002) 035001} [\href{https://arxiv.org/abs/hep-ph/0203270}{{\ttfamily hep-ph/0203270}}].

\bibitem{Williams:2011bu}
K.E.~Williams, H.~Rzehak and G.~Weiglein, \emph{{Higher order corrections to Higgs boson decays in the MSSM with complex parameters}}, \href{https://doi.org/10.1140/epjc/s10052-011-1669-3}{\emph{Eur. Phys. J. C} {\bfseries 71} (2011) 1669} [\href{https://arxiv.org/abs/1103.1335}{{\ttfamily 1103.1335}}].

\end{thebibliography}\endgroup

\end{document}